\newif\ifdraftpaper \draftpaperfalse
\ifdraftpaper
  \documentclass[journal,romanappendices,draftcls,onecolumn,12pt]{IEEEtran} 
  
\else
  \documentclass[journal,romanappendices]{IEEEtran} 
  
\fi
\usepackage[cmex10]{amsmath}
\usepackage{amsfonts}
\usepackage{amssymb}
\usepackage{array}
\usepackage{trace}
\usepackage{units}
\usepackage{cite}
\usepackage{graphicx}
\usepackage[tight,footnotesize]{subfigure} 
\usepackage{url}
\usepackage{bm}
\usepackage{algorithmic}
\usepackage{booktabs}
\usepackage{multirow}
\usepackage{prettyref}
\usepackage{environ}
\ifdraftpaper
  \newcommand{\finalcr}{} 
  \NewEnviron{multlinefinal}{\begin{equation}\BODY\end{equation}}
\else
  \newcommand{\finalcr}{\\}
  \NewEnviron{multlinefinal}{\begin{multline}\BODY\end{multline}}
\fi
\AtBeginDocument{
  \newcommand{\defeq}{\triangleq}
\newcommand{\condmid}{\,|\,} 
\newcommand{\Bigcondmid}{\,\Big|\,}
\newcommand{\e}{e} 
\newcommand{\seq}[1]{\bm{#1}}
\newcommand{\seqs}[1]{\bm{#1}}

\newcommand{\ct}{\tilde{c}} 

\newcommand{\mat}[1]{\bm{#1}}
\newcommand{\abs}[1]{\left\lvert #1 \right\rvert}

\newcommand{\cardinal}[1]{\left\lvert #1 \right\rvert}
\newcommand{\norm}[1]{\left\lVert #1 \right\rVert}
\newcommand{\asymptO}{O} 

\newcommand{\SetR}{\mathbb{R}}
\newcommand{\SetZ}{\mathbb{Z}}

\DeclareMathOperator*{\argmax}{arg\,max}

\newcommand{\bmodI}{_{\mathcal{I}}}
\newcommand{\bmodY}{_{\mathcal{Y}}}

\newcommand{\db}{d_{\node{b}}}
\newcommand{\dc}{d_{\node{c}}}

\newcommand{\wn}[1]{w_{#1}^{(n)}}
\newcommand{\seqw}{\seq{w}}
\newcommand{\seqwn}{\seq{w}^{(n)}}
\newcommand{\Rn}{R^{(n)}}

\newcommand{\node}[1]{\ensuremath{\mathsf{#1}}}
\newcommand{\llr}{\mathfrak{l}}
\newcommand{\dmu}{\bm{\mu}}
\newcommand{\dlambda}{\bm{\lambda}}
\newcommand{\dnu}{\bm{\nu}}
\newcommand{\msg}[2]{\mu^{\node{#1}}_{#2}} 
\newcommand{\msgx}[3]{\mu^{\node{#1}}_{#2}(#3)} 
\newcommand{\msgiter}[3]{\mu^{\node{#1}}_{#2(#3)}}
\newcommand{\msgsiter}[2]{\msgiter{#1}{}{#2}}
\newcommand{\priprob}[2]{\lambda^{\node{#1}}_{#2}}
\newcommand{\priprobx}[3]{\priprob{#1}{#2}(#3)}

\newcommand{\pris}[1]{\lambda^{\node{#1}}_{*}}
\newcommand{\prisx}[2]{\lambda^{\node{#1}}_{\except #2}}
\newcommand{\tpriprob}[2]{\tilde{\lambda}^{\node{#1}}_{#2}}
\newcommand{\tpriprobx}[3]{\tpriprob{#1}{#2}(#3)}
\newcommand{\tpris}[1]{\tilde{\lambda}^{\node{#1}}_{*}}
\newcommand{\tprisx}[2]{\tilde{\lambda}^{\node{#1}}_{\except #2}}
\newcommand{\extprob}[2]{\nu^{\node{#1}}_{#2}}
\newcommand{\textprob}[2]{\tilde{\nu}^{\node{#1}}_{#2}}
\newcommand{\extprobx}[3]{\extprob{#1}{#2}(#3)}
\newcommand{\textprobx}[3]{\textprob{#1}{#2}(#3)}

\newcommand{\extbiu}[1]{\overline{\nu}^{\node{b}}_{i(#1)}}
\newcommand{\extbil}[1]{\underline{\nu}^{\node{b}}_{i(#1)}}
\newcommand{\extaju}[1]{\overline{\nu}^{\node{a}}_{j(#1)}}
\newcommand{\trueextprob}[2]{\nu^{\node{#1}*}_{#2}}
\newcommand{\trueextprobx}[3]{\trueextprob{#1}{#2}(#3)}
\newcommand{\dmsg}[1]{\dmu^{\node{#1}}}
\newcommand{\dpriprob}[1]{\dlambda^{\node{#1}}}

\newcommand{\dtrueextprob}[1]{\dnu^{\node{#1}*}}
\newcommand{\dextbiu}[1]{\overline{\dnu}^{\node{b}}_{i(#1)}}

\newcommand{\dextbuI}[1]{\overline{\dnu}^{\node{b}}_{(\Infopri{b},#1)}}
\newcommand{\dextbuIx}[2]{\overline{\dnu}^{\node{b}}_{(#1,#2)}}
\newcommand{\dextblI}[1]{\underline{\dnu}^{\node{b}}_{(\Infopri{b},#1)}}

\newcommand{\dextbil}[1]{\underline{\dnu}^{\node{b}}_{i(#1)}}
\newcommand{\dextbit}{\dtrueextprob{b}_i}

\newcommand{\dextaju}[1]{\overline{\dnu}^{\node{a}}_{j(#1)}}

\newcommand{\dextauI}[1]{\overline{\dnu}^{\node{a}}_{(\Infopri{a},#1)}}
\newcommand{\dextauIx}[2]{\overline{\dnu}^{\node{a}}_{(#1,#2)}}
\newcommand{\dextajt}{\dtrueextprob{a}_j}

\newcommand{\dmsgz}[1]{\dmsg{#1}_{(0)}}
\newcommand{\dmsgl}[1]{\dmsg{#1}_{(\llr)}}
\newcommand{\dmsgiter}[2]{\dmsg{#1}_{(#2)}}
\newcommand{\constmsg}[1]{\overline{#1}}
\newcommand{\neighbor}[2]{\mathcal{N}^{\node{#1}}_{#2}} 
\newcommand{\Info}[1]{I_{\node{#1}}} 
\newcommand{\Infox}[2]{I_{\node{#1},#2}}
\newcommand{\nextInfo}[1]{I^{+}_{\node{#1}}}
\newcommand{\Ibext}{\Infox{b}{\mathrm{ext}}}
\newcommand{\Ibextu}{\overline{I}_{\node{b},\mathrm{ext}}}
\newcommand{\Ibextl}{\underline{I}_{\node{b},\mathrm{ext}}}
\newcommand{\Iaext}{I_{\node{a},\mathrm{ext}}}
\newcommand{\Iaextu}{\overline{I}_{\node{a},\mathrm{ext}}}
\newcommand{\IbextuI}{\Ibextu^{(\Infopri{b})}}
\newcommand{\IbextlI}{\Ibextl^{(\Infopri{b})}}
\newcommand{\IbextuIx}[1]{\Ibextu^{(#1)}}
\newcommand{\IbextlIx}[1]{\Ibextl^{(#1)}}
\newcommand{\IaextuI}{\Iaextu^{(\Infopri{a})}}
\newcommand{\IaextuIx}[1]{\Iaextu^{(#1)}}
\newcommand{\IbextuIL}{\Ibextu^{(\Infopri{b},L)}}

\newcommand{\IbextlIL}{\Ibextl^{(\Infopri{b},L)}}
\newcommand{\IbextxIL}{\Ibext^{(\Infopri{b},L)}} 

\newcommand{\IaextuIL}{\Iaextu^{(\Infopri{a},L)}}

\newcommand{\Infoiter}[2]{I_{\node{#1}}^{(#2)}}
\newcommand{\Infopriiter}[2]{\Infopri{#1}^{(#2)}}
\newcommand{\InfoinfuI}[1]{\overline{I}_{\node{#1}}^{(\Infopri{b},\infty)}}
\newcommand{\InfoinflI}[1]{\underline{I}_{\node{#1}}^{(\Infopri{b},\infty)}}
\newcommand{\Ibextiter}[1]{I_{\node{b},\mathrm{ext}}^{(#1)}}

\newcommand{\Iuthr}{\Info{u}^{\mathrm{thr}}}

\newcommand{\assign}{\Leftarrow}

\newcommand{\excluding}[1]{\backslash \{#1\}}
\newcommand{\nofootnote}[1]{} 

\newcommand{\nb}{n_{\mathrm{b}}}

\newcommand{\nc}{n_{\mathrm{c}}}

\newcommand{\GF}{\mathrm{GF}}

\newcommand{\oneif}[1]{\mathsf{1}\left[#1 \right]} 

\newcommand{\qed}{\IEEEQED} 
\renewcommand{\Pr}[1]{\mathrm{Pr}\left[#1 \right]} 
\newcommand{\Expe}[1]{\mathrm{E}\left[#1 \right]}
\newcommand{\Expex}[2]{\mathrm{E}_{#1}\left[#2 \right]}
\newcommand{\omegab}[1]{\omega^{\node{b}}_{#1}}
\newcommand{\omegaa}[1]{\omega^{\node{a}}_{#1}}
\newcommand{\omegabs}{\seq{\omega}^{\node{b}}}
\newcommand{\omegaas}{\seq{\omega}^{\node{a}}}
\newcommand{\oplusl}{\bigoplus\limits}
\newcommand{\odotl}{\bigodot\limits}
\newcommand{\opluspow}[2]{#1^{\oplus(#2)}}
\newcommand{\odotpow}[2]{#1^{\odot(#2)}}
\newcommand{\markov}{\,\text{---}\,}
\newcommand{\except}{\sim}
\newcommand{\neighi}{\mathcal{N}_i}
\newcommand{\neighiL}{\neighi^{(L)}}
\newcommand{\neighio}{\neighi^{\circ}}
\newcommand{\neighib}{\neighi^{-}}
\newcommand{\neighix}{\neighi^{+}} 
\newcommand{\neighj}{\mathcal{N}_j}
\newcommand{\neighjL}{\neighj^{(L)}}
\newcommand{\neighjo}{\neighj^{\circ}}
\newcommand{\neighjb}{\neighj^{-}}
\newcommand{\neighjx}{\neighj^{+}}
\newcommand{\lepd}{\preceq} 

\newcommand{\seqb}{\seq{b}}
\newcommand{\seqbr}{\seqb^*} 
\newcommand{\seqa}{\seq{a}}
\newcommand{\seqc}{\seq{c}}
\newcommand{\seqcr}{\seqc^*}
\newcommand{\seqar}{\seqa^*}
\newcommand{\sequ}{\seq{u}}
\newcommand{\sequr}{\sequ^*}
\newcommand{\seqy}{\seq{y}} 

\newcommand{\seqdelta}{\seq{\delta}} 
\newcommand{\seqeps}{\seq{\epsilon}} 
\newcommand{\pyb}{p_{y\condmid b}}
\newcommand{\pzb}{p_{z\condmid b}}
\newcommand{\dz}{\Delta z}
\newcommand{\Zz}{\mathcal{Z}_0}
\newcommand{\Zo}{\mathcal{Z}_1}

\newcommand{\Yz}{\mathcal{Y}_0}
\newcommand{\Yo}{\mathcal{Y}_1}

\newcommand{\dbec}[1]{\mathsf{E}_{#1}}
\newcommand{\Gdbwb}{\mathcal{G}_n(\db, \seqw)}
\newcommand{\SetCW}{\mathcal{C}} 
\newcommand{\SetD}{\mathcal{D}}
\newcommand{\SetCWy}{\SetCW_{\seqy}}
\newcommand{\SetCWt}{\tilde{\mathcal{C}}}
\newcommand{\SetUy}{\mathcal{U}_{\seqy}}
\newcommand{\SetZW}{\mathcal{Z}}

\newcommand{\seqd}{\seq{d}}

\newcommand{\Aebp}{A_{\mathrm{ebp}}}

\newcommand{\seqct}{\seq{\ct}}

\newcommand{\qx}[1]{q^{(#1)}}

\newcommand{\dsym}[1]{\mathsf{D}_{#1}} 
\newcommand{\exi}{\except{i}}
\newcommand{\mur}{\mu^*}
\newcommand{\lambdar}{\lambda^*}
\newcommand{\lambdarexi}{\lambda^*_{\except i}}
\newcommand{\pyu}{p_{y\condmid u}}
\newcommand{\plu}{p_{\lambda\condmid u}}

\newcommand{\SetY}{\mathcal{Y}}
\newcommand{\SetYa}{\SetY_{\alpha}}
\newcommand{\SetS}{\mathcal{S}} 
\newcommand{\Eventa}{E_{\alpha}}
\newcommand{\compose}{\circ}
\newcommand{\lambdaphi}{\lambda \compose \phi}
\newcommand{\invphi}{\phi^{-1}}
\newcommand{\SetG}{\mathbb{G}} 
\newcommand{\SetH}{\mathbb{H}} 
\newcommand{\LatCoarse}{\mathcal{C}_{\mathrm{c}}}
\newcommand{\LatFine}{\mathcal{C}_{\mathrm{f}}}
\newcommand{\seqphi}{\seq{\phi}}
\newcommand{\GdbwK}{\mathcal{G}^K_n(\db, \seqw)} 
\newcommand{\tprisphi}[1]{\tilde{\lambda}^{\node{#1}}_{*} \compose \phi_*}

\newcommand{\Infopri}[1]{\Info{#1}} 
\newcommand{\Infoprix}[2]{\Infox{#1}{#2}}

\newcommand{\IbextuInlLx}[1]{\Ibextu^{(#1,n,l,L)}}
\newcommand{\IbextuInlL}{\IbextuInlLx{\Infopri{b}}}
\newcommand{\IbextlInlLx}[1]{\Ibextl^{(#1,n,l,L)}}
\newcommand{\IbextlInlL}{\IbextlInlLx{\Infopri{b}}}
\newcommand{\IbextmInx}[1]{\Ibext^{*(#1,n)}}
\newcommand{\IbextmIn}{\IbextmInx{\Infopri{b}}}
\newcommand{\IbextmInLx}[1]{\Ibext^{*(#1,n,L)}}
\newcommand{\IbextmInL}{\IbextmInLx{\Infopri{b}}}
\newcommand{\IaextuInlLx}[1]{\Iaextu^{(#1,n,l,L)}}
\newcommand{\IaextuInlL}{\IaextuInlLx{\Infopri{a}}}
\newcommand{\IaextmInx}[1]{\Iaext^{*(#1,n)}}
\newcommand{\IaextmIn}{\IaextmInx{\Infopri{a}}}
\newcommand{\IaextmInLx}[1]{\Iaext^{*(#1,n,L)}}
\newcommand{\IaextmInL}{\IaextmInLx{\Infopri{a}}}

\newcommand{\Iaextl}{\underline{I}_{\node{a},\mathrm{ext}}}
\newcommand{\IbextmIlx}[1]{\Ibextl^{*(#1)}}
\newcommand{\IbextmIl}{\IbextmIlx{\Infopri{b}}}
\newcommand{\IbextmIux}[1]{\Ibextu^{*(#1)}}
\newcommand{\IbextmIu}{\IbextmIux{\Infopri{b}}}
\newcommand{\IaextmIlx}[1]{\Iaextl^{*(#1)}}
\newcommand{\IaextmIl}{\IaextmIlx{\Infopri{a}}}
\newcommand{\IaextmIux}[1]{\Iaextu^{*(#1)}}
\newcommand{\IaextmIu}{\IaextmIux{\Infopri{a}}}

\newcommand{\Ibpril}{\Infopri{b}^-} 
\newcommand{\Ibpriu}{\Infopri{b}^+}
\newcommand{\Ibpriz}{\Infopri{b}^{\circ}}
\newcommand{\Iapriz}{\Infopri{a}^{\circ}}

\newcommand{\GdbwbiL}{\mathcal{G}_n^{i(L)}}
\newcommand{\GdbwbipL}{\mathcal{G}_n^{i'(L)}}
\newcommand{\Gdbwbil}{\mathcal{G}_n^{i(l)}}
\newcommand{\GdbwbjL}{\mathcal{G}_n^{j(L)}}
\newcommand{\Gdbwbjl}{\mathcal{G}_n^{j(l)}}
\newcommand{\PloopbnL}{P^{\mathrm{loop},\node{b}}_{n,L}}
\newcommand{\PloopanL}{P^{\mathrm{loop},\node{a}}_{n,L}}
\newcommand{\dextbitL}{\dtrueextprob{b}_{i(L)}} 
\newcommand{\dextbiitL}{\dtrueextprob{b}_{i'(L)}}
\newcommand{\dextajtL}{\dtrueextprob{a}_{j(L)}}

\newcommand{\IbextlIlx}[1]{\Ibextl^{(#1,l)}}
\newcommand{\IbextuIlx}[1]{\Ibextu^{(#1,l)}}
\newcommand{\IaextuIlx}[1]{\Iaextu^{(#1,l)}}
\newcommand{\IbextlIl}{\IbextlIlx{\Infopri{b}}}
\newcommand{\IbextuIl}{\IbextuIlx{\Infopri{b}}}
\newcommand{\IaextuIl}{\IaextuIlx{\Infopri{a}}}

\newcommand{\Ibprin}{\Infopri{b}^{(n)}} 
\newcommand{\Iaprin}{\Infopri{a}^{(n)}}

\newcommand{\deq}[1]{\left\langle #1\right\rangle} 

\newtheorem{definition}{Definition}
\newtheorem{theorem}{Theorem}
\newtheorem{lemma}[theorem]{Lemma}
\newtheorem{proposition}[theorem]{Proposition}
\newtheorem{example}{Example}
\newtheorem{remark}{Remark}

\newrefformat{obs}{Observation~\ref{#1}}
\newrefformat{con}{Consequence~\ref{#1}}
\newrefformat{sec}{Section~\ref{#1}} 
\newrefformat{app}{Appendix~\ref{#1}}
\newrefformat{tab}{Table~\ref{#1}}
\newrefformat{fig}{Fig.~\ref{#1}}
\newrefformat{alg}{Fig.~\ref{#1}}
\newrefformat{rem}{Remark~\ref{#1}}
\newrefformat{exm}{Example~\ref{#1}}
\newrefformat{def}{Definition~\ref{#1}}
\newrefformat{prop}{Proposition~\ref{#1}}
\newrefformat{thm}{Theorem~\ref{#1}}
\newrefformat{lem}{Lemma~\ref{#1}}
\newrefformat{enum}{\ref{#1})}
\newrefformat{fn}{footnote~\ref{#1}}
\newcommand{\secref}[1]{\prettyref{#1}} 

\ifx\PreviewMacro\undefined
\else
  \PreviewMacro*[]{\FOR}
  \PreviewMacro*[]{\REPEAT}
  \PreviewMacro*[]{\UNTIL}
  \PreviewMacro*[]{\WHILE}
  \PreviewMacro*[]{\COMMENT}
  \PreviewMacro*[]{\subfigure}
  \PreviewMacro*[!]{\raisebox}
  \PreviewEnvironment*[]{proposition} 
\fi

  \newcommand{\mytitle}{Analytical Framework of LDGM-based Iterative Quantization with Decimation}
  \title{\mytitle}%
  \author{Qingchuan~Wang, Chen He, Lingge Jiang%
    \thanks{The authors are with Department of Electronic Engineering,
      Shanghai Jiao Tong University, Shanghai, 200240, China.  E-mail:
      {rainy6144@gmail.com}, {\{chenhe,lgjiang\}@sjtu.edu.cn}.  This paper was supported by National Natural
      Science Foundation of China Grants No.~60772100, 60832009 and 60872017,
      National 863 Program Grant No.~2009AA011505, and by
      Science \& Technology Committee of Shanghai Municipality Grant
      No.~06DZ15013.}%
  } \markboth{Manuscript}%
  {Wang \MakeLowercase{\textit{et al.}}: \mytitle}
}

\begin{document}
\maketitle

\begin{abstract}
  While iterative quantizers based on low-density generator-matrix (LDGM) codes have been shown to
  be able to achieve near-ideal distortion performance with comparatively moderate block length and
  computational complexity requirements, their analysis remains difficult due to the presence of
  decimation steps.  In this paper, considering the use of LDGM-based quantizers in a class of
  symmetric source coding problems, with the alphabet being either binary or non-binary, it is
  proved rigorously that, as long as the degree distribution satisfies certain conditions that can
  be evaluated with density evolution (DE), the belief propagation (BP) marginals used in the
  decimation step have vanishing mean-square error compared to the exact marginals when the block
  length and iteration count goes to infinity, which potentially allows near-ideal distortion
  performances to be achieved.  This provides a sound theoretical basis for the degree distribution
  optimization methods previously proposed in the literature and already found to be effective in
  practice.
\end{abstract}


\section{Introduction}
Near-ideal quantization is important not only in source coding, but also in many channel coding
problems due to e.g.\ signal shaping \cite{close-to-cap-dpc} or compress-and-forward
\cite{cap-theorems-relay-chan} concerns; in particular, in many low-rate source or channel coding
applications, such as dirty-paper coding, small gaps from ideal performance in the quantizer can
translate to a significant percentage loss of the overall code rate
\cite{near-cap-dpc-src-chan-approach}.  For the symmetric cases considered in this paper, where the
shaping gain \cite{lattice-trellis-quant-highrate} is to be maximized and the boundary gain
is not an issue, practical near-ideal quantization methods
include structured trellis-coded quantization (TCQ) \cite{tcq-memoryless-gauss-markov} and polar
codes \cite{channel-polarization, polar-codes-opt-lossy-src-coding}, as well as sparse-graph
constructions mostly based on low-density generator matrix (LDGM) codes \cite{it-quant-codes-graphs,
  lossy-src-comp-ldgm-ana-alg, ld-graph-opt-src-chan-coding-binning-journal}.  Although all three
methods are able to achieve near-ideal distortion performance, as the gap closes, TCQ requires a
large memory length and thus exponential computational complexity, while polar codes are more
severely hampered by the finite block lengths available in practice \cite{empirical-scaling-law-polar,
  flip-ml}, making LDGM-based codes the only choice if performance extremely close to the
theoretical limit, e.g.\ \unit[0.012]{dB} for MSE (mean-square error) quantization \cite{lattices-good-everything}
obtained in \cite{flip-ml}, is to be achieved with reasonable
computational complexity and block lengths.  Such advantage in performance, combined with the high
flexibility and wide applicability of sparse-graph codes in a variety of source and channel coding problems, makes the analysis and design of
LDGM-based constructions for quantization highly important both theoretically and in practice.

In terms of implementation, LDGM-based quantizers require a practical encoding algorithm as well as
optimized degree distributions, and good ones have now been obtained in the literature.  In
particular, the encoding algorithm can be either belief propagation (BP)
\cite{binary-quant-bp-dec-ldgm} or survey propagation (SP) \cite{lossy-src-comp-ldgm-ana-alg}
combined with decimation and preferably also a recovery procedure \cite{flip-ml}, and other
variations such as \cite{modified-bp-codeword-quant} have also been proposed for specific cases.
The degree distribution optimization problem has also been tackled in \cite{ldgm-vq-journal}, although
the duals of optimized low-density parity-check (LDPC) degree distributions used in earlier works, e.g.\
\cite{lossy-src-comp-ldgm-ana-alg}, can often give adequate performance as well.

On the other hand, theoretical analysis of the quantization algorithm remains difficult due
to its iterative nature and use of decimation.  While distortion performance under optimal (MAP)
encoding has been analyzed in \cite{lossy-src-comp-ldgm-ana-alg,
  ld-graph-opt-src-chan-coding-binning-journal} for specific degree distributions using
codeword-counting arguments, good performance under MAP encoding is far insufficient for
guaranteeing good performance under practical BP or SP-based encoding algorithms.  An effective
approach to BP analysis is density evolution (DE), which has been successfully applied to LDPC
decoding \cite{design-cap-approaching-ldpc}; however, while the BP process in LDPC decoding will
converge by itself as long as the decoding threshold is reached, in the LDGM quantizer BP will not
converge without additional decimation steps, and there is no obvious method to make DE work across
decimation steps due to its requirement on the independence of BP messages.  Analysis of similar
decimation steps has been attempted in \cite{solving-csp-bp-dec} for the solution of boolean
satisfiability problems, and \cite{polar-codes-opt-lossy-src-coding} for quantization based on polar
codes, and although both papers offer insights that are valuable to our work, the methods there are
not sufficient for use in LDGM quantization.  Specifically, the successful analysis in
\cite{polar-codes-opt-lossy-src-coding} relies on the availability of exact marginals (or extrinsic
information) during decimation when polar codes are used, allowing them to be viewed as
conditional probabilities corresponding to a known joint probability distribution, but in LDGM
quantization only BP approximations of these marginals are available, whose accuracy remains to
be evaluated; when confronting a more difficult problem where the available marginals are limited to BP approximations as well,
\cite{solving-csp-bp-dec} provides some insights on the application of DE in such situations, but it
still has difficulty accounting for the impact of loops in the factor graph.  Inspired by the works
\cite{maxwell-constr, gen-area-theorem-conseq} attempting to characterize the accuracy of BP
marginals using extrinsic information transfer (EXIT) for LDPC decoding, our previous paper
\cite{ldgm-vq-journal} applies the same method to LDGM quantization, and conjectures that the BP
marginals can be asymptotically accurate when the degree distribution satisfies certain monotonicity
conditions that can be evaluated using DE, in which case the distortion performance can then be
approximated using methods similar to that used for polar codes in
\cite{polar-codes-opt-lossy-src-coding}; although this rough analysis allows the degree distribution
to be optimized that yield good performance, the arguments there are largely heuristic and lack
mathematical rigor, particularly for cases other than binary erasure quantization (BEQ).

Building upon the aforementioned results, this paper attempts to extend the analytical approach of
\cite{ldgm-vq-journal} to a class of ``symmetric'' source coding problems, both binary and
non-binary.  With the introduction of a reference codeword in DE, the properties regarding the
symmetry and degradation relationships among message densities, previously used in LDPC analysis in
\cite{design-cap-approaching-ldpc}, are generalized, and they are then used to relate the actual
densities of BP messages to those obtainable with DE, and to bound the difference between BP and
exact marginals used in decimation with the difference in their mutual information characterized by EXIT curves.  In this way, we are able to
show rigorously that the monotonicity condition used as the optimization criteria in
\cite{ldgm-vq-journal} can indeed lead to good distortion performance in a certain asymptotic sense.
The difficulty in applying DE across decimation steps is side-stepped by considering each decimation
step separately, assuming that exact marginals have been used in all previous decimation steps.  Even
though the actual quantizer can only use BP marginals in all decimation steps, and errors in the
earlier BP marginals can affect subsequent BP marginals in a manner that is difficult to analyze, we believe that
the present results are still able to provide important insights to BP-based quantization
algorithms; in any case, the recovery algorithm in \cite{flip-ml} can greatly alleviate this problem
in practice.

The rest of this paper is organized as follows.  \prettyref{sec:basics} starts from the MSE
quantization problem and introduces a more general class of symmetric lossy compression problems
to be considered in the rest of the paper.  \prettyref{sec:binary} reviews the LDGM code construction and
quantization algorithm that are used to solve such problems, and gives an outline of the
analytical approach.  Our main analytical results are presented in \prettyref{sec:asympt-ana-sync}.
Starting from some basic properties of message densities in the presence of an explicit reference codeword,
the error bounds of BP marginals expressed in terms of DE results are used to justify the
monotonicity conditions for degree distribution optimization, and some more intuitive
results are then given for the special case of BEQ.  Subsequently, \prettyref{sec:nonbinary} briefly
shows how to extend this analytical approach to non-binary constructions, and finally
\prettyref{sec:conclusion} concludes the paper.

\emph{Notational conventions}:
$\SetZ$ and $\SetR$ are respectively the set of integers and real numbers.
$\SetZ_q \defeq \SetZ / q\SetZ$ is the modulo-$q$ additive group.
$\mathcal{A} \backslash \mathcal{B}$ is the difference set containing the elements of set
$\mathcal{A}$ that are not in set $\mathcal{B}$.  $\Expe{\cdot}$ is the expectation operator.
$\norm{\cdot}$ is the Euclidean norm.  $\cardinal{\mathcal{A}}$ is the cardinality of set
$\mathcal{A}$\@.  $\oneif{A}$ is 1 if the condition $A$ is true, 0 otherwise.  $\log(\cdot)$,
entropy and mutual information are computed in base-2, while $\ln(\cdot)$ and $\exp(\cdot)$ are
base-$\e$.  Bold letters denote sequences or vectors whose elements are indicated by subscripts,
e.g.\ %
$\seqy=(y_1,\dotsc,y_n)$, $\seqy_{\except i}$ is the sub-sequence $(y_1,\dotsc,y_{i-1},y_{i+1},\dotsc,y_n)$,
and a sub-sequence with index set $\mathcal{S}$ can be denoted by
$\seqy_{\SetS} = (y_i)_{i\in\SetS}$; note that $y$ itself can denote a scalar variable unrelated to $\seqy$.  Addition and multiplication on sets are element-wise, e.g.\
$\mathcal{U}+2\SetZ^n = \left\{ \sequ + (2d_1,\dotsc,2d_n) \mid \sequ \in \mathcal{U}, d_i \in \SetZ
\right\}$.
$\oplus$ and $\ominus$ denote addition and subtraction in a specific additive abelian group $\SetG$,
but can also denote variants of the check-node operation when applied to probability tuples and densities, as
will be explained in Sections~\ref{sec:binary}, \ref{sec:binary-def-results} and
\ref{sec:prob-tuple-SetG}.  $x \bmod [a,b)$, or simply $(x)_{[a,b)}$, is defined as the unique
element of $(x-(b-a)\SetZ) \cap [a,b)$, and similarly $\seq{x} \bmod [a,b)^n$ or
$(\seq{x})_{[a,b)^n}$ is the unique element of $(\seq{x}-(b-a)\SetZ^n) \cap [a,b)^n$.  The unit
``\unit{b/s}'' means ``bits per symbol''.  For convenience, we do \emph{not} distinguish in notation
between random variables and their possible values, or between the pmfs of discrete random variables
and pdfs of continuous ones, which should be clear from context; for example, $p(b=b')$ or
$p_b(b')$ denotes the probability (density) that \emph{random variable} $b$ takes the \emph{value}
$b'$, while we simply write $p(b)$ if both the random variable and the value are denoted by $b$, or if
it is clear from context what the random variable is.

\section{Problem Formulation and Performance Bounds}
\label{sec:basics}

\subsection{MSE Quantization}
The \emph{mean-squared error (MSE) quantization problem} of $\SetR^n$
\cite[Sec.~II-C]{lattices-good-everything} can be formulated as follows.  Let $\Lambda$ be a
non-empty discrete subset of $\SetR^n$ (the \emph{quantization codebook}, or simply \emph{code}),
and $Q_{\Lambda}: \SetR^n\to\Lambda$ be a quantizer that maps each $\seqy \in \SetR^n$ to a nearby
codeword $Q_{\Lambda}(\seqy) \in \Lambda$.  The mean-square quantization error, averaged over
$\seqy$, is given by
\begin{equation}
  \label{eq:mse}
  \sigma^2 = \limsup_{M\to\infty} \frac{1}{(2M)^n} \cdot \frac{1}{n} \int_{[-M,M]^n} \norm{\seqy
  - Q_{\Lambda}(\seqy)}^2 \, d\seqy.
\end{equation}
The objective is to design $\Lambda$ and a practical quantizer
$Q_{\Lambda}(\cdot)$ such that the scale-normalized MSE
$G(\Lambda)\defeq \sigma^2\rho^{2/n}$ is minimized, where $\rho$ is the codeword density
\begin{equation}
  \label{eq:rho}
  \rho = \limsup_{M\to\infty} \frac{1}{(2M)^n} \cardinal{\Lambda \cap
    [-M,M]^n}.
\end{equation}

It should be noted that \cite{lattices-good-everything} assumes that $\Lambda$ is a lattice, which
ensures that the Voronoi regions corresponding to different codewords in $\Lambda$ differ only by a
translation, and since lattices are closed under addition, such codebooks can often achieve better
performance than unstructured ones in e.g.\ network coding problems involving channels with similar additive structures
\cite{case-struc-random-codes-net-cap}.  On the other hand, in
plain quantization problems, the lattice structure is fairly unimportant, and
indeed trellis codebooks or those generated with a modulation mapping often lack such a structure
and yet still achieve good performance.  Therefore, in this problem formulation we do not constrain
$\Lambda$ to be a lattice, and the definitions in \cite{lattices-good-everything} have been generalized
accordingly.

In this paper we consider asymptotically large dimensionality $n$.  By
a volume argument, it is easy to find an asymptotic lower bound $G^* =
\frac{1}{2\pi\e}$ for $G(\Lambda)$ as $n\to\infty$.  This bound can be approached by
the nearest-neighbor quantizer with a suitable random codebook
\cite{lattices-good-everything}, whose codewords' Voronoi regions
are asymptotically spherical, but such a quantizer has exponential computational
complexity in $n$ and is thus impractical.  The simplest scalar
quantizer $\Lambda_1=\SetZ^n$, on the other hand, has the 1.5329-dB
larger $G_1=G(\Lambda_1)=\frac{1}{12}$, corresponding to the
well-known 1.53-dB loss of scalar quantization.  In general, we call
$10 \log_{10} (G(\Lambda) / G^*)$ the \emph{shaping loss} of a
quantizer, and it is also the gap from the \emph{granular gain} and
\emph{shaping gain} defined in \cite{lattice-trellis-quant-highrate},
for source and channel coding respectively, toward the 1.53-dB limit.

In order to design a practical quantization codebook with a finite alphabet, we consider $\Lambda$
with a periodic structure $\Lambda = \mathcal{U} + M\SetZ^n$, where $\mathcal{U}$ is a set
of $2^{nR}$ codewords from $\SetZ_M^n$ with each $\sequ = \sequ(\seqb) \in \mathcal{U}$ labeled by a
binary sequence $\seqb \in \SetZ_2^{nR}$.  Such a $\Lambda$ is called an \emph{$M$-ary
  rate-$R$ quantization code}, and is also used by TCQ.  Constrained by this $M$-ary structure,
the MSE quantization problem is then equivalent to the lossy compression of an i.i.d.\ uniform
source over $\SetY \defeq [0, M)$ using codebook $\mathcal{U}$ and the modulo-$\mathcal{I}$
($\mathcal{I} \defeq [-\frac{M}{2}, \frac{M}{2})$) distortion measure $d(u,y) = (y-u)\bmodI^2$, and $\sigma^2$ in \eqref{eq:mse} is simply the average distortion and $\rho = 2^{nR}/M^n$;
this equivalent problem is henceforth called \emph{$M$-ary MSE quantization}.  At a given $R$, the $\sigma^2$ corresponding to the bound $G^*$ is
\begin{equation}
  \label{eq:mse-ideal}
  \sigma^2_*(R) \defeq G^* \rho^{-2/n} = (2\pi\e (2^R/M)^2)^{-1}.
\end{equation}
While $\sigma^2_*(R)$ is not exactly achievable at any finite $M$, leaving a gap called the random-coding loss in \secref{sec:random-coding-loss-mse}, this gap can become extremely small as $M$ increases.

\subsection{Symmetric Source Coding Problems over a Finite Abelian Group}
\label{sec:sym-lossy-compr}
$M$-ary MSE quantization is now generalized as follows for uniformity of presentation.

\begin{definition}
  \label{def:sym-lossy-compr}
  Consider the source coding problem involving i.i.d.\ source $y$ taking values in
  $\SetY$ with pmf or pdf $p(y)$, under distortion measure $d(u,y)$; that is, given any block size $n$ and
  rate $R > 0$, we design a codebook $\mathcal{U}$ of size $2^{nR}$ along with encoding and decoding
  functions, which map each possible source sequence $\seqy$ into a reconstructed sequence
  $\sequ(\seqy) \in \mathcal{U}$ with distortion
  $d(\sequ(\seqy),\seqy) \defeq \frac{1}{n} \sum_{j=1}^n d(u_j(\seqy), y_j)$, and the objective is
  to minimize the average distortion $D \defeq \Expe{d(\sequ(\seqy),\seqy)}$ with the expectation
  taken over $p(\seqy) = p_y(y_1)\dotsm p_y(y_n)$.  This is called a \emph{symmetric source coding
    problem over $\SetG$}, if the reconstruction alphabet is a finite abelian group $\SetG$ (i.e.\
  $\mathcal{U}\subseteq \SetG^n$), and if a measure-preserving\footnote{When $p(y)$ is a pdf, we require the group action $\psi$ to be measure-preserving w.r.t.\ the measure over $\SetY$ used to define that pdf, so that the symmetry
    $p(y) = p(\psi_u(y))$ in probability density implies the symmetry in the probability itself.} group action $\psi$ of $\SetG$ exists on
  $\SetY$, such that
  \begin{equation}
    \label{eq:pd-sym-cond}
    p(y) = p(\psi_u(y)) \text{ and } d(u,y) = d(0,\psi_u(y))
  \end{equation}
  for any $y\in\SetY$ and $u\in\SetG$.
\end{definition}

Below are some examples with $\SetG = \SetZ_M$, which may be called \emph{$M$-ary symmetric source coding problems}:\footnote{Not to be confused with source coding \emph{of} $M$-ary symmetric sources, i.e.\ \prettyref{exm:discrete-Mary-hamming} below, which is only a special case.}
\begin{example}
  \label{exm:Mary-mse-quant}
  In $M$-ary MSE quantization, $p(y)$ is uniform over $\SetY = [0,M)$, $d(u,y) = (y-u)\bmodI^2$
  (the $\mathcal{I} = [-\frac{M}{2},\frac{M}{2})$ in the subscript denotes
  modulo operation like above), and $\psi_u(y) = (y-u)\bmodY$.
\end{example}

\begin{example}
  \label{exm:discrete-Mary-hamming}
  In quantization of an $M$-ary discrete source with Hamming
  distortion, $p(y)$ is uniform over $\SetY = \SetZ_M$,
  $d(u,y) = \oneif{y\ne u}$, and $\psi_u(y) = (y-u)\bmod M$.
\end{example}

\begin{example}
  \label{exm:erasure-quant}
  Another well-known example is \emph{$M$-ary erasure quantization},
  where $\SetY = \SetZ_M \cup \{*\}$ ($*$ denotes an
  erased symbol), $p_y(*) = \epsilon$ with $0<\epsilon<1$, $p(y) = (1-\epsilon)/M$ for
  $y\in\SetZ_M$, $d(u,y) = \oneif{y\ne u\ \text{and}\ y\ne *}$,
  while $\psi_u(y) = (y-u)\bmod M$ for $y\in\SetZ_M$ and
  $\psi_u(*)=*$.  This is usually considered in the zero-distortion
  limit, particularly when $M=2$ (known as \emph{binary erasure
    quantization} (BEQ) \cite{it-quant-codes-graphs}), due to its simplicity.
\end{example}

There are also noteworthy symmetric lossy quantization problems with other reconstruction alphabets $\SetG$:
\begin{example}
  \label{exm:lattice-quant}
  MSE quantization can be generalized to $N$ real dimensions per source symbol as follows.  Given
  $N$, let $\LatFine$ be a lattice in $\SetR^N$, i.e.\ a discrete additive subgroup of $\SetR^N$,
  and $\LatCoarse$ be $\LatFine$'s subgroup, which forms a coarser lattice.  Now we make the
  source alphabet $\SetY = \SetR^N / \LatCoarse$ and the reconstruction alphabet
  $\SetG = \LatFine / \LatCoarse$ quotient groups w.r.t.\ $\LatCoarse$, such that each source symbol
  $y$ and reconstruction symbol $u$ can be viewed as an $N$-dimensional vector modulo $\LatCoarse$,
  and $p(y)$ is then the uniform distribution over $\SetY$, $d(u,y) = \norm{(y-u)\bmod\LatCoarse}^2$
  is the squared modulo-$\LatCoarse$ Euclidean distance, and $\psi_u(y) = (y-u) \bmod\LatCoarse$ is
  simply subtraction in the group $\SetY$, of which $\SetG$ is a subgroup.  In particular,
  \prettyref{exm:Mary-mse-quant} is the case that $N=1$, $\LatFine = \SetZ$, and
  $\LatCoarse = M\SetZ$.  This is related to vector precoding
  \cite{vector-perturbation-mod-precoding1} sometimes performed in MIMO systems, especially MIMO
  broadcast channels, that performs spatial signal shaping in order to approach capacity more closely; for example, $\LatFine$ and $\LatCoarse$ can be chosen as respectively
  the lattices $\SetZ^N$ and $M\SetZ^N$ in the receiver-side signal space, transformed to the
  transmitter side using the inverted channel matrix. 
\end{example}

\begin{example}
  \label{exm:erasure-quant-K}
  BEQ can be generalized to $K$ dimensions per source symbol as follows.  Given $K$, we let $\SetG = \SetZ_2^K$ be the
  $K$-dimensional linear space over $\SetZ_2$, and $\SetY$ be the set of all affine subspaces of
  $\SetG$, which can be partitioned by the corresponding vector subspace $x$ into $\cup_{x}\SetY_x$,
  with $x$ ranging over all vector subspaces of $\SetG$ and
  $\SetY_x \defeq \{ x \oplus d \condmid d \in \SetG \}$ being the set of affine subspaces from
  each $x$.  Now let $d(u,y) = \oneif{u\notin y}$ for $u\in\SetG$ and $y\in\SetY$, and constrain $p(y)$ to be uniform over each
  $\SetY_x$, so that \eqref{eq:pd-sym-cond} holds with $\psi_u(y) = y\ominus u$, where
  $\ominus$ is bitwise subtraction in $\SetZ_2^K$ applied element-wise to $y$.  When $K=1$, this
  reduces to BEQ if the affine subspaces $\{0\}$, $\{1\}$ and $\{0,1\}$ of $\SetZ_2$ are identified
  with 0, 1 and $*$ in $\SetY$.
\end{example}

According to rate-distortion theory \cite[Sec.~10.4--10.5]{elements-info-theory}, in the limit of
large $n$, each possible test channel $p(u\condmid y)$ corresponds to an average distortion
$D = \Expe{d(u,y)}$ achievable at rate $R=I(u;y)$ with a random codebook and a quantizer based on
joint typicality, and conversely, any achievable rate can be achieved in this way with some test channel; here $u$ and $y$ are viewed as random variables and $D$ and $R$ are
computed according to joint distribution $p(y)p(u\condmid y)$.  The optimal test channel that
minimizes $D$ at a given $R$ (or vice versa) is straightforward to compute:
\begin{proposition}
  \label{prop:test-channel}
  The optimal test channel for symmetric source coding over $\SetG$ is
  \begin{equation}
    \label{eq:test-channel}
    p(u\condmid y) = \e^{-t d(u,y)} / Q(y), \quad u\in\SetG
  \end{equation}
  where $Q(y) \defeq \sum_u \e^{-t d(u,y)}$ is the normalization factor, and $t$ is the value that
  makes $D_0(t) \defeq \Expe{d(u,y)}$ or $R_0(t) \defeq I(u;y)$ equal to the desired $D$ or $R$; in
  the latter case this $t$ is denoted by $t_0(R)$.
\end{proposition}
\begin{IEEEproof}
  See \prettyref{app:proof-test-channel}.
\end{IEEEproof}

In general, for any $t>0$ (not necessarily equal to $t_0(R)$), we call
$p(u\condmid y) = \e^{-td(u,y)}/Q(y)$ of the above form, or the corresponding $p(y\condmid u)$, a
\emph{test channel} of the symmetry source coding problem.  It is trivial to verify the following symmetry properties of such a test channel:
\begin{proposition}
  \label{prop:test-channel-sym}
  Given the $p(y)$ and $d(u,y)$ from a symmetric source coding problem over $\SetG$, let
  $p(u\condmid y) = \e^{-t d(u,y)} / Q(y)$ with $Q(y) \defeq \sum_{u\in \SetG} \e^{-t d(u,y)}$ for
  some arbitrary $t>0$, then $p(u) \defeq \sum_y p(u\condmid y)p(y)$ is a uniform distribution, and
  $p(y\condmid u) \defeq p(y)p(u\condmid y)/p(u)$ satisfies
  $p_{y\condmid u}(y\condmid u) = p_{y\condmid u}(\psi_u(y)\condmid 0)$.
\end{proposition}

It is also possible to prove that $R_0(t)$ is an increasing function of $t$ while $D_0(t)$ is
decreasing.  Intuitively, given $t$ and the corresponding $p(u\condmid
y)$, for each ``typical'' $\seqy$ w.r.t.\ $p(y)$, the
probability that an independent $\sequ$ typical w.r.t.\ $p(u)$ is jointly typical with $\seqy$ is approximately
$2^{-nI(u;y)} = 2^{-nR_0(t)}$, so on average there are
$2^{n(R-R_0(t))}$ jointly typical sequences $\sequ$ in a random codebook $\mathcal{U}$,
and as long as $R>R_0(t)$ one such $\sequ$ likely exists that will
yield an average distortion close to $D_0(t)$.  In practice, the quantization algorithm is necessarily non-ideal, and the actual rate $R$ and average distortion $D$ could be slightly larger than resp.\ $R_0(t)$ and $D_0(t)$.

\subsection{The Random-Coding Loss of $M$-ary MSE Quantization}
\label{sec:random-coding-loss-mse}
\begin{figure}[!t]
  \centering
  \subfigure[binary code ($M=2$)]{\includegraphics{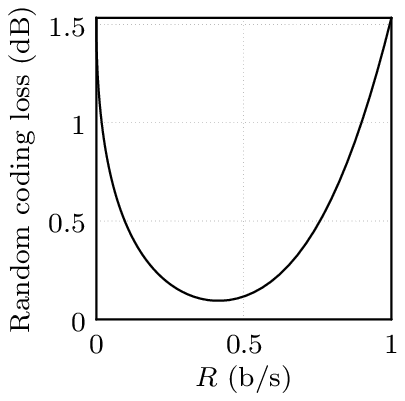}}%
  \hspace{2mm}%
  \subfigure[4-ary code ($M=4$)]{\includegraphics{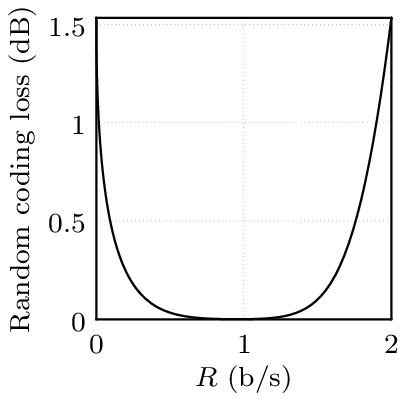}}%
  \caption{Random-coding losses of binary and 4-ary MSE quantization.
    In the binary case, the minimum loss is
    approximately \unit[0.0945]{dB} at $t=3.7114$ and
    $R=R_0(t)=\unit[0.4143]{b/s}$.  In the 4-ary case, the minimum loss is only
    \unit[0.0010]{dB} at approximately $t=2.0053$ and
    $R=R_0(t)=\unit[0.9550]{b/s}$.}%
  \label{fig:power-entropy}%
\end{figure}

\prettyref{prop:test-channel} gives the minimum $G(\Lambda) =
\sigma^2\rho^{2/n} = (2^R/M)^2 D$ achievable with $M$-ary MSE
quantization at each rate $R$.  This is larger than the optimal $G^*$
and we call the corresponding shaping loss
$10\log_{10}(G(\Lambda)/G^*)$ the \emph{random-coding loss} as random
coding is one way to achieve it.  The random-coding loss measures the suboptimality of
the period-$M$ structure of $\Lambda$; as shown in
\prettyref{fig:power-entropy} for $M=2$ and $M=4$, it is very small
for large $M$ and moderate $R$, meaning that $M$-ary MSE quantization
is near-optimal in such cases.

%

\section{The Binary LDGM Quantizer}
\label{sec:binary}
Previous works such as \cite{it-quant-codes-graphs, analysis-ldgm-loss-compression,
  ld-ach-wz-gp-bounds, binary-quant-bp-dec-ldgm, ldgm-vq-globecom07} suggest that LDGM-based code constructions are good candidates for approaching the performance limit in \prettyref{prop:test-channel} for
symmetric source coding problems and, in particular, achieve near-zero shaping loss in MSE
quantization.  In this and the next section, we will carry out a deeper analysis on the use of LDGM
codes with BP in the simpler binary case (i.e.\ $M=2$ and $\SetG = \SetZ_2$), while in
\prettyref{sec:nonbinary} we will consider non-binary constructions that can be applied to more
general symmetric source coding problems and achieve lower random-coding loss in MSE
quantization.

In the quantization algorithm for binary codes, the \textit{a priori} information (priors), extrinsic information and BP messages
are likewise binary and can be viewed as probability distributions of binary random variables.  In this paper, they are mainly
represented by probability tuples, e.g.\ $\mu = (\mu(0), \mu(1))$, $\mu(b)$ being the probability
that the variable equals $b \in \SetZ_2$; the corresponding log-likelihood ratio (LLR) is
$\llr(\mu) \defeq \ln(\mu(0)/\mu(1))$.  For convenience, these tuples are \emph{implicitly
  normalized}; that is, when we write $\mu(b) = q_b$, $b\in\SetZ_2$, we actually make $\mu(b) = q_b/(q_0+q_1)$ so
that $\mu(0) + \mu(1) = 1$, and later appearances of $\mu(b)$ refer to this normalized value.
$\mu \odot \mu' \defeq (\mu(0)\mu'(0), \mu(1)\mu'(1))$ (implicitly normalized) and
$\mu \oplus \mu' \defeq (\mu(0)\mu'(0) + \mu(1)\mu'(1), \mu(0)\mu'(1) + \mu(1)\mu'(0))$ are the
variable-node and check-node operations in LDPC literature, which are associative and thus
immediately applicable to more than two probability tuples.  More generally, if we view $\SetZ_2^m$ as a vector space over field $\SetZ_2$ and let $\SetCW$ be an
affine subspace of it, then given $m-1$ probability tuples
$\lambda_{\except i} \defeq (\lambda_1, \dotsc, \lambda_{i-1}, \lambda_{i+1}, \dotsc, \lambda_m)$,
we may define $\nu(\SetCW; \lambda_{\except i})$ as the probability tuple
$\nu$ with $\nu(b) = \sum_{\seqb\in\SetCW: b_i = b} \prod_{j\ne i} \lambda_j(b_j)$, $b\in\SetZ_2$;
$\odot$ and $\oplus$ are then its special cases with $\SetCW$ being respectively the $(3,1)$
repetition code and the $(3,2)$ single parity-check code.  $\constmsg{0} \defeq (1,0)$,
$\constmsg{1} \defeq (0,1)$ and $\constmsg{*} \defeq (\frac{1}{2},\frac{1}{2})$ are respectively the
``sure-0'', ``sure-1'' and ``unknown'' probability tuples.  We also define
$H(\mu) \defeq H_2(\mu(0))$ and $I(\mu) \defeq 1-H(\mu)$, where $H_2(p) \defeq -p\log p-(1-p)\log(1-p)$ is the binary entropy
function.

\subsection{Outline of the Quantizer and Its Analysis}
\label{sec:outline-quant-ana}
When $\SetG=\SetZ_2$, we use the binary LDGM codebook
\begin{equation}
  \label{eq:SetU-binary}
  \mathcal{U} = \mathcal{U}(\seqa) = \{ \sequ = \sequ(\seqb,\seqa) \defeq
  \seqc \defeq \seqb\mat{G} \oplus \seqa \condmid \seqb \in \SetZ_2^{\nb} \},
\end{equation}
where $\mat{G} = (g_{ij})_{\nb\times \nc}$ is the sparse generator matrix randomly generated
according to the degree distributions optimized below, the matrix multiplication in $\seqb\mat{G}$
as well as $\oplus$ are modulo-2, $\nc \defeq n$, $\nb \defeq nR$, and $R$ is the rate of the LDGM
code.  A fixed \emph{scrambling sequence} $\seqa$ randomly chosen from $\SetZ_2^{\nc}$ has been
introduced in \eqref{eq:SetU-binary}, which ensures that every point $\SetZ_2^n$ is covered by
$2^{nR}$ of the $\mathcal{U}(\seqa)$'s, even though each $\mathcal{U}(\seqa)$ may be
``clumped'' around certain points in $\SetZ_2^n$.  This will be essential in results such as
\prettyref{prop:ref-dist} below.

The quantization algorithm is based on belief propagation, with a \emph{decimation} step that makes
hard decisions in order to help the algorithm converge \cite{ldgm-vq-globecom07, binary-quant-bp-dec-ldgm}.%
\footnote{Unlike LDPC decoding, LDGM quantization will not converge without decimation.  Intuitively speaking,
  when doing LDPC decoding with SNR higher than threshold, the transmitted codeword is normally much
  closer to the received sequence (and thus much more likely) than any other codeword, allowing BP
  to converge to it.  In the case of quantization with LDGM codes, there are usually a large number of
  similarly close codewords to the source sequence, and BP cannot by itself make a decision among
  them.}  Proper analysis of the decimation steps is essential to a good understanding of the algorithm and its performance characteristics, so before
presenting the algorithm in detail, we first outline our analytical approach.  We consider a fixed
$\mat{G}$ for the rest of this section; that is, all probabilities are implicitly conditioned on
$\mat{G}$.  Given the source sequence $\seqy$, we assign a probability to each $\sequ$
according to the test channel $p(u\condmid y) = e^{-t d(u,y)} / Q(y)$, which has the same form as
the optimal one in \prettyref{prop:test-channel} and makes \prettyref{prop:test-channel-sym}
applicable; here $R_0(t) = I(u;y)$ is generally close, but not equal, to $R$ (although we will still
assume that $R_0(t) > 0$), and its choice will be briefly covered in \secref{sec:app-dd-opt}.\@
Ignoring normalization factors depending only on $\seqy$, the probability thus assigned is
\begin{equation}
  \label{eq:quy}
  q(\sequ\condmid \seqy) = \prod_{j=1}^n \e^{-td(u_j, y_j)} = \e^{-nt d(\sequ,\seqy)}.
\end{equation}
As any $\sequ \in \SetZ_2^n$ is equal to $\sequ(\seqb,\seqa)$ for $2^{nR}$ distinct
$(\seqb,\seqa)$'s, \eqref{eq:quy} also gives a joint distribution of $\seqb$ and $\seqa$, which is
$q(\seqb, \seqa \condmid \seqy) = \e^{-nt d(\sequ(\seqb,\seqa), \seqy)}$ without normalization.  If $\seqb$ and $\seqa$
were sampled from this distribution, all $2^n$ possible values of $\sequ$ would be
obtained with probabilities proportional to \eqref{eq:quy}, and the expected distortion would simply
be the $D_0(t)$ from \prettyref{prop:test-channel}.  In reality, $\seqa$ is fixed first, independently from $\seqy$, and given $\seqy$ the
quantizer has to choose a $\seqb$, or equivalently a $\sequ$ from $\mathcal{U}(\seqa)$, but under
certain conditions this will, in a sense, yield the same result as random sampling of $\seqb$ and
$\seqa$ and thus the same distortion $D_0(t)$.

To make this notion of ``same result'' rigorous, prior to the
determination of $\seqa$ and actual quantization, we first generate
two sequences of respectively $\nc = n$ and $\nb$ i.i.d.\ uniform samples
over $[0,1)$, $\omegaas$ and $\omegabs$, as the source of randomness.
The determination of $\seqa$ and $\seqb$ in quantization are then divided respectively into $\nc$
\emph{\node{a}-steps} that determine $a_1, a_2, \dotsc, a_{\nc}$ successively,
followed by $\nb$ \emph{\node{b}-steps} determining $b_1, \dotsc, b_{\nb}$.
In \node{a}-step $j$, we compute a binary probability tuple $\textprob{a}{j}$
and set $a_j = \oneif{\omegaa{j}\ge\textprobx{a}{j}{0}}$, and similarly
in \node{b}-step $i$ probability tuple $\textprob{b}{i}$ is used to
compute $b_i = \oneif{\omegab{i}\ge\textprobx{b}{i}{0}}$.  The two
processes can then be described by the way $\textprob{a}{j}$ and
$\textprob{b}{i}$ are computed:

\begin{definition}
  \label{def:tpq}
  The above quantization process is called the \emph{true probabilistic quantizer} (TPQ),
  if $\textprob{a}{j}$ and $\textprob{b}{i}$ are set to the conditional
  probabilities $\trueextprob{a}{j}$ and $\trueextprob{b}{i}$
  corresponding to $q(\seqb, \seqa \condmid \seqy)$, that is,
  \begin{equation}
    \label{eq:trueexta}
    \trueextprobx{a}{j}{a} \defeq \sum_{\seqa\in \mathcal{A}_j(a)} \sum_{\seqb} q(\seqb, \seqa \condmid \seqy),
  \end{equation}
  where $\mathcal{A}_j(a)$ contains those $\seqa$ with $a_j=a$
  and $a_1,\dotsc,a_{j-1}$ matching the values determined in \node{a}-steps $1,\dotsc,j-1$,
  and
  \begin{equation}
    \label{eq:trueextb}
    \trueextprobx{b}{i}{b} \defeq \sum_{\seqb\in \mathcal{B}_i(b)}
  q(\seqb, \seqa \condmid \seqy),
  \end{equation}
  where $\seqa$ has been determined in the \node{a} steps and $\mathcal{B}_i(b)$
  contains those $\seqb$ with $b_i=b$ and $b_1,\dotsc,b_{i-1}$ matching the
  values determined in the previous \node{b}-steps.
\end{definition}

\begin{definition}
  \label{def:bppq}
  The quantization process is called the \emph{BP probabilistic quantizer} (BPPQ), if it sets
  each $\textprob{a}{j}$ to $\constmsg{*}$ and $\textprob{b}{i}$ to
  $\extprob{b}{i}$, the BP
  approximation of $\trueextprob{b}{i}$ above.  These
  $\textprob{a}{j}$'s, unlike those used by TPQ, do not depend on $\seqy$, so $\seqa$
  can be determined before quantization with a given $\seqy$,
  which is necessary for a useful scheme.
\end{definition}

Clearly, the TPQ yields each possible $\seqb$ and $\seqa$ with
probability proportional to $q(\seqb, \seqa \condmid \seqy)$, so
the average distortion is $D_0(t)$ as stated above.  For each TPQ
instance associated with some $\seqy$, $\omegaas$ and $\omegabs$, if
the \emph{synchronization conditions}
\begin{itemize}
\item $\trueextprob{a}{j} = \constmsg{*}$ for all $j$, and
\item $\trueextprob{b}{i}$ is precisely computed by BP for all $i$,
\end{itemize}
are met in every step, then the corresponding BPPQ instance will also yield the same $\seqa$ and $\seqb$; if this
is true for all TPQ instances, the BPPQ's average distortion will be $D_0(t)$ as well.  Consequently, we can
base our quantization algorithm on the BPPQ, and optimize the degree distributions so that the
synchronization conditions are met asymptotically for large block sizes $n$ and
BP iteration counts $L$, under as high a $t$ (and thus low $D_0(t)$) as possible.  These conditions
cannot be met precisely at finite $n$ and $L$, and the BPPQ will lose synchronization with the TPQ
and yield higher distortion, but a \emph{recovery algorithm} has been proposed in \cite{flip-ml} that can minimize the impact of such synchronization loss.

\subsection{The Quantization Algorithm}
\label{sec:quant-ana-binary}
\begin{figure}[!t]
  \centering
  \subfigure[original form]{\label{fig:binary-fg-a}\includegraphics{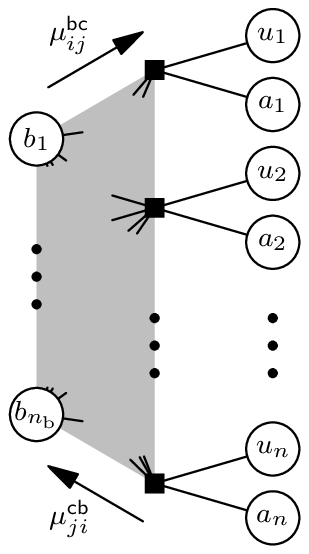}}%
  \hspace{6mm}%
  \subfigure[with the $a_j$'s omitted]{\label{fig:binary-fg-noa}\raisebox{7mm}{\includegraphics{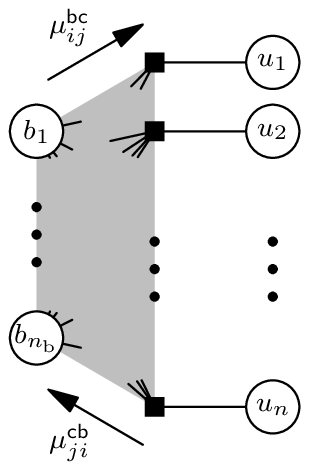}}}%
  \caption{The factor graph of the binary LDGM quantizer.  Circles are variable nodes and black
    squares are factor nodes.  The edges in the gray area are given by $\mat{G}$; specifically, each
    edge from variable node $b_i$ to the $j$-th factor node corresponds to $g_{ij}=1$ in the
    generator matrix $\mat{G}$.  Subfigure (a) shows the full factor graph used in the analysis of
    the quantization algorithm below.  During the actual quantization algorithm, $\seqa$ is constant, so a
    simplified version shown in subfigure (b) suffices.}%
  \label{fig:binary-fg}%
\end{figure}

\prettyref{fig:binary-fg-a} shows the factor graph that can be used to estimate each
$\trueextprob{a}{j}$ and $\trueextprob{b}{i}$ given by
\eqref{eq:trueexta} and \eqref{eq:trueextb}.  The \textit{a priori}
information of each variable $u_j = c_j$, denoted $\priprob{u}{j}$,
is given by
\begin{equation}
  \label{eq:priprobs-u}
  \priprobx{u}{j}{u} = \e^{-t d(u, y_j)},
\end{equation}
which corresponds to a factor in $q(\seqb, \seqa \condmid \seqy)$.  The priors of the $a_j$'s and
$b_i$'s, denoted $\priprob{a}{j}$ and $\priprob{b}{i}$ respectively, are set according to the ranges of
summation in \eqref{eq:trueexta} and \eqref{eq:trueextb}.  That is, when estimating $\trueextprob{a}{j}$, we know from \eqref{eq:trueexta}
that $\priprob{a}{j'} = \constmsg{a_{j'}}$ for $j'<j$ with $a_1,\dotsc,a_{j-1}$ taking the previously determined values,
while the remaining $\priprob{a}{j'}$'s and all the $\priprob{b}{i}$'s are $\constmsg{*}$;
similarly, when estimating $\trueextprob{b}{i}$ in \eqref{eq:trueextb} we let all $\priprob{a}{j} = \constmsg{a_j}$, while
$\priprob{b}{i'}$ is $\constmsg{b_{i'}}$ if $b_{i'}$ has been determined (\emph{decimated}), and
$\constmsg{*}$ otherwise.  The function nodes, shown as black squares in
\prettyref{fig:binary-fg-a}, represent the relationship $\sequ = \seqb\mat{G} \oplus \seqa$, so
similar to LDPC we call them \emph{check nodes}.  In this way, $\trueextprob{a}{j}$ and
$\trueextprob{b}{i}$ are simply the exact marginals (\emph{true extrinsic information}) of variable
$a_j$ and $b_i$ on the factor graph when using those priors, and they can be approximated by respectively $\extprob{a}{j}$ and $\extprob{b}{i}$, the marginals (\emph{BP
  extrinsic information}) computed with the BP (a.k.a.\ sum-product) algorithm.

The quantization algorithm is essentially an implementation of BPPQ: $\seqa$ is chosen randomly, and then in
each \node{b}-step, $\extprob{b}{i}$ is computed with a number of BP iterations as an approximation
of $\trueextprob{b}{i}$, and $b_i$ is decimated to $\oneif{\omegab{i}\ge\extprobx{b}{i}{0}}$.  In
practice, to reduce the number of iterations needed in the entire quantization process, BP message
values from earlier \node{b}-steps are reused, and multiple \node{b}-steps are carried out after
each BP iteration, but this has little impact on the theoretical analysis below.  The algorithm can
thus be outlined in \prettyref{alg:binary} where, apart from the priors $\priprob{u}{j}$ and
$\priprob{b}{i}$, extrinsic information $\extprob{b}{i}$, we also use $\msg{bc}{ij}$ to denote a BP
message from variable node $b_i$ toward check node $j$ (the check node to the left of $u_j$),
and $\msg{cb}{ji}$ for the BP message in the inverse direction, as indicated by the arrows in
\prettyref{fig:binary-fg}, and these BP messages are binary probability tuples here as well;
$\neighbor{bc}{\cdot j}=\neighbor{cb}{j\cdot}$ is the set of indices $i$ for which there exists an
edge between check node $j$ and variable node $b_i$, and
$\neighbor{bc}{i\cdot} = \neighbor{cb}{\cdot i}$ is defined similarly.  To follow BPPQ exactly, in each
decimation step, the bit index $i^*$ is chosen sequentially,\footnote{or randomly among the set of
  undecimated bit indices $\mathcal{E}$, which is equivalent since the LDGM code ensemble is
  symmetric to permutation.} and the decimated value is $b^*\in \SetZ_2$ with probability
$\extprobx{b}{i^*}{b^*}$, which is equivalent to letting $b^* = \oneif{\omegab{i^*}\ge \extprobx{b}{i^*}{0}}$;
this is called the \emph{probabilistic decimator} (PD) and is more
amenable to analysis.\footnote{The PD was previously called the \emph{typical decimator} (TD) in \cite{ldgm-vq-journal} and \cite{flip-ml}, but we find the word ``typical''
  somewhat inaccurate and now consider PD to be the more appropriate name.}  An intuitive alternative is the \emph{greedy decimator}
(GD) which always decimates the ``most certain'' bit, among the set $\mathcal{E}$ of undecimated bit indices, to its most
likely value, i.e.
\begin{equation}
  \label{eq:greedy-dec}
  (i^*, b^*) = \argmax_{(i,b)\in\mathcal{E}\times\SetZ_2} \extprobx{b}{i}{b}.
\end{equation}
As expected, the GD yields better performance than the PD, so it is
more useful in practice, although we will not attempt to analyze it.

\begin{figure}[!t]
  \centering\footnotesize
  \begin{algorithmic}
    \REQUIRE Quantizer parameters $d(\cdot,\cdot)$, $\mat{G}$, $\seqa$, $t$, source sequence $\seqy$
    \ENSURE Quantized codeword $\sequ$ and the corresponding $\seqb$
    \STATE $\priprobx{u}{j}{u} \assign \e^{-td(u, y_j)}$, $j=1,\dotsc,n$, $u=0,1$
    \STATE $\msg{bc}{ij} \assign \constmsg{*}$, $i=1,\dotsc,\nb$, $j \in \neighbor{bc}{i\cdot}$
    \STATE $\priprob{b}{i} \assign \constmsg{*}$, $i=1,\dotsc,\nb$
    \STATE $\mathcal{E} \assign \{1,2,\dotsc,\nb\}$ \COMMENT{the set of bits in $\seqb$ not yet decimated}
    \REPEAT[belief propagation iteration]
      \STATE Adjust the $\priprob{u}{j}$'s with the recovery algorithm
      \FOR[BP computation at check node $j$]{$j=1$ to $n$}
        \STATE $\msg{cb}{ji} \assign (\priprob{u}{j} \oplus \constmsg{a_j})\oplus
        \left(\oplusl_{i'\in\neighbor{bc}{\cdot j} \excluding{i}}
          \msg{bc}{i'j}\right)$, $i\in\neighbor{cb}{j\cdot}$
      \ENDFOR
      \FOR[BP computation at variable node $b_i$]{$i=1$ to $\nb$}
        \STATE $\msg{bc}{ij} \assign \priprob{b}{i} \odot
        \left(\odotl_{j'\in\neighbor{cb}{\cdot i} \excluding{j}}
          \msg{cb}{j'i}\right)$, $j\in\neighbor{bc}{i\cdot}$
        \STATE $\extprob{b}{i} \assign \odotl_{j'\in\neighbor{cb}{\cdot i}} \msg{cb}{j'i}$
      \ENDFOR
      \WHILE{$\mathcal{E} \ne \emptyset$ and more decimation is necessary in this iteration}
        \STATE Choose the bit index $i^*$ to decimate and its value $b^*$
        \STATE $\priprob{b}{i^*} \assign \constmsg{b^*}$,
        $\msg{bc}{i^*j} \assign \constmsg{b^*}$, $j\in\neighbor{bc}{i^*\cdot}$ \COMMENT{decimate $b_i$ to $b^*$}
        \STATE $\mathcal{E} \assign \mathcal{E} \excluding{i^*}$
      \ENDWHILE
      \vspace{0.5mm}
    \UNTIL{$\mathcal{E}=\emptyset$}
    \STATE $b_i \assign 0$ (resp. $1$) if $\priprob{b}{i}=\constmsg{0}$ (or $\constmsg{1}$), $i=1,\dotsc,\nb$
    \STATE $\sequ \assign \seqb \mat{G} \oplus \seqa$
  \end{algorithmic}
  \caption{The binary LDGM quantization algorithm}
  \label{alg:binary}
\end{figure}

In practice, it is important to control the amount of decimation in each iteration (which
we call the \emph{pace of decimation}), so that distortion performance can be optimized under a
limited number of iterations.  Moreover, the recovery algorithm mentioned at the end of
\secref{sec:outline-quant-ana} is also necessary for good performance.  However, these issues can safely be ignored in the theoretical analysis in this paper, and thus will not be considered in detail here; practical algorithms for them have been proposed in \cite{ldgm-vq-journal} and \cite{flip-ml}.

\section{Asymptotic Analysis of the Synchronization Conditions}
\label{sec:asympt-ana-sync}
Compared to the analysis of LDPC decoding via density evolution, the analysis of the LDGM quantizer
is complicated by its use of decimation based on extrinsic information, as well as the lack of a
natural reference codeword corresponding to the all-zero codeword in LDPC analysis.  To solve these
problems, we have introduced the TPQ, the BPPQ and the synchronization conditions, and in this
section we will show that TPQ gives a reference codeword that allows the synchronization conditions
to be analyzed with density evolution methods, for asymptotically large block length $n$ and
iteration count $L$.


We use LDGM codes that are regular at variable nodes $b_i$ and irregular at the check nodes for quantization, as suggested by the
LDGM-LDPC duality in \cite{it-quant-codes-graphs}.  The degree distribution is described by $\db \ge 2$,
the number of 1's in each of the $\nb$ rows of $\mat{G}$, as well as the $w_d$'s, each of which representing the
fraction of columns in $\mat{G}$ with $d$ 1's; we also use $v_d \defeq d w_d/(R \db)$ to
denote the fraction of 1's residing in these columns among the $nR\db$ 1's in the entire $\mat{G}$.  All degrees are assumed to be at least 1.  These degree
distribution parameters satisfy the constraints
\begin{equation}
  \label{eq:dd-constr-bin}
  \sum_d w_d = 1,\quad \sum_d v_d = 1,\quad w_d\ge 0 \text{\ for\ } d=1,2,\dotsc.
\end{equation}
Strictly speaking, a given degree distribution cannot be followed exactly at arbitrary block lengths
$n$ since $nR$ and the $nw_d$'s are not necessarily integers.  To avoid this problem, for each $n$
we pick $\Rn$ and $\wn{d}$ such that $n\Rn$ and all $n\wn{d}$'s are integers, and at the same time
$\Rn \to R$ and $\wn{d} \to w_d$ as $n\to\infty$.  Denoting by $\seqw$ and $\seqwn$ the vector
comprised of respectively the $w_d$'s and the $\wn{d}$'s, we can now redefine $\nb \defeq n\Rn$ and let $\Gdbwb$ be the set of
$\mat{G}$'s with rate $\Rn$ and degree distribution given by $(\db, \seqwn)$.

At each $n$, let $\mat{G}$ be uniformly distributed in $\Gdbwb$, and we then have an ensemble of TPQ
and corresponding BPPQ instances, with one for each $(\mat{G}, \seqy, \omegaas, \omegabs)$ tuple;
when $\mat{G}$, $\seqy$, $\omegaas$ and $\omegabs$ are viewed as random variables, so are the resulting $\seqa$ and $\seqb$ from
either quantizer, as well as the BP priors, messages and extrinsic information.  During the analysis
of the synchronization conditions below, all random variables will be defined over the TPQ ensemble.
In other words, the bits in $\seqa$ used as input for the quantization algorithm are chosen
sequentially as $a_j = \oneif{\omegaa{j}\ge \trueextprobx{a}{j}{0}}$, $j=1,\dotsc,\nc$, and the BP priors, messages and extrinsic information in each iteration
are then defined by following the algorithm in \prettyref{alg:binary}, except that the sequential decimation of each
$b_i$ in $\seqb$ uses $\trueextprob{b}{i}$ from the TPQ formula $b_i = \oneif{\omegab{i}\ge\trueextprobx{b}{i}{0}}$ instead of the BP extrinsic information $\extprob{b}{i}$,
thus yielding the $\seqb$ from TPQ at the end, and we then say the quantization algorithm \emph{follows TPQ}.  In this way, we can
investigate the synchronization conditions when all previous \node{a}- and \node{b}-steps have
yielded TPQ's decimation result, i.e.\ whether the BPPQ will remain synchronized with the
TPQ if it is previously so.  We denote the $\seqb$ and $\seqa$ from TPQ by $\seqbr$ and $\seqar$
respectively, and use them or the corresponding $\sequr \defeq \seqcr \defeq \seqbr\mat{G} \oplus \seqar$ as
the \emph{reference codeword} for density evolution.  Conditioned on a fixed $\mat{G}$, the joint
distribution of $\seqbr$, $\seqar$ and $\sequr$ can be obtained following the discussion in
\prettyref{sec:outline-quant-ana}, as follows:
\begin{proposition}
  \label{prop:ref-dist}
  Conditioned on a fixed $\mat{G}$ (omitted in the conditional probabilities below),
  $(\seqbr, \seqar) \markov \sequr \markov \seqy$ as well as $(\seqbr, \seqar) \markov u^*_j \markov y_j$ for any $j$ form Markov chains,
  $p(\seqbr, \seqar \condmid \sequr) = 2^{-\nb}$ (i.e.\ uniform) for any $(\seqbr, \seqar)$
  satisfying $\seqbr\mat{G}\oplus \seqar = \sequr$, while
  $p(\sequr \condmid \seqy) = \prod_j p_{u\condmid y}(u_j^* \condmid y_j)$ and
  $p(\seqy) = \prod_j p_y(y_j)$ with $p(u\condmid y) = \e^{-t d(u,y)} / Q(y)$ being the test channel
  chosen in \prettyref{sec:outline-quant-ana} and $p(y)$ being the source pdf.  Consequently,
  $p(\sequr) = \prod_j p_u(u_j^*)$ is uniform because $p(u)$ is so according to
  \prettyref{prop:test-channel-sym}, while
  $p(\seqy \condmid \sequr) = \prod_j p_{y\condmid u}(y_j \condmid u_j^*)$, and
  $p(\seqbr, \seqar) = 2^{-(n+\nb)}$ is uniform as well.
\end{proposition}
\begin{IEEEproof}
  See \prettyref{app:proof-ref-dist}.
\end{IEEEproof}

The need to have an explicit reference codeword in density evolution necessitates the
use of some new notations; first of all, we will introduce these
notations and express some known results in terms of them.

\subsection{Review of Binary Message Densities and Their Properties}
\label{sec:binary-def-results}
Given the reference codeword, each variable node $b_i$, $u_j$ or $a_j$ then corresponds to a bit in
the reference codeword, namely $b^*_i$, $u^*_j$ or $a^*_j$, which is a binary random variable.  Consequently, each probability tuple
involved in BP can also be assigned such a bit from the reference codeword as its \emph{reference bit} according to the
associated variable node.  In particular, for binary LDGM quantization, the reference bit of each
$\priprob{b}{i}$, $\extprob{b}{i}$, $\trueextprob{b}{i}$, $\msg{bc}{ij}$ and $\msg{cb}{ji}$ is
$b_i^*$, while that of $\priprob{u}{j}$ and $\priprob{a}{j}$ are $u_j^*$ and $a_j^*$ respectively.

A \emph{message density} (or simply \emph{density}) is a conditional probability distribution of
a probability tuple (itself a random variable) given its reference bit, and is usually shown in bold; for example, the density of
$\msg{bc}{ij}$ (with reference bit $b_i^*$) can be denoted by $\dmsg{bc}$, and we then write
$\msg{bc}{ij}\condmid b_i^* \sim \dmsg{bc}$.  Such a density $\dmu$ can be concretely represented by
the conditional pdf or pmf of $\mu(0)$ or the LLR $\llr(\mu)$ given $b$ when we let
$\mu \condmid b \sim \dmu$, and they are respectively denoted $\dmsgz{}(p\condmid b)$ and
$\dmsgl{}(\llr \condmid b)$.  We also formally write $\dmu(\mu\condmid b)$ as the
conditional pdf if the actual representation of the probability tuple is not of concern, so that
$\mu \condmid b \sim \dmu$ implies $p(\mu\condmid b) = \dmu(\mu\condmid b)$.

Unless otherwise noted, the distributions of all the random variables here, particularly the densities
of probability tuples, are defined with respect to the entire ensemble of TPQ and BPPQ instances
involving all $\mat{G} \in \Gdbwb$.  Sometimes we will also limit our consideration to those
instances involving a specific $\mat{G}$ or subset of $\mat{G}$'s (e.g.\ those with certain loop-free neighborhoods), and obtain the \emph{conditional}
distributions and message densities over this sub-ensemble denoted by e.g.\ $\mathcal{E}$; for
example, if the conditional probability density $p(\mu \condmid b, \mathcal{E})$ can be represented
by message density $\dmu$, then we may write $\mu \condmid b, \mathcal{E} \sim \dmu$.  The
properties of message densities given below are clearly applicable to such conditional densities as
well.

%

The symmetry condition of message densities plays an important role in both LDPC analysis
\cite{design-cap-approaching-ldpc} and here.  Based on the above definitions, symmetry can be defined as
follows:
\begin{definition}
  \label{def:sym-density}
  A message density $\dmu$ is said to be \emph{symmetric} if
  \begin{align}
    \label{eq:msg-sym-01}
    \dmsgz{}(p\condmid 0) &= \dmsgz{}(1-p\condmid 1), \\
    \label{eq:msg-sym-consistency}
    (1-p)\cdot \dmsgz{}(p\condmid 0) &= p\cdot \dmsgz{}(1-p\condmid 0),
  \end{align}
  for all $p\in[0,1]$.  If $\mu \condmid b \sim \dmu$, we then say the
  random probability tuple $\mu$ \emph{has a symmetry density} (\emph{is symmetric})
  with respect to (w.r.t.) $b$; if not stated explicitly, the reference bit $b$ refers to
  that of $\mu$ defined above.
\end{definition}

A message density $\dmu$ can be viewed as a binary-input channel $\dmu(\mu\condmid b)$ with the
reference bit $b$ as input and the probability tuple $\mu$ as output.  Under this view,
\eqref{eq:msg-sym-01} is simply a kind of input symmetry of this channel, commonly
used in LDPC literature when they assume that the correct codeword used as reference is all-zero.
Condition \eqref{eq:msg-sym-consistency} is about the ``consistency'' of the density, i.e.\ whether
each possible channel output $(p,1-p)$ has its likelihood ratio
$\dmsgz{}(p\condmid 0) / \dmsgz{}(p \condmid 1)$ equal to $p/(1-p)$, which can also be formally
expressed as $\dmu(\mu\condmid 0) / \dmu(\mu \condmid 1) = \mu(0) / \mu(1)$ for any $\mu$.  In this paper,
$p(b)$ is often uniform over $\SetZ_2$; if so, then when $\mu$ has a symmetric density
w.r.t.\ $b$, i.e.\ $p(\mu\condmid b=0) / p(\mu\condmid b=1) = \mu(0) / \mu(1)$, we have
\begin{equation}
  \label{eq:msg-sym-p}
  p(b \condmid \mu) \propto p(\mu \condmid b) \propto \mu(b), \text{ i.e.\ } p(b\condmid \mu) = \mu(b),
\end{equation}
where $\propto$ denotes equality up to a factor not containing $b$.
In LLR form \eqref{eq:msg-sym-consistency} becomes $\dmsgl{}(\llr) /
\dmsgl{}(-\llr) = \e^{\llr}$, which is exactly the symmetry condition
in LDPC literature.

Naturally, for any symmetric binary-input channel, its likelihood
function has a symmetric density:
\begin{proposition}
  \label{prop:sym-chan-sym-likelihood}
  Let $b$ be a binary random variable, $y$ be another
  random variable taking values in $\SetY$ and with conditional pmf or pdf $p(y\condmid b)$, and
  $\mu$ be the probability tuple giving the likelihood of $y$, i.e.\
  $\mu(b') = p(y\condmid b')$.  If there exists an measure-preserving group action
  $\psi_b(\cdot)$ of $\SetZ_2$ on $\SetY$, such that
  $\pyb(y\condmid b) = \pyb(\psi_b(y)\condmid 0)$, then
  $\mu\condmid b \sim \dmu$ is a symmetric density.
\end{proposition}
\begin{IEEEproof}
  Theorem~4.27 in \cite{modern-coding-theory} is a proof for the case
  $\SetY = \SetR$ and $\psi_b(\cdot)$ being $\psi_1(y) = -y$.
  This generalization is proved similarly; see
  \prettyref{app:proof-sym-chan-sym-likelihood}.
\end{IEEEproof}

Given a symmetric density $\dmu$, if we let $b$ be an equiprobable
binary random variable and $\mu$ satisfying $\mu\condmid b \sim \dmu$, then for any possible value $\mu'$
of $\mu$, we have $p_{b\condmid \mu}(b\condmid \mu') = \mu'(b)$,
so the entropy $H(b\condmid \mu = \mu') = H(\mu')$; taking the
expectation over $\mu$ we get $H(b\condmid \mu) = \Expe{H(\mu)}$ and
$I(b;\mu) = \Expe{I(\mu)}$.  We thus define $H(\dmu) \defeq
\Expe{H(\mu)}$ and $I(\dmu) \defeq \Expe{I(\mu)}$, and call them
respectively the \emph{entropy} and \emph{mutual information (MI)} of the
symmetric density $\dmu$.

Given densities $\dmu_1, \dotsc, \dmu_m$ and weights $\alpha_1, \dotsc, \alpha_m \in [0,1]$
with $\sum_i \alpha_i = 1$, we can straightforwardly define the \emph{convex combination}
$\dmu = \sum_i \alpha_i\dmu_i$ e.g.\ by making
$\dmu_{(0)}(p\condmid b) = \sum_i \alpha_i \dmu_{(0)}^{i}(p\condmid b)$.  This definition can
naturally be extended to an arbitrary family $(\dmu_I)_{I\in\mathcal{X}}$ of densities weighted
by a probability distribution over $\mathcal{X}$.  Specifically, let $I$ be a random variable taking values in set $\mathcal{X}$ and independent from the reference bit $b$,
and $\mu$ be random probability tuple that depend on both $b$ and $I$, then over the sub-ensemble with a specific $I$, the conditional message density
$\mu \condmid b, I \sim \dmu_I$ may be called the density of $\mu$ \emph{conditioned on $I$}, while
the message density over the entire ensemble $\mu \condmid b \sim \dmu$ is called $\mu$'s density
\emph{(averaged) over all $I\in\mathcal{X}$}; in this case, $\dmu$ is a convex combination of
$(\dmu_I)_{I\in\mathcal{X}}$ weighted by the pmf or pdf of $I$.

Convex combinations of symmetric densities remain symmetric (a more general result, \prettyref{prop:convex-comb-symm}, will be proved in detail).  Conversely, for any $q \in [0,1]$, we may
let $\qx{0} \defeq q$ and $\qx{1} \defeq 1-q$, and define $b$ and $\mu$ such that given $b\in\SetZ_2$, $\mu = (q,1-q)$ with probability $\qx{b}$ and is $(1-q,q)$
otherwise, i.e.\ the conditional pmf
\begin{equation}
  \label{eq:dsym-q}
  p(\mu \condmid b) = \sum_{e \in \SetZ_2} \qx{b \oplus e} \cdot \oneif{\mu(b') = \qx{b' \oplus e}, b'=0,1},
\end{equation}
then the density $\mu \condmid b \sim \dsym{q} = \dsym{1-q}$ is symmetric, and any symmetric density can be expressed as a convex combination of the family
$(\dsym{q})_{q\in[0,1/2]}$.  In this way, many results need only to be proved for $\dsym{q}$, and
they can then be applied to symmetric densities by linearity.

The $\nu(\cdot; \cdot)$ operator for probability tuples defined in \prettyref{sec:binary}, which
includes $\odot$ and $\oplus$ as special cases, can naturally be applied to densities using the following definition:
\begin{definition}
  \label{def:nu-density}
  Given a deterministic affine subspace $\SetCW$ of $\SetZ_2^m$ and $(m-1)$ message densities denoted
  by
  $\dlambda_{\except i} \defeq (\dlambda_1, \dotsc, \dlambda_{i-1}, \dlambda_{i+1}, \dotsc,
  \dlambda_m)$,
  we let $\seqb = (b_1, \dotsc, b_m)$ be uniformly distributed over $\SetCW$, construct $m-1$ random
  binary probability tuples $\lambda_{\except i}$ such that for any $j\ne i$, $\lambda_j$ depends
  only on $b_j$ with $\lambda_j\condmid b_j \sim \dlambda_j$, then the distribution of the
  probability tuple $\nu(\SetCW; \lambda_{\except i})$ conditioned on the reference $b_i$ is the
  message density denoted by $\nu(\SetCW; \dlambda_{\except i})$.
\end{definition}

The properties of this $\nu(\cdot;\cdot)$ operator are reviewed below, and they are also applicable to
$\odot$ and $\oplus$.
\begin{proposition}
  \label{prop:linear-sym}
  If $\dlambda_{\except i}$ are $m-1$ symmetric densities, then
  $\dnu \defeq \nu(\SetCW; \dlambda_{\except i})$ is also symmetric.  Moreover, the
  $\nu \defeq \nu(\mathcal{C}; \lambda_{\except i})$ in \prettyref{def:nu-density} forms a Markov chain
  $\seqb \markov b_i \markov \nu$, so the distribution of $\nu$
  conditioned on $\seqb$ is fully described by $\dnu$.
\end{proposition}
\begin{IEEEproof}
  This is essentially a restatement of
  \cite[Theorem~4.30]{modern-coding-theory} using our definitions and
  notation.  We will prove the more general \prettyref{prop:linear-sym-suff-stat-2K} in \prettyref{app:proof-linear-sym-suff-stat-2K}.
\end{IEEEproof}

\begin{proposition}
  \label{prop:nu-suff-stat}
  Let $\SetCW$ be a deterministic affine subspace of $\SetZ_2^m$, $\seqb$ be
  a random vector uniformly distributed over $\SetCW$,
  $\lambda_{\except i}$ be $(m-1)$ random binary probability tuples with $\lambda_j$ depending only on $b_j$ and
  $\lambda_j\condmid b_j \sim \dlambda_j$ being symmetric for $j\ne i$,
  and $\nu_i = \nu(\SetCW; \lambda_{\except i})$.  Then $\nu_i$
  is a sufficient statistic for $b_i$ given $\lambda_{\except i}$,
  i.e.\ $b_i \markov \nu_i \markov \lambda_{\except i}$ forms a Markov
  chain.
\end{proposition}
\begin{IEEEproof}
  The more general \prettyref{prop:linear-sym-suff-stat-2K} will be proved in \prettyref{app:proof-linear-sym-suff-stat-2K}.
\end{IEEEproof}

When $b \markov \mu_1 \markov \mu_2$ forms a Markov chain, we say $\mu_2$ is a
\emph{physically degraded} version of $\mu_1$ with respect to $b$,
denoted by $\mu_2 \lepd \mu_1$ when the reference bit
$b$ is unambiguous.  In particular, we always have
$\constmsg{*} \lepd \mu_1 \lepd \constmsg{b}$.  Given two densities
$\dmu_1$ and $\dmu_2$, if random probability tuples $\mu_1$ and
$\mu_2$ can be constructed for an arbitrary binary random variable $b$
such that $\mu_1 \condmid b \sim \dmu_1$,
$\mu_2 \condmid b \sim \dmu_2$ and $\mu_1 \lepd \mu_2$ w.r.t.\ $b$, we
say $\dmu_2$ is a \emph{degraded} version of $\dmu_1$ and write
$\dmu_2 \lepd \dmu_1$.  By the data processing inequality, if
$\dmu_2 \lepd \dmu_1$ are both symmetric, then
$I(\dmu_2) \le I(\dmu_1)$ because this is equivalent to
$I(b;\mu_2) \le I(b;\mu_1)$ for an equiprobable $b$.
(Physical) degradation relationships among symmetric densities are also preserved by convex combinations (recall that the index variable must be independent from the reference bit),
as well as the $\nu(\cdot; \cdot)$ (and thus $\odot$ and $\oplus$) operations:
\begin{proposition}
  \label{prop:pdeg-convex-comb}
  Let $I$ be an arbitrary random variable, $b$ be uniformly distributed over $\SetZ_2$ and independent from $I$,
  and $\mu$ and $\nu$ be random binary probability tuples that, when conditioned on $I$, are symmetric w.r.t.\ $b$ and satisfy
  $\nu \lepd \mu$ w.r.t.\ $b$.  In this case, after averaging over all $I$, we still have $\nu \lepd \mu$ w.r.t.\ $b$.
\end{proposition}
\begin{IEEEproof}
  A generalized version \prettyref{prop:pdeg-convex-comb-2K} will be proved in \secref{sec:prob-tuple-SetG}.
\end{IEEEproof}
\begin{proposition}
  \label{prop:linear-pdeg}
  Let $\SetCW$ be a deterministic affine subspace of $\SetZ_2^m$, $\seqb$ be a
  random vector uniformly distributed over $\SetCW$, and
  $\lambda_{\except i}$ and $\lambda'_{\except i}$ each be $m-1$
  random binary probability tuples such that for each $j \ne i$,
  \begin{itemize}
  \item $\lambda_j$ and $\lambda'_j$ depend only on bit $b_j$ in $\seqb$,
    with $\lambda_j\condmid b_j \sim \dlambda_j$ and $\lambda'_j \condmid b_j \sim \dlambda'_j$ both being symmetric densities,
  \item $\lambda'_j \lepd \lambda_j$ w.r.t.\ $b_j$.
  \end{itemize}
  Now let $\nu_i = \nu(\SetCW; \lambda_{\except i})$ and $\nu'_i =
  \nu(\SetCW; \lambda'_{\except i})$, then $\nu'_i \lepd \nu_i$ w.r.t.\ $b_i$.
\end{proposition}
\begin{IEEEproof}
  Similar to \cite[Lemma~4.82]{modern-coding-theory}; we will give the proof of the more general
  \prettyref{prop:linear-pdeg-2K} in \prettyref{app:proof-linear-pdeg-2K}.  Note that $\nu_i$ being
  a sufficient statistic is important; the result would not hold if $\nu_i$ loses too much
  information from $\lambda_{\except i}$ that $\nu'_i$ happens to retain.
\end{IEEEproof}
\begin{proposition}
  \label{prop:linear-ddeg}
  Let $\SetCW$ be a deterministic affine subspace of $\SetZ_2^m$,
  and $\dlambda_{\except i}$ and $\dlambda'_{\except i}$ each be
  $m-1$ symmetric densities with $\dlambda'_j \lepd \dlambda_j$ for
  all $j \ne i$, then
  $\nu(\SetCW; \dlambda'_{\except i}) \lepd \nu(\SetCW;
  \dlambda_{\except i})$.
\end{proposition}
\begin{IEEEproof}
  This is an obvious corollary to \prettyref{prop:linear-pdeg}.
\end{IEEEproof}

Physical degradation relationships
enable us to prove the closeness of individual probability tuples from the
synchronization conditions by comparing the average MIs:
\begin{proposition}
  \label{prop:msg-compare-mi}
  Given an equiprobable binary random variable $b$ and two random probability tuples
  $\mu_1$ and $\mu_2$ such that $\mu_2 \lepd \mu_1$ w.r.t.\ $b$.  If $\mu_1\condmid b\sim \dmu_1$ and $\mu_2\condmid b \sim
  \dmu_2$ are both symmetric densities, then
  \begin{equation}
    \label{eq:msg-compare-mi}
    \Expe{(\mu_1(0) - \mu_2(0))^2} \le \frac{\ln 2}{2} (I(\dmu_1) - I(\dmu_2)).
  \end{equation}
  This implies that $I(\dmu_2) \le I(\dmu_1)$, which is also obvious
  from the data processing inequality.
\end{proposition}
\begin{IEEEproof}
  Similar to \cite[Lemma 15]{gen-area-theorem-conseq}; see
  \prettyref{app:proof-msg-compare-mi}.
\end{IEEEproof}
Conversely, we have the following result:
\begin{proposition}
  \label{prop:msg-compare-mi-converse}
  For any $\epsilon > 0$ there exists a $\delta > 0$ such that, given an equiprobable binary random
  variable $b$ and two random probability tuples $\mu_1$ and $\mu_2$ with
  $\mu_1 \condmid b \sim \dmu_1$ and $\mu_2 \condmid b \sim \dmu_2$ being symmetric densities, if
  $\abs{I(\dmu_1) - I(\dmu_2)} \ge \epsilon$, then $\Expe{(\mu_1(0) - \mu_2(0))^2} \ge \delta$.
\end{proposition}
\begin{IEEEproof}
  Since $\Expe{\abs{I(\mu_1) - I(\mu_2)}} \ge \abs{I(\dmu_1) - I(\dmu_2)} \ge \epsilon$, and
  $\abs{I(\mu_1) - I(\mu_2)} \le 1$ with probability 1, we have
  \begin{equation}
    \Pr{\abs{I(\mu_1) - I(\mu_2)} \ge \epsilon/2} \ge \epsilon/2.
  \end{equation}
  Now $I(\mu_1)$ is a continuous function of $\mu_1(0)$ over $[0,1]$ and thus uniformly continuous,
  so there exists a $\delta' > 0$ such that $\abs{I(\mu_1) - I(\mu_2)} \ge \epsilon/2$ implies that
  $\abs{\mu_1(0) - \mu_2(0)} \ge \delta'$.  Therefore, letting
  $\delta = (\delta')^2 \cdot \epsilon/2$ leads to the desired result.
\end{IEEEproof}

An important class of symmetric densities is the \emph{erasure-like}
densities defined as follows:
\begin{definition}
  \label{def:erasure-like-density}
  For $x\in[0,1]$, let $b$ be a binary random variable, and $\mu$ be
  a random probability tuple that equals $\constmsg{b}$ with
  probability $x$ and $\constmsg{*}$ with probability $1-x$, then we
  define the resulting density $\mu\condmid b \sim \dbec{x}$ and
  call such densities \emph{erasure-like}.
  In particular, $\dbec{0}$ and $\dbec{1}$ are respectively the \emph{always-unknown} and
  \emph{always-sure} densities.
\end{definition}

Erasure-like densities are thus similar to binary erasure channels
(BECs) and have the following simple properties, whose proofs are omitted here:
\begin{proposition}
  \label{prop:erasure-like-density}
  For any $x, x_1, x_2 \in [0,1]$,
  \begin{itemize}
  \item $\dbec{x}$ is symmetric with $I(\dbec{x}) = x$;
  \item $\dbec{x_1} \odot \dbec{x_2} = \dbec{x}$ where
    $x=1-(1-x_1)(1-x_2)$;
  \item $\dbec{x_1} \oplus \dbec{x_2} = \dbec{x_1x_2}$;
  \item $\sum_i \alpha_i \dbec{x_i} = \dbec{x}$, where $x = \sum_i \alpha_i x_i$;
  \item If $x_1 \le x_2$, then $\dbec{x_1} \lepd \dbec{x_2}$.
  \end{itemize}
\end{proposition}

Moreover, the $\nu(\SetCW; \cdot)$ operator preserves erasure-like densities:
\begin{proposition}
  \label{prop:erasure-nu}
  If $\SetCW$ is a deterministic affine subspace of $\SetZ_2^m$ and
  $\dlambda_{\except i}$ are $m-1$ erasure-like message densities, then
  $\dnu \defeq \nu(\SetCW; \dlambda_{\except i})$ is also erasure-like.
\end{proposition}
\begin{IEEEproof}
  See \prettyref{app:proof-erasure-nu}.
\end{IEEEproof}

\subsection{Synchronization at \node{b}-steps}
\label{sec:sync-b}
We now analyze the synchronization condition at the $i$-th \node{b}-step of the TPQ; namely,
assuming that all \node{a}-steps and the previous \node{b}-steps have followed TPQ to yield
$\seqa = \seqar$ and $b_{i'} = b^*_{i'}$ for all $i'<i$, whether the $\extprob{b}{i}$ obtained in
\node{b}-step $i$ approaches $\trueextprob{b}{i}$ after a large number of BP iterations, so that
BPPQ can maintain synchronization with the TPQ after this \node{b}-step.

BPPQ in the actual quantization algorithm starts with the $\msg{bc}{i'j}$'s being all-$\constmsg{*}$ and updates them with BP
across all \node{b}-steps.  To simplify the analysis of one specific \node{b}-step
here, we instead assume that the $\msg{bc}{i'j}$'s are reinitialized to all-$\constmsg{*}$ at the
beginning of this \node{b}-step, BP is carried out for $L$ iterations, and the resulting
$\extprob{b}{i}$ is used as the $\textprob{b}{i}$ in decimation.  While such treatment is
inefficient in practice, it is straightforward to prove via physical degradation arguments that, in
terms of whether the synchronization condition is asymptotically satisfied (in the sense of
\prettyref{prop:sync-lm} below), it is equivalent to the actual algorithm.  This $\extprob{b}{i}$
obtained from $L$ BP iterations starting from all-$\constmsg{*}$ $\msg{bc}{i'j}$'s is henceforth
denoted by $\extbil{L}$; on the other hand, if every $\msg{bc}{i'j}$ is hypothetically initialized to hard decision
$\constmsg{b_{i'}^*}$ before the $L$ BP iterations, the resulting $\extprob{b}{i}$ is denoted by
$\extbiu{L}$.

\begin{figure*}[!t]
  \centering
  \subfigure[One layer of the neighborhood]{\label{fig:neigh-binary-layer}\includegraphics{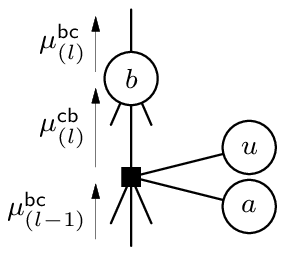}}\qquad
  \subfigure[Neighborhood $\neighiL$ around variable node $b_i$ ($L=2$)]{\label{fig:neighi}\includegraphics{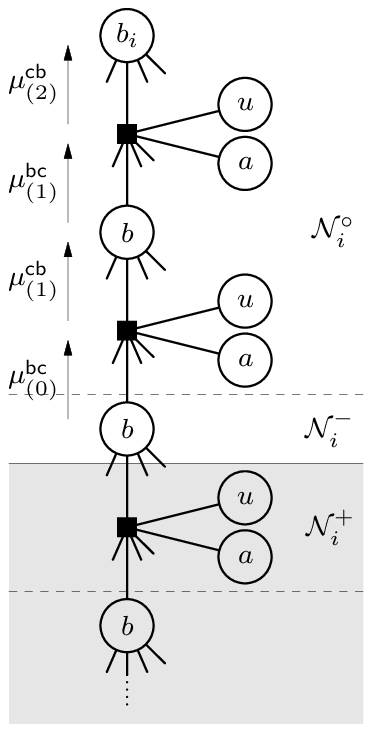}}\qquad
  \subfigure[Neighborhood $\neighjL$ around variable node $a_j$ ($L=1$)]{\label{fig:neighj}\includegraphics{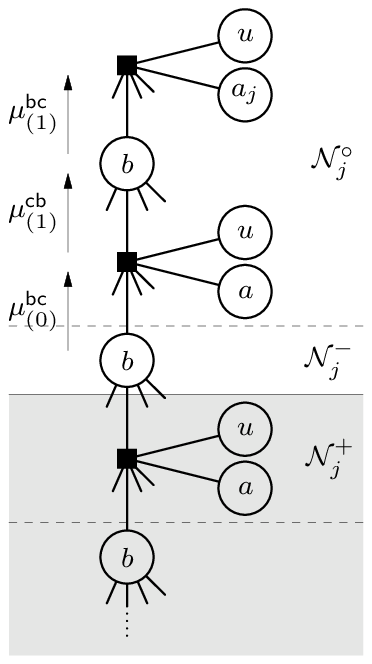}}%
  \caption{Neighborhoods of variable nodes $b_i$ and $a_j$ involved in $L$ BP iterations.
    Subscripts have been omitted except those of the central nodes $b_i$ and $a_j$
    themselves.}%
  \label{fig:neighij}%
\end{figure*}

In the factor graph \prettyref{fig:binary-fg-a}, these $L$ BP iterations involve a
\emph{neighborhood} $\neighi = \neighiL$ of the variable node $b_i$, which can be further divided
into the \emph{interior part} $\neighio$ and the \emph{border part} $\neighib$.
\prettyref{fig:neighi} illustrates the structure of the factor graph around variable node $b_i$,
with $\neighi$ being the entire unshaded region, in which each layer shown in
\prettyref{fig:neigh-binary-layer} corresponds to one BP iteration.  Only the priors of the variable
nodes in $\neighio$, and the initial BP messages from variable nodes in $\neighib$ to check nodes in
$\neighio$ (labeled with $\msgsiter{bc}{0}$ in \prettyref{fig:neighi}), affect $\extbil{L}$ and
$\extbiu{L}$.  Below we will use e.g.\ $\node{b}_{i'} \in \neighio$ to express that the variable
node $b_{i'}$ (denoted by $\node{b}_{i'}$ to avoid confusion with the value of $b_{i'}$) is in the
neighborhood $\neighio$.

Analysis of the BP process frequently requires $\neighiL$ to be loop-free.  Given the degree distributions,
$L$, $n$ and $i$, we use $\GdbwbiL$ to denote the sub-ensemble of $\Gdbwb$ with a loop-free $\neighiL$.
If $\mat{G}$ is uniformly distributed over $\Gdbwb$, the probability that $\mat{G} \notin \GdbwbiL$
obviously does not vary with $i$, and is thus denoted by $\PloopbnL$.  Using the methods employed in
LDPC analysis (e.g.\ the proof of Theorem~1 in \cite{improved-ldpc-irregular}), it is possible to
prove that
\begin{equation}
  \label{eq:Ploopb-lim}
  \lim_{n\to\infty} \PloopbnL = 0.
\end{equation}
for any degree distribution and $L$.

Now consider a fixed $\mat{G} \in \GdbwbiL$.  Define
\begin{equation}
  \label{eq:SetCW}
  \SetCW \defeq \{ (\seqb, \seqa, \sequ) \condmid \sequ = \seqb\mat{G} \oplus \seqa \},
\end{equation}
then $\SetCW$ is a linear (and thus affine) subspace of $\SetZ_2^{\nb+\nc+n}$, and by
\prettyref{prop:ref-dist}, $(\seqbr, \seqar, \sequr)$ is uniformly distributed over it when conditioned on $\mat{G}$.  Given any
priors $\tpris{a} \defeq (\tpriprob{a}{1}, \dotsc, \tpriprob{a}{\nc})$,
$\tpris{u} \defeq (\tpriprob{u}{1}, \dotsc, \tpriprob{u}{n})$,
$\tprisx{b}{i} \defeq (\tpriprob{b}{1}, \dotsc, \tpriprob{b}{i-1}, \tpriprob{b}{i+1}, \dotsc,
\tpriprob{b}{\nb})$,
the result of $\nu(\SetCW; \tprisx{b}{i}, \tpris{a}, \tpris{u}) \defeq \nu$ is then
\begin{equation}
  \nu(b) = \sum_{\substack{(\seqb, \seqa, \sequ) \in \SetCW \\ b_i = b}}
  \prod_{i'\ne i} \tpriprobx{b}{i'}{b_{i'}}
  \prod_{j} \tpriprobx{a}{j}{a_j} \prod_{j} \tpriprobx{u}{j}{u_j},
\end{equation}
which is also the true extrinsic information at $b_i$ on the factor graph in
\prettyref{fig:binary-fg-a} given those priors.  In particular, if the priors are those used in the
quantization algorithm, i.e.\ $\priprobx{u}{j}{u} = \e^{-t d(u,y_j)}$ and
$\priprob{a}{j} = \constmsg{a^*_j}$ for any $j$, $\priprob{b}{i'}=\constmsg{b^*_{i'}}$ for
$i'<i$ (the decimated positions) and $\constmsg{*}$ for $i'>i$, then
$\nu(\SetCW; \prisx{b}{i}, \pris{a}, \pris{u}) = \trueextprob{b}{i}$.  Now we will prove that
$\extbil{L}$ and $\extbiu{L}$ can be expressed in the form of $\nu(\SetCW; \cdot)$ as well.
\begin{proposition}
  \label{prop:ext-nu-b}
  If $\neighiL$ is loop-free, then $\extbil{L}$ and $\extbiu{L}$ are both equal to
  $\nu(\SetCW; \tprisx{b}{i}, \tpris{a}, \tpris{u})$, with the priors $\tprisx{b}{i}$, $\tpris{a}$
  and $\tpris{u}$ given in \prettyref{tab:ext-nu-b-priors}.
\end{proposition}

\begin{table}[!t]
  \centering
  \caption{The priors corresponding to $\protect\extbil{L}$, $\trueextprob{b}{i}$ and $\extbiu{L}$}
  \label{tab:ext-nu-b-priors}
  \begin{tabular}{cccc}
    \toprule
    & $\extbil{L}$ & $\trueextprob{b}{i}$ & $\extbiu{L}$ \\
    \midrule
    $\tpriprob{b}{i'},\ \node{b}_{i'}\in\neighio$    & $\priprob{b}{i'}$ & $\priprob{b}{i'}$ & $\priprob{b}{i'}$ \\
    $\tpriprob{b}{i'},\ \node{b}_{i'}\notin\neighio$ & $\constmsg{*}$    & $\priprob{b}{i'}$ & $\constmsg{b^*_{i'}}$ \\
    $\tpriprob{a}{j},\ \node{a}_j\in\neighio$     & $\constmsg{a^*_j}$ & $\constmsg{a^*_j}$ & $\constmsg{a^*_j}$ \\
    $\tpriprob{a}{j},\ \node{a}_j\notin\neighio$  & $\constmsg{*}$    & $\constmsg{a^*_j}$ & $\constmsg{a^*_j}$ \\
    $\tpriprob{u}{j},\ \node{u}_j\in\neighio$     & $\priprob{u}{j}$  & $\priprob{u}{j}$  & $\priprob{u}{j}$ \\
    $\tpriprob{u}{j},\ \node{u}_j\notin\neighio$  & $\constmsg{*}$    & $\priprob{u}{j}$  & $\priprob{u}{j}$ \\
    \bottomrule
  \end{tabular}
\end{table}

\begin{IEEEproof}
  $\neighi$ forms a loop-free subgraph of the factor graph, so the true extrinsic information at
  $b_i$ on it can be obtained exactly with BP, and it is just $\extbil{L}$ or $\extbiu{L}$ depending
  on whether the priors at the variable nodes $b_{i'}$ in $\neighib$ are $\constmsg{*}$ or
  $\constmsg{b^*_{i'}}$.  What we need to prove now is that
  $\nu(\SetCW; \tprisx{b}{i}, \tpris{a}, \tpris{u})$, being the true extrinsic information at $b_i$
  on the complete factor graph, is also equal to $\extbil{L}$ or $\extbiu{L}$.  It is thus necessary
  to show that the loop-free part $\neighi$ can be separated from the remaining, usually loopy, part
  of the factor graph, so the latter does not affect the true extrinsic information at $b_i$ apart
  from a normalization factor.

  For $\extbil{L}$, we will remove the part of the factor graph labeled by $\neighix$ in
  \prettyref{fig:neighi}, thus separating $\neighi$ from the rest of the factor graph.  We note that each check
  node $j$ in $\neighix$ correspond to a factor
  \begin{equation}
    f_j(u_j, a_j, \seqb) \defeq \oneif{u_j \oplus a_j \oplus (\seq{b}\mat{G})_j = 0},
  \end{equation}
  and variable $a_j$ only occurs in this factor and the prior $\tpriprob{a}{j}$ (correspondingly,
  the variable node $a_j$ is only connected to check node $j$).  By definition, the true
  extrinsic information at $b_i$ over the complete factor graph is given by the product of the
  factors corresponding to the function nodes and to the priors at variable nodes other than
  $b_i$, then summed over all variables other than $b_i$.  Here the summation over $a_j$ involves just the two factors
  $f_j(u_j, a_j, \seqb) \tpriprobx{a}{j}{a_j}$ it appears in, and when we let
  $\tpriprob{a}{j} = \constmsg{*}$, since $\tpriprobx{a}{j}{a_j}$ is always
  $\frac{1}{2}$ while $f_j(u_j, a_j, \seqb)$ is once 0 and once 1 as $a_j$ varies over $\{0,1\}$, this summation over
  $a_j$ also gives a constant $\frac{1}{2}$, thus eliminating the factor $f_j(u_j, a_j, \seqb)$; in
  other words, the check node $j$ and variable $a_j$ in the factor graph can be removed without
  affecting the true extrinsic information at $b_i$, and what remains is $\neighi$ along with a
  subgraph disconnected from it, so the true extrinsic information at $b_i$ on the entire factor
  graph can equivalently be computed on just $\neighi$, giving $\extbil{L}$.

  For $\extbiu{L}$, we note that any variable node $b_{i'}$ in $\neighib$ now has prior
  $\tpriprob{b}{i'} = \constmsg{b^*_{i'}}$, so in the summation formula yielding the true extrinsic
  information at $b_i$, any non-zero term has $b_{i'} = b^*_{i'}$.  For any check node $j$ in
  $\neighix$ connected to $b_{i'}$, the corresponding factor $f_j(u_j, a_j, \seqb)$ can then have
  $b^*_{i'}$ substituted for $b_{i'}$, thus breaking the edge between check node $j$ and variable
  node $b_{i'}$.  In this way $\neighi$ also gets separated from the rest of the factor graph.
\end{IEEEproof}

\begin{remark}
  The scrambling sequence $\seqa$ plays an important role in eliminating the impact of the possibly
  loopy part of the factor graph beyond $\neighiL$, thus allowing us to relate $\trueextprob{b}{i}$,
  which involves the entire factor graph, to its BP counterparts involving $\neighiL$ only.
  Incidentally, the closely related result about the relationship between exact and BP extrinsic
  information in LDPC decoding, \cite[Theorem~9]{gen-area-theorem-conseq}, apparently requires
  similar treatment as well: the BP estimate of a transmitted bit $X_i$ there is not simply
  $E[X_i \condmid Y_{\sim i}^{(l)}]$, as stated in that paper, and instead the parity constraints
  beyond the $l$-iteration neighborhood must be ignored when taking that expectation.  Introducing a
  scrambling sequence there and making its bits at those parity constraints have
  $\constmsg{*}$-priors seems to be an effective way to achieve this, as demonstrated in the above
  proof regarding $\extbil{L}$.
\end{remark}

Given $i'$ and $j$ and conditioned on a fixed $\mat{G}$, all the $\tpriprob{b}{i'}$'s and $\tpriprob{a}{j}$'s in \prettyref{tab:ext-nu-b-priors}
are deterministic given their respective reference bits, $b^*_{i'}$ and $a^*_j$, and their densities are either $\dbec{0}$ or $\dbec{1}$, which
are also symmetric.  As for $\priprob{u}{j}$ (and thus $\tpriprob{u}{j}$), since it is a function of $y_j$, and $y_j \markov u^*_j \markov (\seqbr,\seqar)$ is a Markov chain by \prettyref{prop:ref-dist}, we see that it depends only on $u^*_j$ among $(\seqbr,\seqar,\sequr)$.  It is easy to prove that
the density of $\priprob{u}{j}$ w.r.t.\ $u^*_j$ conditioned on $\mat{G}$ is symmetric as well:
\begin{proposition}
  \label{prop:priprob-u-sym}
  Given a binary symmetric source coding problem, generator matrix $\mat{G}$ and parameter $t>0$,
  the $\priprob{u}{j}$ as defined in \eqref{eq:priprobs-u} has a symmetric density w.r.t.\ $u^*_j$
  conditioned on $\mat{G}$, and this density does not vary with $\mat{G}$ or $j$.
\end{proposition}
\begin{IEEEproof}
  We know from \prettyref{prop:ref-dist} that the conditional pdf
  $p(y_j\condmid u_j^*) = p_{y\condmid u}(y_j \condmid u_j^*)$ is given by the test channel
  $p(y\condmid u)$, which by \prettyref{prop:test-channel-sym} satisfies
  $p_{y\condmid u}(y\condmid 1) = p_{y\condmid u}(\psi_1(y)\condmid 0)$, so
  \prettyref{prop:sym-chan-sym-likelihood} can be applied to prove that the likelihood function of
  $y=y_j$ has a symmetric density w.r.t.\ $u=u^*_j$.  Now $\priprobx{u}{j}{u} = \e^{-t d(u,y_j)}$,
  so by the definition of the test channel, when viewed as a function of $u$ it is proportional to
  $p_{u\condmid y}(u\condmid y_j)$ and thus $p_{y\condmid u}(y_j \condmid u)$, i.e.\
  $\priprob{u}{j}$ is exactly the said likelihood function of $y_j$ on the test channel.  Therefore,
  $\priprob{u}{j}$ has a symmetric density w.r.t.\ $u^*_j$, and this density is determined by the
  test channel only, so it does not vary with $\mat{G}$ and $j$.
\end{IEEEproof}

Combining the results of \prettyref{prop:ext-nu-b} and \prettyref{prop:priprob-u-sym}, we can
immediately apply \prettyref{prop:linear-sym} and \prettyref{prop:linear-pdeg} to see that, when
conditioned on a fixed $\mat{G} \in \GdbwbiL$ (which is also in $\Gdbwbil$ for any $l\le L$) and
using $b^*_i$ as the reference bit (which is 0 or 1 with equal probability independent from $\mat{G}$ by
\prettyref{prop:ref-dist}), we have
\begin{equation}
  \label{eq:extbi-degradation}
  \extbil{1} \lepd \dotsb \lepd \extbil{L} \lepd \trueextprob{b}{i}
  \lepd \extbiu{L} \lepd \dotsb \lepd \extbiu{1},
\end{equation}
and all these probability tuples have symmetric densities.
By \prettyref{prop:msg-compare-mi}, the mean-square differences among these probability tuples can be upper-bounded
with the MI differences of their densities.

Averaging over all $\mat{G}$ with loop-free $\neighiL$, we obtain the densities defined over
$\GdbwbiL$ for iteration counts $l=1,2,\dotsc,L$,
\begin{equation}
  \begin{split}
    \trueextprob{b}{i} \condmid b_i^*, \mat{G} \in \GdbwbiL &\sim
    \dextbitL, \\
    \extbil{l}\condmid b_i^*, \mat{G} \in \GdbwbiL &\sim \dextbil{l;L}, \\
    \extbiu{l}\condmid b_i^*, \mat{G} \in \GdbwbiL &\sim \dextbiu{l;L}.
  \end{split}
\end{equation}
These densities, being convex combinations of the densities conditioned on individual $\mat{G}$'s (note that we need $b_i^*$ and $\mat{G}$ to be independent when taking the convex combination, which is true since $p(b_i^* \condmid \mat{G}) = 1/2$ for any $b_i^*$ and $\mat{G}$ due to \prettyref{prop:ref-dist}), clearly remain
symmetric.  Using \prettyref{prop:pdeg-convex-comb}, the physical degradation relationships are also preserved, thus
\begin{equation}
  \label{eq:dextbi-degradation}
  \dextbil{1;L} \lepd \dotsb \lepd \dextbil{L;L} \lepd \dextbitL
  \lepd \dextbiu{L;L} \lepd \dotsb \lepd \dextbiu{1;L},
\end{equation}
and the bound from \prettyref{prop:msg-compare-mi} can likewise be averaged to yield
\begin{align}
  &\quad \frac{2}{\ln 2}\Expe{(\extbil{l}(0) - \trueextprobx{b}{i}{0})^2 \Bigcondmid \mat{G} \in \GdbwbiL} \\
  \label{eq:b-sync-bound-t}
  &\le I(\dextbitL) - I(\dextbil{l;L}) \\
  \label{eq:b-sync-bound}
  &\le I(\dextbiu{l;L}) - I(\dextbil{l;L})
\end{align}
for any $l\le L$, which bounds the amount of synchronization error at the $i$-th
\node{b}-step.  The hard-to-compute $\dextbitL$ has been eliminated from this bound, leaving only $\dextbiu{l;L}$ and $\dextbil{l;L}$,
which in the $n\to\infty$ limit can be obtained via DE.

\subsection{Synchronization at \node{a}-steps}
\label{sec:sync-a}
Now we analyze the synchronization condition at the $j$-th
\node{a}-step of the TPQ, namely whether $\trueextprob{a}{j}$ is close
to $\constmsg{*}$.  To make analysis feasible, analogous to
the $\extbiu{L}$ above, we define an ``upper bound'' of
$\trueextprob{a}{j}$ denoted by $\extaju{L}$ by hypothetically running BP for $L$ iterations starting with all
$\msg{bc}{ij'} = \constmsg{b^*_i}$, as shown in \prettyref{alg:binary-a}.

\begin{figure}[!t]
  \centering\footnotesize
  \begin{algorithmic}
    \REQUIRE $\mat{G}$, $\seqbr$ as well as $\priprob{u}{j'}$ and $\priprob{a}{j'}$ for all $j'$ ($\priprob{b}{i}=\constmsg{*}$ for all $i$)
    \ENSURE $\extaju{L}$
    \STATE $\msg{bc}{ij'} \assign \constmsg{b^*_i}$, $i=1,\dotsc,\nb$, $j' \in \neighbor{bc}{i\cdot}$
    \FOR[$L$ iterations]{$l=1$ to $L$}
      \FOR[BP computation at check node $j'$]{$j'=1$ to $\nc$}
        \STATE $\msg{cb}{j'i} \assign (\priprob{u}{j'} \oplus \priprob{a}{j'})\oplus
        \left(\oplusl_{i'\in\neighbor{bc}{\cdot j'} \excluding{i}}
          \msg{bc}{i'j'}\right)$, $i\in\neighbor{cb}{j'\cdot}$
      \ENDFOR
      \FOR[BP computation at variable node $b_i$]{$i=1$ to $\nb$}
        \STATE $\msg{bc}{ij'} \assign \odotl_{j''\in\neighbor{cb}{\cdot i} \excluding{j'}}
          \msg{cb}{j''i}$, $j'\in\neighbor{bc}{i\cdot}$
      \ENDFOR
    \ENDFOR
    \STATE $\extaju{L} \assign \priprob{u}{j} \oplus
        \left(\oplusl_{i'\in\neighbor{bc}{\cdot j}} \msg{bc}{i'j}\right)$
  \end{algorithmic}
  \caption{An algorithmic definition of $\extaju{L}$.}
  \label{alg:binary-a}
\end{figure}

Again, the computation of $\extaju{L}$ only involves a neighborhood of variable node $a_j$ in
the factor graph, as shown in \prettyref{fig:neighj} and denoted by $\neighj = \neighjL$,
and it can be further divided into the interior part $\neighjo$ and the border part $\neighjb$,
with each repetition unit in \prettyref{fig:neigh-binary-layer} corresponding to one
iteration.  Given the degree distribution, $L$, $n$ and $j$, the set of $\mat{G} \in \Gdbwb$ with a
loop-free $\neighjL$ is denoted by $\GdbwbjL$; the probability that a uniformly distributed
$\mat{G}$ over $\Gdbwb$ lies outside $\GdbwbjL$ is again independent of $j$, and can be denoted by
$\PloopanL$ that satisfies $\lim_{n\to\infty} \PloopanL = 0$.

For any priors $\tpris{b}$, $\tprisx{a}{j}$ and $\tpris{u}$,
$\nu \defeq \nu(\SetCW; \tpris{b}, \tprisx{a}{j}, \tpris{u})$ is now the true extrinsic information
corresponding to these priors at variable node $a_j$ in the factor graph in
\prettyref{fig:binary-fg-a}.  In particular, $\trueextprob{a}{j}$ as defined in \eqref{eq:trueexta}
is equal to $\nu(\SetCW; \pris{b}, \prisx{a}{j}, \pris{u})$, where $\priprob{b}{i} = \constmsg{*}$
for all $i$, $\priprob{a}{j'} = \constmsg{a^*_{j'}}$ for $j'<j$ (i.e.\ at the positions decimated in
previous \node{a}-steps) and is $\constmsg{*}$ for $j'>j$, and
$\priprobx{u}{j'}{u} = \e^{-t d(u,y_{j'})}$ for all $j'$.  When $\neighjL$ is loop-free, similar to
\prettyref{prop:ext-nu-b}, we can prove that $\extaju{L}$ can be expressed in this form as well:
\begin{proposition}
  \label{prop:ext-nu-a}
  If $\neighjL$ is loop-free, then $\extaju{L} = \nu(\SetCW; \tpris{b}, \tprisx{a}{j}, \tpris{u})$,
  with the priors $\tpris{b}$, $\tprisx{a}{j}$ and $\tpris{u}$ given by
  \prettyref{tab:ext-nu-a-priors}.
\end{proposition}

\begin{table}[!t]
  \centering
  \caption{The priors corresponding to $\trueextprob{a}{j}$ and $\extaju{L}$}
  \label{tab:ext-nu-a-priors}
  \begin{tabular}{ccc}
    \toprule
    & $\trueextprob{a}{j}$ & $\extaju{L}$ \\
    \midrule
    $\tpriprob{b}{i},\ \node{b}_i\in\neighjo$    & $\constmsg{*}$ & $\constmsg{*}$ \\
    $\tpriprob{b}{i},\ \node{b}_i\notin\neighjo$ & $\constmsg{*}$ & $\constmsg{b^*_i}$ \\
    $\tpriprob{a}{j'},\ \node{a}_{j'}\in\neighjo$     & $\priprob{a}{j'}$ & $\priprob{a}{j'}$ \\
    $\tpriprob{a}{j'},\ \node{a}_{j'}\notin\neighjo$  & $\priprob{a}{j'}$ & $\constmsg{a^*_{j'}}$ \\
    $\tpriprob{u}{j'},\ \node{u}_{j'}\in\neighjo$     & $\priprob{u}{j'}$ & $\priprob{u}{j'}$ \\
    $\tpriprob{u}{j'},\ \node{u}_{j'}\notin\neighjo$  & $\priprob{u}{j'}$ & $\priprob{u}{j'}$ \\
    \bottomrule
  \end{tabular}
\end{table}

\begin{IEEEproof}
  Let $\nu \defeq \nu(\SetCW; \tpris{b}, \tprisx{a}{j}, \tpris{u})$, where the priors are those for $\extaju{L}$ in the table.
  $\nu$ is then the true extrinsic information at $a_j$ in the factor
  graph corresponding to these priors.  Similar to the treatment of $\extbiu{L}$ in the proof of \prettyref{prop:ext-nu-b}, since
  the variable nodes $b_i$ in $\neighjb$ have prior $\tpriprob{b}{i} = \constmsg{b^*_i}$, they can
  be disconnected from the check nodes in $\neighjx$ and have $b^*_i$ substituted into the
  corresponding factors.  After such a transformation, the $\neighj$ part of the factor graph
  becomes disconnected from the rest, and the true extrinsic information at $a_j$ on this tree-like
  part of the factor graph, which is still equal to $\nu$, can now be exactly computed with BP using
  the algorithm in \prettyref{alg:binary-a}.
\end{IEEEproof}

Combining \prettyref{prop:ext-nu-a} and \prettyref{prop:priprob-u-sym} with
Propositions~\ref{prop:linear-sym} and \ref{prop:linear-pdeg}, we again find that, conditioned on a
fixed $\mat{G} \in \GdbwbjL$ (which is thus also in $\Gdbwbjl$ for any $l\le L$) and using $a^*_j$
as the reference bit, we have the physical degradation relationships
\begin{equation}
  \label{eq:extaj-degradation}
  \constmsg{*} \lepd \trueextprob{a}{j} \lepd \extaju{L} \lepd \dotsb \lepd \extaju{1},
\end{equation}
with all these probability tuples having symmetric densities, so \prettyref{prop:msg-compare-mi} can
still be applied to bound the mean-square difference between $\trueextprobx{a}{j}{0}$ and
$\frac{1}{2}$.  Now we define for $l=1,2,\dotsc,L$ the average densities over $\GdbwbjL$, namely
\begin{equation}
  \begin{split}
  \trueextprob{a}{j} \condmid a_j^*, \mat{G} \in \GdbwbjL &\sim \dextajtL,\\
  \extaju{l} \condmid a_j^*, \mat{G} \in \GdbwbjL &\sim \dextaju{l;L},
  \end{split}
\end{equation}
then they remain symmetric and satisfy
\begin{equation}
  \label{eq:dextaj-degradation}
  \constmsg{*} \lepd \dextajtL \lepd \dextaju{L;L} \lepd \dotsb \lepd \dextaju{1;L},
\end{equation}
and the bound from \prettyref{prop:msg-compare-mi} can also be averaged to yield, for any $l\le L$,
\begin{align}
  \label{eq:a-sync-bound-t}
  \frac{2}{\ln 2}\Expe{\left(\trueextprobx{a}{j}{0} - 1/2 \right)^2 \Bigcondmid \mat{G} \in \GdbwbjL}
  & \le I(\dextajtL) - I(\dbec{0}) \\
  \label{eq:a-sync-bound}
  & \le I(\dextaju{l;L}).
\end{align}
Eq.~\eqref{eq:a-sync-bound} now bounds the amount of synchronization error at the $j$-th
\node{a}-step in terms of $I(\dextaju{l;L})$, a quantity computable with
DE in the $n\to\infty$ limit.

\subsection{The Asymptotic Synchronization Conditions in terms of DE Results}
\label{sec:sync-de}
We now introduce some notations for DE results.  We use
$\opluspow{\dmu}{d} \defeq \dmu\oplus\dotsb\oplus\dmu$ to denote the result of the $\oplus$
operation on $d$ independent message densities (with $\opluspow{\dmu}{0} \defeq \dbec{1}$),
$\odotpow{\dmu}{d}$ for the $\odot$ operation with $\odotpow{\mu}{0} \defeq \dbec{0}$, and $\sum$ to
denote the convex combination operation in \secref{sec:binary-def-results}.  The density
$\dpriprob{u}$ is that of each $\priprob{u}{j}$ w.r.t.\ $u^*_j$, which does not vary with $\mat{G}$
or $j$ due to \prettyref{prop:priprob-u-sym}, and its MI is
$\Infopri{u} \defeq I(\dpriprob{u}) = I(u;y) = R_0(t) > 0$.  We also let
$\dpriprob{b} \defeq \dbec{\Infopri{b}}$, where $\Infopri{b} \in [0,1]$ can be understood as the
fraction of bits in $\seqb$ decimated in previous \node{b}-steps.  Now, corresponding to the $L$ BP
iterations that yield $\extbil{L}$, we can let $\dmsgiter{bc}{0} \defeq \dbec{0}$ and define iteratively
\begin{align}
  \label{eq:dextb-cb}
  \dmsgiter{cb}{l} &\defeq \dpriprob{u} \oplus \left( \sum_{d} v_d \cdot \opluspow{(\dmsgiter{bc}{l-1})}{d-1} \right), \\
  \label{eq:dextb-bc}
  \dmsgiter{bc}{l} &\defeq \dpriprob{b} \odot \odotpow{(\dmsgiter{cb}{l})}{\db-1},\quad l=1,\dotsc,L,
\end{align}
which finally yields
\begin{equation}
  \label{eq:dextblI}
  \dextblI{L} = \odotpow{(\dmsgiter{cb}{L})}{\db}.
\end{equation}
If the above process instead starts from $\dmsgiter{bc}{0} \defeq \dbec{1}$, the result is then
denoted by $\dextbuI{L}$.  Since the $\constmsg{a_j} = \constmsg{a_j^*}$ used in BP has the
``always-sure'' density $\dbec{1}$, the $\oplus$ operation with it has no effect and has been
omitted from \eqref{eq:dextb-cb}.

Similarly, during the \node{a}-steps, if we let $\Infopri{a} \in [0,1]$ be the fraction of bits in
$\seqa$ decimated in the previous steps and let
$\dpriprob{a} \defeq \dbec{\Infopri{a}}$, then the density $\dextauI{L}$ corresponding to the
process in \prettyref{alg:binary-a} can be defined as follows:
\begin{align}
  \dmsgiter{bc}{0} &= \dbec{1}, \\
  \label{eq:dexta-cb}
  \dmsgiter{cb}{l} &= (\dpriprob{u} \oplus \dpriprob{a}) \oplus \left( \sum_{d} v_d \cdot \opluspow{(\dmsgiter{bc}{l-1})}{d-1} \right), \\
  \label{eq:dexta-bc}
  \dmsgiter{bc}{l} &= \odotpow{(\dmsgiter{cb}{l})}{\db-1}, \quad l=1,\dotsc,L,\\
  \label{eq:dextauI}
  \dextauI{L} &= \dpriprob{u} \oplus \sum_d w_d \cdot \opluspow{(\dmsgiter{bc}{L})}{d}.
\end{align}


Now compare the DE result $\dextblI{l}$ defined above to the $\dextbil{l;L}$ defined in \secref{sec:sync-b} for a given $l\le L$.  As
$n\to\infty$, we make $i$ a function of $n$ that causes the fraction of decimated bits in
$\seqb_{\exi}$, $(i-1)/(\nb-1)$, to converge to some $\Infopri{b}$.  $\dextbil{l;L}$ is an average
over $\mat{G} \in \GdbwbiL$, and over this ensemble, the degrees of different nodes in $\neighiL$,
as well as their decimatedness (i.e.\ whether the node's index $i'$ is above or below $i$), are
asymptotically independent as $n\to\infty$,\footnote{\label{fn:indep} Technically, they are not
  exactly independent because the total number of nodes of some degree $d$ and the number of
  decimated nodes in the entire factor graph are fixed, so one node in the neighborhood having a
  certain degree makes another node less likely to have the same degree, but this has negligible
  impact when $n$ is large enough that $\neighiL$ is only a vanishing fraction of it, and can be
  dealt with using conventional bounding techniques.} and the probability that a given $b_{i'}$ has been
decimated is $(i-1)/(\nb-1)$, which approaches $\Infopri{b}$ as well.  Comparing the definitions of
$\extbil{l}$ and its density $\dextbil{l;L}$ to the above DE result $\dextblI{l}$, and noting that
each DE step in \eqref{eq:dextb-bc}, \eqref{eq:dextb-cb} and \eqref{eq:dextblI} is obviously
continuous with respect to convergence in distribution, we can conclude that $\dextbil{l;L}$
converges in distribution to $\dextblI{l}$ as $n\to\infty$.  Similarly, $\dextbiu{l;L}$ converges in
distribution to $\dextbuI{l}$, and if $j$ is made to vary with $n$ such that
$\lim_{n\to\infty} (j-1)/(\nc-1) = \Infopri{a} \in [0,1]$, then $\dextaju{l;L}$ also converges in
distribution to $\dextauI{l}$ as $n\to\infty$.

The above discussion involves the densities of the BP extrinsic information $\extprob{b}{i}$ and
$\extprob{a}{j}$ and the corresponding DE results.  For the true extrinsic information
$\trueextprob{b}{i}$, we have defined in \secref{sec:sync-b} its density over $\mat{G} \in \GdbwbiL$
as $\dextbitL$.  The density of $\trueextprob{b}{i}$ over all $\mat{G} \in \Gdbwb$, including those with loopy
neighborhoods, will be denoted by $\dextbit$, and its symmetry can still be established with
\prettyref{prop:linear-sym}.  Similarly, $\dextajt$ is defined as the density of
$\trueextprob{a}{j}$ over all $\mat{G} \in \Gdbwb$.

The aforementioned densities can all be characterized by their MIs.  For the DE results, we define
$\IbextlIL \defeq I(\dextblI{L})$, $\IbextuIL \defeq I(\dextbuI{L})$ and
$\IaextuIL \defeq I(\dextauI{L})$.  For the densities of BP extrinsic information over those $\mat{G}$
with a loop-free neighborhood, namely $\dextbil{l;L}$, $\dextbiu{l;L}$ and $\dextaju{l;L}$, the
corresponding MIs are denoted by $\IbextlInlL$, $\IbextuInlL$ and $\IaextuInlL$, where
$\Infopri{b} = (i-1)/(\nb-1)$, $\Infopri{a} = (j-1)/(\nc-1)$, and linear interpolation is performed
to extend their definitions to all $\Infopri{b}$ and $\Infopri{a}$ in $[0,1]$.  Since the MI of a
probability tuple is a bounded and continuous function, the above convergence in distribution
results immediately lead to the convergence of MI due to the portmanteau theorem; specifically, for
any $\Infopri{b}, \Infopri{a} \in [0,1]$ and $l\le L$, we have
\begin{multline}
  \label{eq:bp-limit-n}
  \lim_{n\to\infty} \IbextlInlL = \IbextlIl, \quad
  \lim_{n\to\infty} \IbextuInlL = \IbextuIl, \\
  \lim_{n\to\infty} \IaextuInlL = \IaextuIl.
\end{multline}
Note that the limits depend only on $l$ but not $L$, as long as $L\ge l$.

For the densities of the true extrinsic information, namely $\dextbitL$, $\dextbit$, $\dextajtL$ and
$\dextajt$, their MIs are likewise denoted by $\IbextmInL$, $\IbextmIn$, $\IaextmInL$ and
$\IaextmIn$ respectively, where $\Infopri{b} = (i-1)/(\nb-1)$ and $\Infopri{a} = (j-1)/(\nc-1)$ and
can again be linearly interpolated onto $[0,1]$.  However, unlike those of the BP extrinsic
information, it is generally difficult to prove that the densities or MIs of the true extrinsic
information converge as $n\to\infty$ \cite[Sec.~III-A]{maxwell-constr}, except when BP bounds can be used, e.g.\
when $\IbextlIL = \IbextuIL$.  Therefore, we instead define the limit inferior/superior
\begin{equation}
  \label{eq:map-limit-n}
  \IbextmIl \defeq \liminf_{n\to\infty} \IbextmIn, \quad
  \IbextmIu \defeq \limsup_{n\to\infty} \IbextmIn
\end{equation}
for $\IbextmIn$, and similarly $\IaextmIl$ and $\IaextmIu$ for $\IaextmIn$.  As $\IbextmInL$ and
$\IbextmIn$ differ only in the treatment of $\mat{G}$ with loopy neighborhoods, their difference is
upper-bounded by $\PloopbnL$ which vanishes as $n\to\infty$, so $\IbextmIl$ and $\IbextmIu$ are
also the limits of $\IbextmInL$, and similarly $\IaextmIl$ and $\IaextmIu$ are the limits of
$\IaextmInL$, and all these limits are independent from $L$.

For any finite $L$, using the continuity of each DE step w.r.t.\ convergence in distribution, it is
clear that $\IbextuIL$ and $\IbextlIL$ are continuous functions of each $\Infopri{b}$ and the degree
distribution $\seqw$.  However, their $L\to\infty$ limits $\IbextuI$ and $\IbextlI$ defined below
are not necessarily so, and neither are the $n\to\infty$ MIs of the true extrinsic information,
$\IbextmIl$ and $\IbextmIu$.  On the other hand, the finite-$n$ MIs such as $\IbextlInlL$ are
trivially continuous w.r.t.\ $\Infopri{b}$ due to them being linear interpolations.

\begin{figure}[!t]
  \centering
  \includegraphics{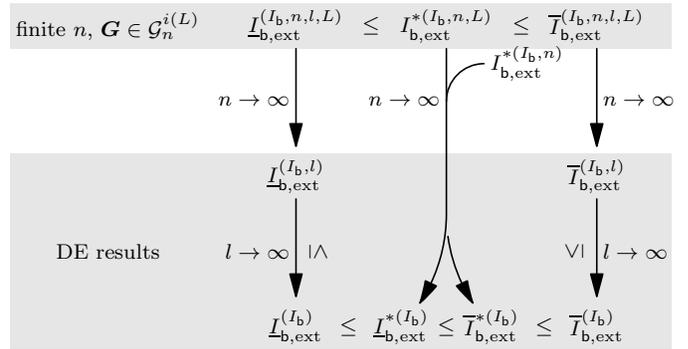}
  \caption{The relationship among the MIs involved in \node{b}-steps}
  \label{fig:mi-relationship-b}
\end{figure}

The relationships among the above MIs are given by the following result.  For the MIs
involved in \node{b}-steps, these relationships can be visualized by
\prettyref{fig:mi-relationship-b}, and the relationships among the MIs in \node{a}-steps are
similar.

\begin{proposition}
  \label{prop:exit-prop}
  The MIs above satisfy the following results:
  \begin{enumerate}
  \item\label{enum:Ibext-order-L} Given $\Infopri{b} \in [0,1]$ and $n>0$, $L>0$, then as long as
    $l\le L$, $\IbextlInlL$ is increasing (not necessarily strictly so; same below) and
    $\IbextuInlL$ is decreasing with $l$, with $\IbextlInlL \le \IbextmIn \le \IbextuInlL$.
    Consequently, the $n\to\infty$ limits $\IbextlIl$ and $\IbextuIl$ are likewise respectively
    increasing and decreasing functions of $l$, whose $l\to\infty$ limits $\IbextlI$ and $\IbextuI$
    thus exist and satisfy
    \begin{equation}
      \label{eq:Ibext-order}
      \IbextlIl \le \IbextlI \le \IbextmIl \le \IbextmIu \le \IbextuI \le \IbextuIl,\ \forall l.
    \end{equation}
  \item\label{enum:Iaext-order-L} Given $\Infopri{a} \in [0,1]$ and $n>0$, $L>0$, then as long as
    $l\le L$, $\IaextuInlL$ is a decreasing function of $l$ and satisfies
    $\IaextmInL \le \IaextuInlL$, so its $n\to\infty$ limit $\IaextuIl$ is also decreasing with $l$,
    and a further $l\to\infty$ limit can be taken to yield $\IaextuI$ that satisfies
    \begin{equation}
      \label{eq:Iaext-order}
      0 \le \IaextmIl \le \IaextmIu \le \IaextuI \le \IaextuIl,\ \forall l.
    \end{equation}
  \item\label{enum:Ibext-order-Ib} For any $n$ and $l\le L$, $\IbextmIn$, $\IbextmInL$,
    $\IbextlInlL$ and $\IbextuInlL$ are increasing functions of $\Infopri{b}$; consequently, so are
    $\IbextmIl$, $\IbextmIu$, $\IbextlIl$, $\IbextuIl$, as well as $\IbextlI$ and $\IbextuI$.
  \item\label{enum:Iaext-order-Ia} For any $n$ and $l\le L$, $\IaextmIn$, $\IaextmInL$ and
    $\IaextuInlL$ are increasing functions of $\Infopri{a}$, and consequently $\IaextmIl$,
    $\IaextmIu$, $\IaextuIl$ and $\IaextuI$ are so as well.
  \end{enumerate}
\end{proposition}
\begin{IEEEproof}
  See \prettyref{app:proof-exit-prop}.
\end{IEEEproof}

By \prettyref{prop:msg-compare-mi} and \prettyref{prop:msg-compare-mi-converse}, the synchronization
conditions should hold in an asymptotic sense if and only if $\IaextmIu = 0$ for all
$\Infopri{a} \in [0,1]$ (actually the $\Infopri{a}=1$ case is sufficient due to monotonicity) and
$\IbextlI = \IbextmIu$ for all $\Infopri{b} \in [0,1]$.  This is expressed formally with the following proposition:
\begin{proposition}
  \label{prop:sync-lm}
  Given a degree distribution, if
  \begin{gather}
    \label{eq:b-sync-lm}
    \IbextlI = \IbextmIu, \quad \forall \Infopri{b} \in [0,1], \\
    \label{eq:a-sync-lm}
    \IaextmIu = 0, \quad \forall \Infopri{a} \in [0,1],
  \end{gather}
  then
  \begin{enumerate}
  \item For any sequence of $i = i(n) \in \{1,\dotsc,\nb\}$ indexed by $n$, as long as
    $\Ibprin \defeq (i-1)/(\nb-1)$ has an $n\to\infty$ limit $\Ibpriz$ at which $\IbextlI$ is continuous w.r.t.\ $\Infopri{b}$, then
    \begin{equation}
      \label{eq:asympt-sync-b}
      \lim_{l\to\infty} \limsup_{n\to\infty} \Expe{(\extbil{l}(0) - \trueextprobx{b}{i}{0})^2} = 0;
    \end{equation}
  \item For any sequence of $j = j(n) \in \{1,\dotsc,\nc\}$ indexed by $n$, as long as
    $\Iaprin \defeq (j-1)/(\nc-1)$ has an $n\to\infty$ limit $\Iapriz$, then
    \begin{equation}
      \label{eq:asympt-sync-a}
      \limsup_{n\to\infty} \Expe{\left(\trueextprobx{a}{j}{0} - \frac{1}{2} \right)^2} = 0. 
    \end{equation}
  \end{enumerate}
  When these two results hold, we say \emph{the synchronization conditions are asymptotically
    satisfied}.  Conversely,
  \begin{enumerate}
  \item If \eqref{eq:b-sync-lm} fails to hold, then there exists $\epsilon > 0$ such that, for any
    $l$ and $n_0$, there always exist $n\ge n_0$ and $i\in \{1,\dotsc,\nb\}$ (where $\nb = n\Rn$
    as explained at the beginning of this section) that satisfy
    \begin{equation}
      \label{eq:asympt-nsync-b}
      \Expe{(\extbil{l}(0) - \trueextprobx{b}{i}{0})^2} \ge \epsilon;
    \end{equation}
  \item If \eqref{eq:a-sync-lm} fails to hold, then there exists $\epsilon > 0$ such that, for any
    $n_0$, there always exist $n\ge n_0$ and $j \in \{1,\dotsc,\nc\}$ ($\nc=n$) that satisfy
    \begin{equation}
      \label{eq:asympt-nsync-a}
      \Expe{\left(\trueextprobx{a}{j}{0} - \frac{1}{2} \right)^2} \ge \epsilon.
    \end{equation}
  \end{enumerate}
  In either case, we say \emph{the synchronization conditions are asymptotically unsatisfied}.
\end{proposition}
\begin{IEEEproof}
  See \prettyref{app:proof-sync-lm}.  As the convergence of $\IbextlIl$ to $\IbextlI$ as
  $l\to\infty$ may not be uniform w.r.t.\ $\Infopri{b}$, it seems necessary to introduce the
  continuity condition at $\Ibpriz$ in the direct part; however, since $\IbextlI$ is a monotonic
  function of $\Infopri{b}$, it has at most countably many discontinuities, and its continuity can
  be checked numerically anyway.  The \node{a}-step result \eqref{eq:asympt-sync-a} does not require
  such a continuity condition because the counterpart of $\IbextlI$ is constant zero, which is always continuous.
\end{IEEEproof}

Similar to \cite{maxwell-constr}, we may plot $\IbextlI$ and $\IbextuI$ against $\Infopri{b}$ and
call the resulting curves the \emph{lower} and \emph{upper BP EXIT curves}, which can be obtained
with DE methods.  On the other hand, the curves of $\IbextmIl$ and $\IbextmIu$ versus $\Infopri{b}$
can be called the \emph{MAP (maximum a posteriori) EXIT curves}, which are difficult to obtain
directly, but by \eqref{eq:Ibext-order}, they always lie between the BP EXIT curves, and an example
will be given in \prettyref{fig:ebp} below.

We will now present a sufficient
condition for the synchronization conditions to be asymptotically satisfied, in terms of the BP
curves only.  For this purpose we need the following lemma:
\begin{lemma}
  \label{lem:Ibext-b0a1}
  Let $\IbextuIx{0}$ be the value of $\IbextuI$ at $\Infopri{b}=0$, and $\IaextuIx{1}$ be the value of $\IaextuI$ at $\Infopri{a}=1$,
  then for any degree distribution, $\IbextuIx{0}=0$ if and only if $\IaextuIx{1}=0$.
\end{lemma}
\begin{IEEEproof}
  Comparing \eqref{eq:dextb-cb} and \eqref{eq:dextb-bc} with
  \eqref{eq:dexta-cb} and \eqref{eq:dexta-bc}, we note that $\dextauIx{1}{L}$ and $\dextbuIx{0}{L}$
  have the same $\dmsgiter{cb}{l}$'s and $\dmsgiter{bc}{l}$'s in their iterative definitions, and
  they can respectively be expressed as
  \begin{align}
    \dextauIx{1}{L} &= \dpriprob{u} \oplus \left( \sum_d w_d \cdot \opluspow{(\odotpow{(\dmsgiter{cb}{L})}{\db-1})}{d} \right), \\
    \dextbuIx{0}{L} &= \odotpow{(\dmsgiter{cb}{L})}{\db}.
  \end{align}
  Since we have assumed that $I(\dpriprob{u}) = \Infopri{u} = R_0(t)$ is strictly positive, $\db \ge 2$ and all degrees are non-zero,
  we can use the results in \cite{bounds-info-combining} regarding the MI combining
  behavior of the $\odot$ and $\oplus$ operators to show that $I(\dextauIx{1}{L})$ and $I(\dextbuIx{0}{L})$
  go to zero as $L\to\infty$ if and only if $I(\dmsgiter{cb}{L})$ does.  Consequently, $\IaextuIx{1}=0$ if and only if $\IbextuIx{0}=0$.
\end{IEEEproof}

Using \prettyref{lem:Ibext-b0a1} and the monotonicity of $\IaextuI$ w.r.t.\ $\Infopri{a}$ in \prettyref{prop:exit-prop}, we can immediately obtain the
following sufficient condition from \prettyref{prop:sync-lm}:
\begin{theorem}
  \label{thm:sync-ul}
  Given a degree distribution, if
  \begin{gather}
    \label{eq:a-sync-b0}
    \IbextuIx{0} = 0, \\
    \label{eq:b-sync-ul}
    \IbextlI = \IbextuI, \quad \forall \Infopri{b} \in [0,1],
  \end{gather}
  then the synchronization conditions are asymptotically satisfied.
\end{theorem}

Although the MAP curves themselves are difficult to compute, they are known to satisfy the following \emph{area theorem}:
\begin{proposition}
  \label{prop:map-area}
  For any degree distribution and $n$, we have
  \begin{equation}
    \label{eq:map-area-n}
    \sum_{j=1}^{\nc} \IaextmInx{\Infoprix{a}{j}} + \sum_{i=1}^{\nb} \IbextmInx{\Infoprix{b}{i}} = n\Infopri{u},
  \end{equation}
  where $\Infoprix{a}{j} \defeq (j-1)/(\nc-1)$, $\Infoprix{b}{i} \defeq (i-1)/(\nb-1)$.  Note that
  \eqref{eq:map-area-n} uses the average MI $\IbextmIn$ over all $\mat{G}$, including those with loopy neighborhoods.

  Consequently, as $n\to\infty$, any degree distribution satisfies the area theorem
  \begin{multline}
    \label{eq:map-area}
    \int_0^1 \IaextmIl \,d\Infopri{a} + R \int_0^1 \IbextmIl \,d\Infopri{b} \le \Infopri{u} \\
    \le \int_0^1 \IaextmIu \,d\Infopri{a} + R \int_0^1 \IbextmIu \,d\Infopri{b}.
  \end{multline}
\end{proposition}
\begin{IEEEproof}
  See \prettyref{app:proof-map-area}.  This can be regarded as a special case of
  \cite[Theorem~1]{maxwell-constr}, where the reference codeword $(\seqbr, \seqar)$ corresponds to
  $X$ there, the $\priprob{b}{i}$'s and $\priprob{a}{j}$'s are the $Y$, and the $\priprob{u}{j}$'s
  (or $\seqy$) are the additional observation $\Omega$.
\end{IEEEproof}

This immediately leads to the following necessary condition for the synchronization conditions to be satisfied:
\begin{theorem}
  \label{thm:sync-area}
  For any degree distribution,
  \begin{equation}
    \label{eq:area-b-ineq}
    \int_0^1 \IbextlI \,d\Infopri{b} \le \Infopri{u}/R.
  \end{equation}
  Moreover, equality holds in \eqref{eq:area-b-ineq} when the synchronization conditions are asymptotically satisfied.
\end{theorem}
\begin{IEEEproof}
  Application of \eqref{eq:Ibext-order} and \eqref{eq:Iaext-order} in the first inequality of \eqref{eq:map-area} gives
  \begin{equation}
    R\int_0^1 \IbextlI \,d\Infopri{b} \le \int_0^1 \IaextmIl \,d\Infopri{a} + R \int_0^1 \IbextmIl \,d\Infopri{b} \le \Infopri{u},
  \end{equation}
  which leads to \eqref{eq:area-b-ineq}.

  When the synchronization conditions are asymptotically satisfied, i.e.\ \eqref{eq:b-sync-lm} and \eqref{eq:a-sync-lm} hold, the second inequality of \eqref{eq:map-area} becomes
  \begin{equation}
    \Infopri{u} \le \int_0^1 \IaextmIu \,d\Infopri{a} + R \int_0^1 \IbextmIu \,d\Infopri{b} = R \int_0^1 \IbextlI \,d\Infopri{b},
  \end{equation}
  so equality holds in \eqref{eq:area-b-ineq}.
\end{IEEEproof}

\subsection{The Case of Binary Erasure Quantization}
\label{sec:beq-de}
As an important and intuitive special case, we consider the BEQ problem as defined in
\prettyref{exm:erasure-quant} at $t\to\infty$.  Given $\mat{G}$ and a source sequence $\seqy$, we
say a certain $(\seqb, \seqa)$ or the corresponding $\sequ = \sequ(\seqb, \seqa)$ is consistent with
it if $d(\seqy, \sequ) = 0$ (i.e.\ $y_j = *$ or $y_j = u_j$ for all $j$), and the set of such
$(\seqb, \seqa)$'s, which is non-empty due to the freedom in the choice of $\seqa$, is denoted
$\SetCWy$.  For a BEQ problem with a definite $\seqa$, it is said to have a solution if
there is some $(\seqb, \seqa) \in \SetCWy$.  According to the discussion in
\prettyref{sec:outline-quant-ana}, the reference codeword $(\seqbr, \seqar)$ yielded by the TPQ is
uniformly distributed over $\SetCWy$, and the joint distribution of $\seqbr$, $\seqar$, $\sequr$ and
$\seqy$ is given by \prettyref{prop:ref-dist}, with
\begin{gather}
  p(u^*_j \condmid y_j) = p_{u\condmid y}(u_j^* \condmid y_j) =
  \begin{cases}
    1/2, & y_j = *; \\
    \oneif{y_j = u^*_j}, & y_j = 0,1;
  \end{cases} \\
  p(y_j) = p_y(y_j) =
  \begin{cases}
    \epsilon, & y_j = *; \\
    (1-\epsilon)/2, & y_j = 0,1
  \end{cases}
\end{gather}
for all $j$.  We thus have
\begin{equation}
  \label{eq:pyu-beq}
  p(y_j \condmid u^*_j) =
  \begin{cases}
    \epsilon, & y_j = *; \\
    (1-\epsilon) \cdot \oneif{y_j = u^*_j}, & y_j = 0,1.
  \end{cases}
\end{equation}

Each $\priprob{u}{j}$ is a function of $y_j$; according to \eqref{eq:priprobs-u}, it is $\constmsg{*}$ when $y_j=*$ and
$\constmsg{y_j}$ when $y_j$ is 0 or 1.  Combined with \eqref{eq:pyu-beq}, we have
\begin{equation}
  p(\priprob{u}{j} \condmid u^*_j) =
  \begin{cases}
    \epsilon, & \priprob{u}{j} = \constmsg{*}; \\
    1-\epsilon, & \priprob{u}{j} = \constmsg{u^*_j}.
  \end{cases}
\end{equation}
In other words, $\priprob{u}{j} \condmid u^*_j \sim \dpriprob{u}$ is simply $\dbec{1-\epsilon}$, a
symmetric density independent of $j$, and $\Infopri{u} = I(\dpriprob{u}) = 1-\epsilon$.  These properties are consistent with the above discussion such as \prettyref{prop:sym-chan-sym-likelihood}.

By \prettyref{prop:erasure-like-density} and \prettyref{prop:erasure-nu}, all the densities involved in the DE steps in
\prettyref{sec:sync-de}, as well as those of the true extrinsic information, $\dextajt$ and
$\dextbit$, are erasure-like, and can thus be uniquely determined by their MIs, so the conditions
\eqref{eq:a-sync-b0} and \eqref{eq:b-sync-ul} can be evaluated as follows.  Let
$\Infoiter{cb}{l} \defeq I(\dmsgiter{cb}{l})$ and $\Infoiter{bc}{l} \defeq I(\dmsgiter{bc}{l})$,
then \eqref{eq:dextb-cb} and \eqref{eq:dextb-bc} can respectively be expressed as
\begin{align}
  \label{eq:exit-binary-c}
  \Infoiter{cb}{l} &= \Infopri{u} \sum_d v_d \cdot (\Infoiter{bc}{l-1})^{d-1}, \\
  \label{eq:exit-binary-b}
  \Infoiter{bc}{l} &= 1 - (1-\Infopri{b}) (1-\Infoiter{cb}{l})^{\db-1},
\end{align}
while \eqref{eq:dextblI} becomes
\begin{equation}
  \label{eq:Ibext}
  \IbextxIL = 1- (1-\Infoiter{cb}{L})^{\db},
\end{equation}
where the resulting $\IbextxIL$ is $\IbextlIL = I(\dextblI{L})$ when starting with $\Infoiter{bc}{0} = 0$, and
$\IbextuIL = I(\dextbuI{L})$ when $\Infoiter{bc}{0} = 1$.

For conciseness of presentation, we introduce functions $f(\cdot)$, $g(\cdot)$ and
$h(\cdot)$ which, in the case of BEQ, are defined as
\begin{align}
  \label{eq:exit-f-beq}
  f(x) &\defeq \sum_d v_d x^{d-1}, \\
  \label{eq:exit-g-beq}
  g(y) &\defeq y^{\db - 1}, \\
  \label{eq:exit-h-beq}
  h(y) &\defeq y^{\db},
\end{align}
so that we can write
\begin{align}
  \label{eq:exit-binary-c-de}
  \Infoiter{cb}{l} &= \Infopri{u} \cdot f(\Infoiter{bc}{l-1}), \\
  \label{eq:exit-binary-b-de}
  \Infoiter{bc}{l} &= 1-(1-\Infopriiter{b}{l}) \cdot g(1-\Infoiter{cb}{l}), \\
  \label{eq:Ibext-de}
  \Ibextiter{l} &= 1-h(1-\Infoiter{cb}{l}).
\end{align}
When all the $\Infoiter{b}{l}$'s are equal to the same $\Infopri{b}$, the resulting $\Ibextiter{L}$ is the
$\IbextxIL$ above ($\IbextlIL$ or $\IbextuIL$ depending on the initial $\Infoiter{bc}{0}$.

Now we combine \eqref{eq:exit-binary-c-de} and \eqref{eq:exit-binary-b-de} to yield a mapping
$\nextInfo{bc}$ such that $\Infoiter{bc}{l} = \nextInfo{bc}(\Infoiter{bc}{l-1}; \Infopri{u},
\Infopri{b})$.
$\nextInfo{bc}(\cdot; \cdot, \cdot)$ is an increasing function of all three variables in $[0,1]$ and
its result is also in $[0,1]$; therefore, given fixed $\Infopri{u}$ and $\Infopri{b}$ and starting with
$\Infoiter{bc}{0} = 0$ (resp.\ $1$), iterative application of
$\nextInfo{bc}(\cdot) \defeq \nextInfo{bc}(\cdot; \Infopri{u}, \Infopri{b})$ gives an increasing (resp.\
decreasing) sequence $(\Infoiter{bc}{l})_{l=0}^{\infty}$, whose limit as $l\to\infty$ always exists
and can be denoted $\InfoinflI{bc}$ and $\InfoinfuI{bc}$.  Taking the $l\to\infty$ limit of
\eqref{eq:Ibext-de} and \eqref{eq:exit-binary-c-de} and using continuity, we can finally express
$\IbextlI$ and $\IbextuI$ in \eqref{eq:a-sync-b0} and \eqref{eq:b-sync-ul} in terms of $\InfoinflI{bc}$ and $\InfoinfuI{bc}$, as
\begin{equation}
  \IbextlI = \Ibext(\InfoinflI{bc}), \quad \IbextuI = \Ibext(\InfoinfuI{bc}),
\end{equation}
where
\begin{equation}
  \label{eq:Ibext-Ibc}
  \Ibext(x) \defeq 1 - h(1 - \Infopri{u} \cdot f(x))
\end{equation}
is a strictly increasing function of $x$.

Both $\InfoinflI{bc}$ and $\InfoinfuI{bc}$ are clearly fixed points of $\nextInfo{bc}(\cdot)$ at the given $\Infopri{u}$ and $\Infopri{b}$; indeed,
due to monotonicity of $\nextInfo{bc}(\cdot)$ to prove that they are the minimum and the maximum fixed points among the possibly
multiple ones at such $\Infopri{u}$ and $\Infopri{b}$.  In the case of BEQ, it is actually straightforward to obtain all the fixed points by
equating $\Infoiter{bc}{l}$ and $\Infoiter{bc}{l-1}$ in \eqref{eq:exit-binary-c-de} and
\eqref{eq:exit-binary-b-de}.  Denoting $x \defeq \Infoiter{bc}{l-1} = \Infoiter{bc}{l}$, we get
\begin{equation}
  \label{eq:Ib}
  \Infopri{b} = 1-\frac{1-x}{g(1-\Infopri{u} \cdot f(x))};
\end{equation}
therefore as we vary $x$ over $[0,1]$, if the $\Infopri{b}$ given by \eqref{eq:Ib} is also within
$[0,1]$, then $x$ is a fixed point of $\nextInfo{bc}(\cdot)$ at this $\Infopri{b}$, and all fixed
points can be obtained in this way (note that the denominator in \eqref{eq:Ib} cannot be zero as
long as $\Infopri{u} < 1$).  Each fixed point $x$ can equivalently be expressed in terms of
$\Ibext \defeq \Ibext(x)$.  We can now define the \emph{EBP EXIT curve} (or simply the \emph{EBP
  curve}), original proposed in \cite{maxwell-constr} for LDPC decoding over BEC, as the parametric
$\Infopri{b}$ vs.\ $\Ibext$ curve given by \eqref{eq:Ib} and \eqref{eq:Ibext-Ibc} for $x\in[0,1]$.

While $\Ibext$ is a strictly increasing function of $x$, $\Infopri{b}$ is not necessarily so.  However,
with simple algebra we can still obtain the following properties of the EBP curve:
\begin{proposition}
  \label{prop:ebp-prop}
  The EBP curve for any degree distribution under BEQ satisfies
  \begin{gather}
    \label{eq:Ib-range}
    \Infopri{b} |_{x=0} \le 0, \quad \Infopri{b} |_{x=1} = 1, \\
    \label{eq:Ibext0}
    \Ibext |_{x=0} = 1-(1-\Infopri{u} v_1)^{\db}, \\
    \label{eq:ebp-area}
    \db\Infopri{u} v_1 + \int_0^1 (1-\Infopri{b}) \frac{d\Ibext}{dx} \,dx = \frac{\Infopri{u}}{R}.
  \end{gather}
  In \eqref{eq:Ib-range} equality holds if and only if $v_1 = 0$.
\end{proposition}
\begin{IEEEproof}
  See \prettyref{app:proof-ebp-prop}.
\end{IEEEproof}

\begin{figure}[!t]
  \centering
  \includegraphics{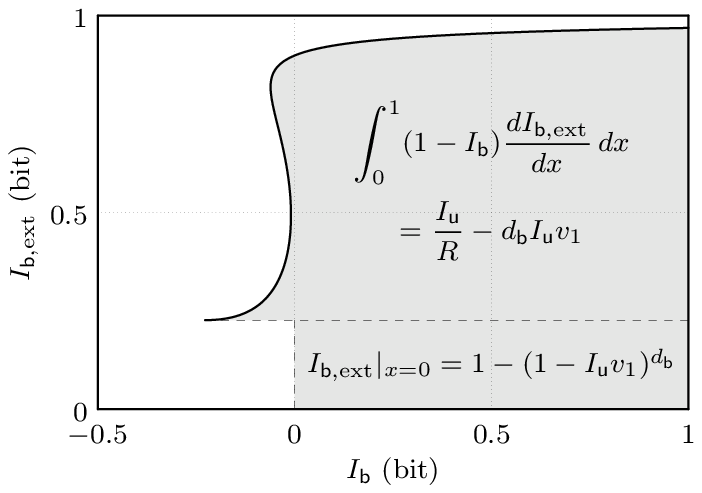}%
  \caption{The area under the EBP curve (the thick solid curve) when
    $v_1>0$.  In such cases the EBP curve does not start from
    $(0,0)$, and we define the area $\Aebp$ under it as the total area of the
    two gray regions, whose respective areas are shown in the figure.}%
  \label{fig:ebp-area}%
\end{figure}

Eq.~\eqref{eq:ebp-area} can be visualized as an area result in \prettyref{fig:ebp-area}: if we define the
total shaded area as the \emph{area under the EBP curve}
\begin{equation}
  \label{eq:Aebp}
  \Aebp \defeq \Ibext |_{x=0} + \int_0^1 (1-\Infopri{b}) \frac{d\Ibext}{dx} \,dx,
\end{equation}
then since
\begin{equation}
  \label{eq:Ibext-bound}
  \Ibext|_{x=0} = 1-(1-\Infopri{u}v_1)^{\db} \le \db\Infopri{u}v_1,
\end{equation}
by \eqref{eq:ebp-area} we have
\begin{equation}
  \label{eq:Aebp-bound}
  \Aebp \le \Infopri{u}/R,
\end{equation}
where equality holds if and only if $v_1 = 0$; actually, from \eqref{eq:Ibext-bound} we see that
the difference is only $\asymptO(v_1^2)$.

Every crossing the EBP EXIT curve makes with a constant-$\Infopri{b}$ vertical line corresponds to a
fixed point at this $\Infopri{b}$.  As stated above, the minimum and maximum fixed points at each
$\Infopri{b}$ are at $x=\InfoinflI{bc}$ and $x=\InfoinfuI{bc}$, or equivalently at $\Ibext = \IbextlI$
and $\Ibext = \IbextuI$, respectively, so the lower and upper BP EXIT curves defined in
\prettyref{sec:sync-de} are simply the lower and upper envelopes of the EBP EXIT curve.  The
conditions \eqref{eq:b-sync-ul} and \eqref{eq:a-sync-b0} can now be
expressed in terms of the monotonicity of the EBP EXIT curve, which can be easily computed from the degree distribution;
this is formally expressed by the following theorem:
\begin{theorem}
  \label{thm:sync-mono}
  For BEQ, if the EBP EXIT curve given by \eqref{eq:Ib} satisfies the following \emph{monotonicity conditions}\footnote{Note that
    this has nothing to do with monotonicity with respect to a class of channels, as discussed in
    LDPC literature \cite{cap-ldpc-msgpassing-dec}.}
  \begin{align}
    \label{eq:mono-cond1}
    \Infopri{b} |_{x=0} &= 0, \\
    \label{eq:mono-cond2}
    \frac{d\Infopri{b}}{dx} &> 0, \quad x\in [0,1],
  \end{align}
  then the synchronization conditions are asymptotically satisfied.  Conversely, if
  $\Infopri{b}|_{x=0} < 0$, or $d\Infopri{b}/dx < 0$ for any $x\in [0,1]$, then the synchronization conditions
  are asymptotically unsatisfied.
\end{theorem}
\begin{IEEEproof}
  \emph{Direct part}: Condition \eqref{eq:mono-cond2} implies that $\Infopri{b}$ is a strictly
  increasing function of $x$, so $x$ and thus $\Ibext$ are also uniquely defined and strictly
  increasing functions of $\Infopri{b}$, and by \eqref{eq:Ib-range} they are defined for all $\Infopri{b} \in [0,1]$.
  Therefore, at each $\Infopri{b}$, $\nextInfo{bc}(\cdot)$ has a unique fixed point corresponding to
  this $\Ibext$, so $\IbextlI$ and $\IbextuI$ will both be equal to this value, thus
  \eqref{eq:b-sync-ul} holds.  Condition \eqref{eq:mono-cond1} implies that the fixed point is at
  $\Ibext = 0$ when $\Infopri{b} = 0$, so \eqref{eq:a-sync-b0} holds as well.  \prettyref{thm:sync-ul}
  can thus be applied to obtain the desired result.

  \emph{Converse part}: Since the lower-BP curve is the lower envelope of the EBP curve, the area
  under it never exceeds $\Aebp$ in \prettyref{fig:ebp-area}, and a finite difference will
  exist if $d\Infopri{b}/dx < 0$ for any $x \in [0,1]$ (note that $\Ibext$ is strictly increasing with
  respect to $x$).  On the other hand, we have found that $\Aebp \le \Infopri{u}/R$
  and is strictly smaller when $v_1 > 0$ or equivalently $\Infopri{b} |_{x=0} < 0$.  Combining the two
  results, we can see from \prettyref{thm:sync-area} that the synchronization conditions will be
  asymptotically unsatisfied in either case.
\end{IEEEproof}

\begin{figure*}[!t]
  \centering
  \subfigure[EBP curves of $(4,2)$ regular LDGM code]{\label{fig:ebp42}\includegraphics{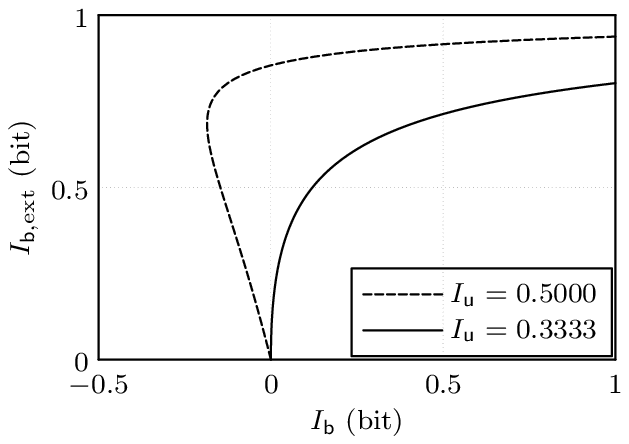}}%
  \hspace{1mm}%
  \subfigure[EBP curves of $(5,3)$ regular LDGM code]{\label{fig:ebp53}\includegraphics{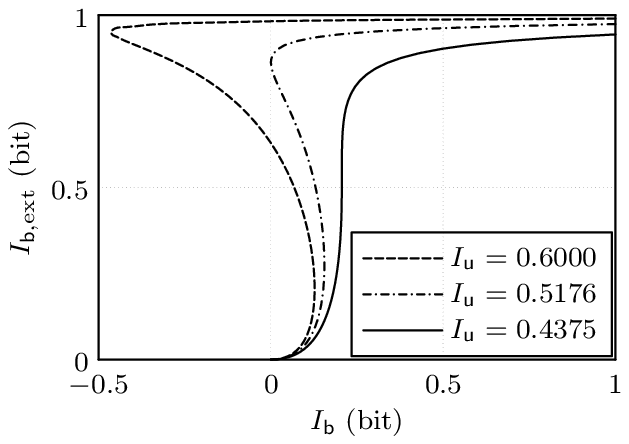}}%
  \hspace{1mm}%
  \subfigure[Comparison of EBP, BP and MAP]{\label{fig:ebp-map}\includegraphics{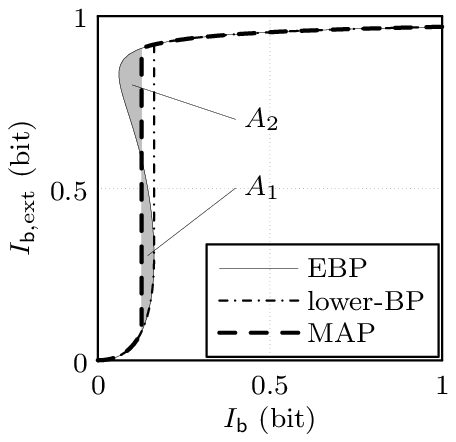}}%
  \caption{The EBP curves of some $(\db,\dc)$ regular LDGM codes, i.e.\ those with the given $\db$
    and $v_d = \oneif{d=\dc}$.}%
  \label{fig:ebp}%
\end{figure*}

Finally, we give as examples in \prettyref{fig:ebp} the EXIT curves under BEQ of some $(\db,\dc)$ regular LDGM codes
at different values of $t$, or equivalently, $\Infopri{u}=R_0(t)$.

\prettyref{fig:ebp42} shows the EBP curves of the $(4,2)$ regular code with rate $R=1/2$.  When
$1/2 > \Infopri{u} > \Iuthr \defeq 1/3$, part of the EBP curve lies in the $\Infopri{b}<0$
half-plane, but once $\Infopri{b}$ becomes positive, it is monotonically increasing.  Therefore, for
any $\Infopri{b} > 0$, $\nextInfo{bc}(\cdot)$ has a unique fixed point with the corresponding
$\Ibext$ equal to $\IbextlI = \IbextmIl = \IbextmIu = \IbextuI$, thus the synchronization conditions
are asymptotically satisfied in \node{b}-steps.  On the other hand, corresponding to the fixed point
with the largest MI at $\Infopri{b}=0$, we have $\IbextuIx{0}>0$ and consequently $\IaextuIx{1}>0$;
in fact, since $\Aebp=\Infopri{u}/R$ and the left-hand side of \eqref{eq:area-b-ineq} corresponds to a strictly smaller area,
the necessary condition \eqref{eq:area-b-ineq} is unsatisfied,
so the synchronization conditions must fail to hold ($\IaextmIu>0$) at some $\Infopri{a}$, so a non-vanishing mean-square difference will exist between some
$\trueextprob{a}{j}$ and $\constmsg{*}$ as $n\to\infty$, implying that the corresponding BEQ
problems usually have no solutions.  When $\Infopri{u} < 1/3$, the synchronization conditions are
asymptotically satisfied in both \node{b}- and \node{a}-steps due to \prettyref{thm:sync-mono}.

\prettyref{fig:ebp53} is for the $(5,3)$ regular code with rate $R=0.6$.  When $\Infopri{u}$ is
reduced below 0.5176, the EBP curve no longer extends into the $\Infopri{b}\le 0$ half-plane, so
both $\IbextuIx{0}$ and $\IaextuIx{1}$ are zero, and consequently all $\IaextmIu$ are zero as well,
implying that the BEQ problems have solutions in an asymptotic sense.  However, unless $\Infopri{u}$
is further reduced below $\Iuthr = 7/16 = 0.4375$, the EBP curve is still not monotonic, therefore
the BP fixed points are not unique at some values of $\Infopri{b}$, where $\IbextlI < \IbextuI$ and
$\IbextmIl$ and $\IbextmIu$ lie between them.  Indeed, \eqref{eq:area-b-ineq} is again unsatisfied because its left-hand side is strictly smaller than $\Aebp=\Infopri{u}/R$,
so the synchronization conditions also fail to hold at some $\Infopri{b}$, i.e.\ the BP result $\extprob{b}{i}$ will fail
to converge to $\trueextprob{b}{i}$ in a mean-square sense as $n\to\infty$; in other words, the solutions of
the BEQ problems can usually not be obtained with BP\@.  Only when $\Infopri{u} < 0.4375$ will the EBP
curve become monotonic, allowing the synchronization conditions to be asymptotically satisfied.

\prettyref{fig:ebp-map} is a comparison of the EBP and the lower-BP curves of the $(5,3)$ regular
code at $\Infopri{u} = 0.5$, as well as postulated MAP curves based on the monotonicity results
in \prettyref{prop:exit-prop}, the area results in \prettyref{prop:map-area} and
\prettyref{prop:ebp-prop}, and the analysis of the similar EXIT curves arising in LDPC decoding over
BEC in \cite{maxwell-constr}.  BEQ is actually quite similar to LDPC decoding over BEC considered in
\cite{maxwell-constr}, as both involve a system of linear equations over $\SetZ_2$.  If the results in \cite{maxwell-constr} remain true,
we may conjecture that $\IbextmIl = \IbextmIu$ for all $\Infopri{b}$, and the MAP curve formed by it looks like the dashed line in \prettyref{fig:ebp-map}.  Note that the area under this MAP curve is $\Infopri{u}/R$ according to \eqref{eq:map-area}, which is also equal to $\Aebp$, so the two regions between the EBP and MAP curves necessarily have the same area $A_1 = A_2$.
The area $A_1$ to the right of the MAP curve
represents the $b_i$'s whose $\trueextprob{b}{i} = \constmsg{b_i^*}$ but
$\extprob{b}{i} = \constmsg{*}$ and thus violate the synchronization condition; that is, the values
of these bits are determined by previous decimation results but not available from BP at the time,
and they are apparently ``guesses'' until they are ``confirmed'' by an equal number of equations
encountered later represented by $A_2$.  That $A_2 = A_1$ intuitively means that confirmations constrain earlier guesses rather than $\seqa$, so the BEQ problem does
have a solution in an asymptotic sense.  This is not the case for e.g.\ the $(4,2)$ regular code at
$\Infopri{u} = 0.5$ in \prettyref{fig:ebp42}: there the MAP and the lower-BP curves overlap with the
EBP curve in the $\Infopri{b}\ge 0$ half-plane but does not extend to the left, and the area between
the EBP curve and the $\Infopri{b} = 0$ axis represent ``confirmations'' that, having no earlier
guesses, become constraints on $\seqa$.

\subsection{Application in Degree Distribution Optimization}
\label{sec:app-dd-opt}
We may summarize the above analysis as follows:
\begin{itemize}
\item The quantization algorithm using PD, being an implementation of BPPQ, can reach the distortion
  $D_0(t)$ of the TPQ if the synchronization condition is satisfied exactly.
\item The synchronization condition is satisfied asymptotically, as the block length $n$ and the
  iteration count $L$ goes to infinity, if the degree distribution satisfies the conditions in
  \prettyref{thm:sync-ul} or (in case of BEQ) \prettyref{thm:sync-mono} at the chosen $t$ (or $\Infopri{u}$).
\end{itemize}
These results suggest that the asymptotic synchronization condition, which can be evaluated
numerically with DE for any specific degree distribution, can be used as the constraint for LDGM
degree distribution optimization.  For ordinary symmetric source coding problems, we want to maximize $t$ such
that $D_0(t)$ is minimized, while for BEQ, $t$ is fixed at infinity with $D_0(t) = 0$, and we want
to find the source with the minimum $\epsilon$ that can still be encoded at a given $R$.  This is thus equivalent to the maximization of $\Infopri{u}$,
which is $R_0(t)$ in the former case and $1-\epsilon$ in the latter.  Alternatively, the optimization problem can also be
formulated as the minimization of $R$ at a given $t$ or $\Infopri{u}$.

The details of this optimization have been tackled in \cite{ldgm-vq-journal}.  The method starts
from the degree distribution optimized for BEQ using \prettyref{thm:sync-mono}, i.e.\ the erasure
approximation (EA) result, due to the availability of an explicit formula for the EBP curve in this case;
numerical DE is then performed on this degree distribution and the results are used to derive a
correction factor $r(x)$ for use in the next iteration of the optimization process.  As the degree
distribution resulting from this iterative process can be numerically verified to satisfy the
asymptotic synchronization condition, our analysis above suffices as a theoretical
justification for this approach.

It should be noted that asymptotic satisfaction of the synchronization condition does not imply its
exact satisfaction, particularly since both the block length $n$ and the iteration count $L$ are necessarily
finite in practice.  While this residual synchronization error can be effectively tackled with the
recovery algorithm in \cite{ldgm-vq-journal} and \cite{flip-ml}, this also suggests that making the
synchronization condition asymptotically satisfied might not be optimal, as allowing for a small
asymptotic synchronization error might lower $D_0(t)$ at the same
$R$ by a larger amount than the extra distortion caused by the synchronization error; indeed, an improved optimization method for finite $L$ has been proposed in
\cite{ldgm-vq-journal}.  However, these improvements can still be regarded as variations of the
method based on the asymptotic synchronization condition.

\section{Extension to Non-Binary Constructions}
\label{sec:nonbinary}
We now consider non-binary LDGM-based code constructions that are necessary in many source coding
problems.  For example, it has been shown in \secref{sec:random-coding-loss-mse} that the shaping
loss of binary MSE quantization is lower-bounded by \unit[0.0945]{dB} due to the random-coding loss,
and this loss can be greatly reduced if a larger alphabet is used; this issue has also been noted in
e.g.\ \cite{near-cap-dpc-tcq-ira} in the context of shaping for dirty-paper coding.  In general, a
symmetric source coding problem over a finite abelian group $\SetG$ ($\SetG = \SetZ_M$ in $M$-ary
MSE quantization) can be solved using LDGM codes in either of the following two ways:
\begin{itemize}
\item When $\cardinal{\SetG}=2^K$, binary LDGM codes may be used, with every $K$ bits from an LDGM codeword modulated into a reconstructed
  symbol, similar to bit-interleaved coded modulation (BICM) in channel coding \cite{bicm,bicm-id};
\item Use an $\cardinal{\SetG}$-ary LDGM code directly, similar to the use of trellis-coded modulation (TCM)
  \cite{chan-coding-multi-level-phase} and non-binary LDPC codes in channel coding, or TCQ in source
  coding.
\end{itemize}
The latter approach has been attempted in e.g.\ \cite{quant-sig-sp-gen-fg-codes}, but degree
distribution optimization and convergence issues have not been tackled there and will be more
difficult than the binary case; a notable issue is that many possible $\SetG$'s, such as
$\SetG = \SetZ_M$ with $M=2^K>2$ used in $M$-ary MSE quantization, cannot be given a field
structure, so the LDGM code has to be defined on a field, usually $\GF(M)$, with a different
additive group structure, which is no more natural than the simpler former approach.  Therefore, in
the previous work \cite{ldgm-vq-globecom07} as well as this paper we adopt the former BICM-like
approach, which allows near-ideal codes to be designed with relative ease; such an approach has also been used in
other works such as \cite{robust-multires-coding}.  Of course, if linearity is a concern, e.g.\ in some problems
involving network coding, it would be necessary to adopt the TCM-like approach, usually with $\SetG$
possessing a field structure (e.g.\ $\SetG=\SetZ_p$ with $p$ being a prime number) and with the LDGM
code defined on it; such code constructions will not be considered in this paper, but can be
analyzed with largely the same method.

\subsection{Probability Tuples over a Finite Abelian Group}
\label{sec:prob-tuple-SetG}
For symmetric source coding over a finite abelian group $\SetG$, the proposed non-binary LDGM
quantizer will make use of probability distributions over either $\SetG$ or $\SetZ_2^K$; as
$\SetZ_2^K$ is itself a finite abelian group under component-wise addition and can thus be regarded as a special case, it suffices to consider
distributions over $\SetG$, which can be viewed as nonnegative-valued functions defined on
$\SetG$ and represented by \emph{probability tuples over finite abelian group $\SetG$}.  Similar to
the binary case, each component of such a probability tuple $\lambda$ is denoted by $\lambda(u)$
($u\in\SetG$), whose sum is implicitly normalized to 1, and various definitions can also be extended
in a straightforward manner as follows:
\begin{itemize}
\item Given $u \in \SetG$, $\constmsg{u}$ is the
  sure-$u$ probability tuple with
  $\constmsg{u}(u) = 1$ and all other components being zero, while $\constmsg{*}$ is the
  ``unknown'' probability tuple with all components being $1/\cardinal{\SetG}$;
\item The entropy of a probability tuple $\lambda$ over $\SetG$ is
  $H(\lambda) \defeq -\sum_{u\in\SetG} \lambda(u) \log \lambda(u)$, while its MI
  $I(\lambda) \defeq \log\cardinal{\SetG} - H(\lambda)$.
\item The $\odot$ operation on two probability tuples does pairwise multiplication of the $\cardinal{\SetG}$
  components and then normalizes the result;
\item The $\oplus$ operation on two probability tuples are defined according to the addition
  operator on $\SetG$, also denoted by $\oplus$; specifically, given two
  probability tuples
  $\lambda_1$ and $\lambda_2$ over $\SetG$, $\lambda = \lambda_1 \oplus \lambda_2$ is defined as
  \begin{equation}
    \label{eq:oplus-msgMK}
    \lambda(u) = \sum_{\substack{u_1, u_2 \in \SetG \\ u_1 \oplus u_2 = u}} \lambda_1(u_1) \lambda_2(u_2), \quad u\in\SetG.
  \end{equation}
  The $\ominus$ operator
  is defined similarly for subtraction over $\SetG$.  In particular,
  $\lambda \oplus \constmsg{u}$ is simply $\lambda$ with its components permuted, and
  $\constmsg{u_1} \oplus \constmsg{u_2} = \constmsg{u_1 \oplus u_2}$.
\item More generally, let $(\SetZW_i)_{i=1}^m$ be $m$ finite abelian groups, and
  \begin{equation}
    \label{eq:SetZW}
    \SetZW \defeq \SetZW_1 \times \SetZW_2 \times \dotsb \times \SetZW_m
  \end{equation}
  be their direct product (thus also an abelian group under element-wise $\oplus$ addition).
  Now let $\SetCW$ be a subset (usually a subgroup or its coset) of $\SetZW$,
  $i \in \{1,\dotsc,m\}$, $\lambda_{\except i}$ be $m-1$ probability tuples with each $\lambda_j$ defined
  over $\SetZW_j$, we then define $\nu(\SetCW; \lambda_{\except i})$ as the probability tuple $\nu$ over $\SetZW_i$
  with
  \begin{equation}
    \nu(u_i) = \sum_{\sequ' \in \SetCW: u'_i = u_i} \prod_{j\ne i}
    \lambda_j(u'_j),
  \end{equation}
  where $\sequ' = (u'_1, \dotsc, u'_m)$.  $\odot$, $\oplus$ and $\ominus$ then refer to the case with
  $i=m=3$, $\SetZW_1 = \SetZW_2 = \SetZW_3 = \SetG$, and $\SetCW$ being
  respectively $\{ (u, u, u) \condmid u \in \SetG \}$,
  $ \{ (u_1, u_2, u_1 \oplus u_2) \condmid u_1, u_2 \in \SetG \}$ and $\{ (u_1, u_2, u_1 \ominus u_2) \condmid u_1, u_2 \in \SetG \}$, which are all subgroups of $\SetZW$.
\end{itemize}

If $\lambda$ is a random probability tuple over $\SetG$, i.e.\ each possible value of $\lambda$ is a (deterministic) probability tuple over $\SetG$, we can assign to it
a random variable $u$ whose value lies in $\SetG$ as its \emph{reference variable}, and the conditional distribution of $\lambda$
given $u$ is again called its \emph{density}.  Since the set of possible values of $\lambda$ is the unit $(\cardinal{\SetG}-1)$-simplex (which is no longer one-dimensional when $\cardinal{\SetG}>2$), the probability distribution of $\lambda$ is a probability measure over this simplex, and can be represented by its pdf
w.r.t.\ the Hausdorff measure with a suitable dimensionality depending on the discreteness of the distribution ($\cardinal{\SetG}-1$ in the fully
continuous case and zero in the fully discrete case).  When we write e.g.\ $p(\lambda)$, it will
refer to such a pdf.  In the binary case, we have used bold greek letters to represent the densities
themselves; while such notations remain usable here, e.g.\ $\lambda \condmid u \sim \dlambda$,
we usually prefer to talk about ``the density of $\lambda$ w.r.t.\ $u$'' directly.

Like the binary case, given a random variable $u$ in $\SetG$ and random probability tuples
$\lambda_1$ and $\lambda_2$ over $\SetG$, if $u \markov \lambda_1 \markov \lambda_2$ forms a Markov
chain, we say $\lambda_2$ is a \emph{physically degraded} version of $\lambda_1$ w.r.t.\ $u$, and
write $\lambda_2 \lepd \lambda_1$.

Based on the binary case in \prettyref{def:sym-density} and the discussion that follows, the notion
of \emph{symmetric message densities} can likewise be extended as follows:
\begin{definition}
  \label{def:sym-density-MK}
  Let $\lambda$ be a random probability tuple over finite abelian group $\SetG$ and $u$ be a random variable in
  $\SetG$, then we say $\lambda$ \emph{has a symmetric density} w.r.t.\ $u$ if, for any deterministic
  $u', u'' \in \SetG$ and probability tuple $\lambda'$ over $\SetG$,
  \begin{align}
    \label{eq:msg-sym-MK}
    \plu(\lambda' \condmid u') &= \plu(\lambda' \oplus \constmsg{u''} \condmid u' \oplus u''), \\
    \label{eq:msg-consistent-MK}
    \plu(\lambda' \condmid u') &= C(\lambda') \cdot \lambda'(u'),
  \end{align}
  where the normalization factor $C(\lambda')$ does not vary with $u'$.
\end{definition}

To facilitate further discussion involving symmetric densities, we now
define for any probability tuple $\lambda$ over $\SetG$,
\begin{equation}
  \label{eq:deq}
  \deq{\lambda} \defeq \{ \lambda \oplus \constmsg{u} \condmid u\in\SetG \}
\end{equation}
as a kind of orbit containing $\lambda$, and view its $\cardinal{\SetG}$ elements as
distinct for convenience; we then use e.g.\ $p(\deq{\lambda})$ to denote the
probability that $\lambda$ (as a random variable) lies in a certain (deterministic) $\deq{\lambda}$.
This allows each symmetric density to be reduced to a probability distribution of $\deq{\lambda}$ through the following proposition:
\begin{proposition}
  \label{prop:symm-orbits}
  Let $u$ be a random variable over a finite abelian group $\SetG$ and $\lambda$ be a random probability tuple over it, then
  $\lambda$ has a symmetric density w.r.t.\ $u$ if and only if $p(\lambda \condmid u)$ satisfies
  \begin{equation}
    \label{eq:symm-deq}
    p(\lambda \condmid u) = p(\deq{\lambda}) \cdot \lambda(u).
  \end{equation}
\end{proposition}
\begin{IEEEproof}
  For any random probability tuple $\lambda$ with a symmetric density w.r.t.\ $u$, we
  have $p(\deq{\lambda}\! \condmid u) = p(\deq{\lambda})$ due to \eqref{eq:msg-sym-MK}, and
  \begin{equation}
    \label{eq:consistency-deq}
    \begin{split}
      p(\lambda \condmid \!\deq{\lambda}, u)
      &= \frac{p(\lambda \condmid u)}{p(\deq{\lambda}\! \condmid u)} \\
      &= \frac{p(\lambda \condmid u)}{\sum_{u'\in\SetG} p_{\lambda \condmid u}(\lambda\oplus\constmsg{u'} \condmid u)} \\
      &= \frac{p(\lambda \condmid u)}{\sum_{u'\in\SetG} p_{\lambda \condmid u}(\lambda \condmid u\ominus u')} \\
      &= \frac{\lambda(u)}{\sum_{u'\in\SetG} \lambda(u\ominus u')}
      = \lambda(u).
    \end{split}
  \end{equation}
  Consequently,
  \begin{equation}
    p(\lambda \condmid u) = p(\deq{\lambda}\! \condmid u) p(\lambda \condmid \!\deq{\lambda}, u)
    = p(\deq{\lambda}) \cdot \lambda(u).
  \end{equation}
  Conversely, any $p(\lambda \condmid u)$ in the form of \eqref{eq:symm-deq} obviously satisfies \eqref{eq:msg-sym-MK}
  and \eqref{eq:msg-consistent-MK} and thus makes $\lambda$ symmetric w.r.t.\ $u$.
\end{IEEEproof}

Convex combinations of symmetric densities can be defined just like the binary case: let the index variable $I$ be an arbitrary random variable and the reference variable $u$ be a random variable over $\SetG$ that is independent from $I$, then the density of a random probability tuple $\lambda$ over $\SetG$ w.r.t.\ $u$, represented by $p(\lambda \condmid u)$, is regarded as
a convex combination of densities conditioned on $I$ represented by $p(\lambda \condmid u,I)$.  In particular, using the independence of $u$ from $I$, we have
\begin{equation}
  \label{eq:lambda-u-convex-comb}
  p(\lambda \condmid u) = \sum_I p(I\condmid u) p(\lambda \condmid u,I) = \sum_I p(I) p(\lambda \condmid u,I).
\end{equation}
From \prettyref{prop:symm-orbits}, it is easy to obtain the following results regarding symmetric densities and their convex combinations.
Firstly, convex combinations of symmetric densities remain symmetric:
\begin{proposition}
  \label{prop:convex-comb-symm}
  Let $I$ be an arbitrary random variable, $u$ be a random variable over a finite abelian group $\SetG$ that is independent from $I$,
  and $\lambda$ be a random probability tuple over $\SetG$.  If $\lambda$ has a symmetric density w.r.t.\ $u$ conditioned on each possible $I$, then it is symmetric w.r.t.\ $u$ unconditionally (i.e.\ when averaged over $I$).
\end{proposition}
\begin{IEEEproof}
  By \prettyref{prop:symm-orbits}, each $p(\lambda \condmid u,I)$ has a corresponding $p(\deq{\lambda}\! \condmid I)$ that satisfies
  \begin{equation}
    p(\lambda \condmid u,I) = p(\deq{\lambda}\! \condmid I) \lambda(u).
  \end{equation}
  Substitution into \eqref{eq:lambda-u-convex-comb} gives $p(\lambda \condmid u) = p(\deq{\lambda}) \lambda(u)$ with
  $p(\deq{\lambda}) = \sum_I p(I) p(\deq{\lambda}\! \condmid I)$, so $\lambda$ has a symmetric density w.r.t.\ $u$.
\end{IEEEproof}

Secondly, similar to $\dsym{q}$ in the binary case, we can construct a set of ``minimal'' symmetric densities such that all symmetric densities are convex combinations of them,
allowing many properties satisfied by such densities to be applicable to all symmetric densities by linearity.
\begin{proposition}
  Given any deterministic probability tuple $\lambdar$ over finite abelian group $\SetG$, we define conditional pmf
  (which can be regarded as a pdf w.r.t.\ the zero-dimensional Hausdorff measure)
  \begin{equation}
    \label{eq:sym-basis-MK}
    p(\lambda \condmid u) = \sum_{u' \in \SetG} \lambdar(u \ominus u') \cdot
    \oneif{\lambda = \lambdar \oplus \constmsg{u'}},
  \end{equation}
  then $\lambda$ has a symmetric density w.r.t\ $u$.  Moreover, all symmetric densities are convex combinations of such densities with various values of $\lambdar$.
\end{proposition}
\begin{IEEEproof}
  It is easy to verify that the $p(\lambda \condmid u)$ in \eqref{eq:sym-basis-MK} can be expressed in the form of \eqref{eq:symm-deq} with pmf
  $p(\deq{\lambda}) = \oneif{\deq{\lambda} = \deq{\lambdar}}$, so $\lambda$ is symmetric w.r.t.\ $u$.  Convex combinations of such densities can then yield any possible
  $p(\deq{\lambda})$ and thus any symmetric density.
\end{IEEEproof}

For symmetric densities, physical degradation relationships are still preserved after taking convex combinations:
\begin{proposition}
  \label{prop:pdeg-convex-comb-2K}
  Let $I$ be an arbitrary random variable, $u$ be uniformly distributed over a finite abelian group
  $\SetG$ and independent from $I$, and $\mu$ and $\nu$ be random probability tuples over $\SetG$
  that, when conditioned on $I$, are symmetric w.r.t.\ $u$ and satisfy $\nu \lepd \mu$, then after
  averaging over all $I$, we still have $\nu \lepd \mu$ w.r.t.\ $u$.
\end{proposition}
\begin{IEEEproof}
  We need to prove that
  \begin{equation}
    p(\nu \condmid \mu, u) = \sum_{I} p(\nu \condmid \mu, u, I) p(I\condmid \mu, u)
  \end{equation}
  does not vary with $u$.  Given that $\nu \lepd \mu$ conditioned on $I$,
  $p(\nu \condmid \mu, u, I)$ is already independent from $u$, so it suffices to prove that
  $p(I\condmid \mu, u)$ does not vary with $u$ either.  Using the independence between $u$ and $I$ as well as the symmetry
  of $\mu$ w.r.t.\ $u$ conditioned $I$ (and consequently, when averaged over $I$), we can find that
  \begin{equation}
    p(I\condmid \mu, u) = \frac{p(I\condmid u) p(\mu\condmid u,I)}{p(\mu\condmid u)}
    = \frac{p(I) p(\deq{\mu}\! \condmid I) \mu(u)}{p(\deq{\mu}) \mu(u)},
  \end{equation}
  which indeed does not vary with $u$.
\end{IEEEproof}

$\nu(\SetCW; \cdot)$ can be applied to densities over finite abelian groups, like
\prettyref{def:nu-density}, as follows:
\begin{definition}
  \label{def:nu-density-M}
  Let $(\SetZW_i)_{i=1}^m$ be $m$ finite abelian groups, $\SetZW$ be their direct product, $\SetCW$
  be a deterministic subset of $\SetZW$, $i \in \{1,\dotsc,m\}$, and
  $\dlambda_{\except i} \defeq (\dlambda_1, \dotsc, \dlambda_{i-1}, \dlambda_{i+1}, \dotsc,
  \dlambda_m)$
  be $(m-1)$ message densities, with each $\dlambda_j$ defined over $\SetZW_j$.  Now make
  $\sequ = (u_1, \dotsc, u_m)$ uniformly distributed over $\SetCW$, construct $(m-1)$ random
  probability tuples $\lambda_{\except i}$ such that for any $j\ne i$, $\lambda_j$ is over
  $\SetZW_j$, depends only on $u_j$, and has $\lambda_j\condmid u_j \sim \dlambda_j$, then the
  distribution of the probability tuple $\nu(\SetCW; \lambda_{\except i})$ conditioned on the
  reference $u_i$ is the message density denoted by $\nu(\SetCW; \dlambda_{\except i})$.
\end{definition}

Similar to the binary case (\prettyref{prop:linear-sym} and \prettyref{prop:linear-pdeg}), we can
prove that $\nu(\SetCW; \cdot)$ on symmetric densities preserves symmetry and physical degradation
relationships, provided that $\SetCW$ is a subgroup of $\SetZW$ or a coset thereof.  When $\SetZW_i = \SetZ_2^{K_i}$, $i=1,\dotsc,m$, the direct product
$\SetZW = \SetZ_2^K$ ($K = \sum_i K_i$) can also be regarded as a vector space over $\SetZ_2$, and
it is then equivalent to require that $\SetCW$ be an affine subspace of $\SetZW$.
\begin{proposition}
  \label{prop:linear-sym-suff-stat-2K}
  Let $(\SetZW_i)_{i=1}^m$ be $m$ finite abelian groups, $\SetZW$ be their direct product, $\SetCW$
  be a deterministic subgroup or coset of $\SetZW$, and $\sequ = (u_1, \dotsc, u_m)$ be
  uniformly distributed over $\SetCW$.  Now given $(m-1)$ random probability tuples
  $\lambda_{\except i}$, each $\lambda_j$ defined over $\SetZW_j$, depending only on $u_j$ and
  having a symmetric density with respect to it, the probability tuple $\nu \defeq \nu(\SetCW; \lambda_{\except i})$
  over $\SetZW_i$ then satisfies the follows:
  \begin{itemize}
  \item $\nu$ has a symmetric density w.r.t.\ $u_i$;
  \item $\nu$ depends only on $u_i$, and is also a sufficient statistic for $u_i$ given
    $\lambda_{\exi}$, i.e.\ $\sequ \markov u_i \markov \nu \markov \lambda_{\exi}$ forms a Markov chain.
  \end{itemize}
\end{proposition}
\begin{IEEEproof}
  See \prettyref{app:proof-linear-sym-suff-stat-2K}.
\end{IEEEproof}

\begin{proposition}
  \label{prop:linear-pdeg-2K}
  Let $(\SetZW_i)_{i=1}^m$ be $m$ finite abelian groups, $\SetZW$ be their direct product, $\SetCW$
  be a deterministic subgroup or coset of $\SetZW$, $\sequ = (u_1, \dotsc, u_m)$ be
  uniformly distributed over $\SetCW$, and
  $\lambda_{\except i}$ and $\lambda'_{\except i}$ each be $(m-1)$
  random probability tuples such that for each $j \ne i$,
  \begin{itemize}
  \item $\lambda_j$ and $\lambda'_j$ are probability tuples over $\SetZW_j$, depend only on
    $u_j$ in $\sequ$, and have symmetric densities w.r.t.\ $u_j$;
  \item $\lambda'_j \lepd \lambda_j$ w.r.t.\ $u_j$.
  \end{itemize}
  Now let $\nu_i = \nu(\SetCW; \lambda_{\except i})$ and $\nu'_i =
  \nu(\SetCW; \lambda'_{\except i})$, then $\nu'_i \lepd \nu_i$ w.r.t.\ $u_i$.
\end{proposition}
\begin{IEEEproof}
  See \prettyref{app:proof-linear-pdeg-2K}.
\end{IEEEproof}

When the test channel has the form of \eqref{eq:test-channel} in
\prettyref{prop:test-channel-sym}, analogous to \prettyref{prop:sym-chan-sym-likelihood}, the
likelihood function used as the BP priors has a symmetric density over $\SetG$:
\begin{proposition}
  \label{prop:sym-chan-sym-likelihood-M}
  Let $u$ be a random variable in $\SetG$, $y\in\SetY$ be another
  random variable with conditional pmf or pdf $p(y\condmid u)$, and
  $\lambda$ be a probability tuple over $\SetG$ determined by $y$ with
  $\lambda(u) = p(y\condmid u)$ before normalization.  If there exists an measure-preserving group action
  $\psi_u(\cdot)$ of $\SetG$ on $\SetY$, such that
  \begin{equation}
    \label{eq:chan-sym-M}
    \pyu(y\condmid u) = \pyu(\psi_u(y)\condmid 0),
  \end{equation}
  then $\lambda$ has a symmetric density w.r.t.\ $u$.
\end{proposition}
\begin{IEEEproof}
  See \prettyref{app:proof-sym-chan-sym-likelihood-M}.
\end{IEEEproof}
However, what we actually need is symmetry over $\SetZ_2^K$ after the priors pass through a
modulation mapping, so such mappings are investigated in detail below.


Given $\SetG$ with $\cardinal{\SetG}=2^K$, we define a \emph{modulation mapping} $\phi(\cdot)$ as a possibly random
bijection from $\SetZ_2^K$ to $\SetG$, which can thus map between probability tuples over $\SetZ_2^K$ and
those over $\SetG$ as well.  In particular, since a probability tuple $\lambda$ over $\SetG$ is a
real-valued function over $\SetG$, the corresponding probability tuple over $\SetZ_2^K$ is simply a
function composition $\lambdaphi$.  In general, a random probability tuple $\lambda$'s symmetry
w.r.t.\ random variable $u \in \SetG$ does not necessarily imply $\lambdaphi$'s symmetry w.r.t.\
$\invphi(u)$, nor vice versa; similar to the case of non-binary LDPC coding \cite{ldpc-st-wireless},
dithering is necessary to maintain symmetry.
\begin{proposition}
  \label{prop:sym-M-sym-2K}
  Let $\phi(\cdot)$ be a deterministic modulation mapping from $\SetZ_2^K$ to $\SetG$, $u$ be a random variable
  uniformly distributed in $\SetG$ and $\lambda$ be a random probability tuple over $\SetG$ with a symmetric density w.r.t.\ $u$.
  Now define a random modulation mapping $\phi_1(\cdot)$ with
  $\phi_1(\seqct') \defeq \phi(\seqct' \oplus \seqeps)$ for any vector $\seqct' \in \SetZ_2^K$, where $\seqeps$ is uniformly distributed over
  $\SetZ_2^K$ and independent from $\lambda$ and $u$, then $\lambdaphi_1$ has a symmetric density w.r.t.\ $\invphi_1(u)$.
\end{proposition}
\begin{IEEEproof}
  See \prettyref{app:proof-sym-M-sym-2K}.
\end{IEEEproof}
Conversely, if we want to preserve symmetry when converting a density over $\SetZ_2^K$ into one over
$\SetG$, dither should be introduced on the $\SetG$-side:
\begin{proposition}
  \label{prop:sym-2K-sym-M}
  Let $\phi(\cdot)$ be a deterministic modulation mapping from $\SetZ_2^K$ to $\SetG$, $\seqct$ be a
  random vector uniformly distributed in $\SetZ_2^K$ and $\mu$ be a random probability tuple over
  $\SetZ_2^K$ with a symmetric density w.r.t.\ $\seqct$.  Now define a random modulation mapping
  $\phi_1(\cdot)$ with $\phi_1(\seqct) \defeq \phi(\seqct) \oplus \delta$, where $\delta$ is
  uniformly distributed over $\SetG$ and independent from $\mu$ and $\seqct$, then
  $\mu \compose \invphi_1$ has a symmetric density w.r.t.\ $\phi_1(\seqct)$.
\end{proposition}
\begin{IEEEproof}
  Similar to the proof of \prettyref{prop:sym-M-sym-2K}; see \prettyref{app:proof-sym-2K-sym-M}.
\end{IEEEproof}

In light of these results, our code construction below will perform dithering over both $\SetZ_2^K$
and $\SetG$ by using the modulation mapping
$\phi_1(\seqct) = \phi(\seqct \oplus \seqeps) \oplus \delta$.  In this way, the symmetry of the
priors over $\SetG$ from \prettyref{prop:sym-chan-sym-likelihood-M} can be promoted to
symmetry over $\SetZ_2^K$, and through straightforward generalizations to \prettyref{prop:ext-nu-b}
and \prettyref{prop:ext-nu-a}, the BP messages and extrinsic information are also appropriately
symmetric when a loop-free neighborhood is available, allowing their errors to be bounded using physical degradation relationships just like the
binary case.  At the same time, the extrinsic information of $u_j$, denoted by $\extprob{u}{j}$ in
\cite{flip-ml}, will also have a symmetric density, enabling the recovery algorithm there to be
used.

Finally, for probability tuples over $\SetZ_2^K$, the definition of the entropy $H(\cdot)$ can be extended as follows
for use in the analysis below.  Given a deterministic probability tuple $\mu$ over
$\SetZ_2^K$, it can be viewed as the probability distribution of some random vector
$\seqct \in \SetZ_2^K$, i.e.\ $\Pr{\seqct = \seqct'} = \mu(\seqct')$ for any
$\seqct' \in \SetZ_2^K$.  Now for any $\SetS \subseteq \{ 1,\dotsc,K \}$, we can define
$H_{\SetS}(\mu) \defeq H(\seqct_{\SetS}) \defeq H((\ct_k)_{k\in\SetS})$ as the joint entropy of the
corresponding subset of bits in $\seqct$, and over the $(2^K-1)$ possible non-empty choices of
$\SetS$, the $(2^K-1)$-dimensional vector of $H_{\SetS}(\mu)$'s can be called the \emph{entropy
  function} \cite{on-char-entropy-func-info-ineq} of $\mu$.  For convenience, for any
non-intersecting subsets $\SetS$ and $\SetS'$ of $\{1,\dotsc,K\}$, we also define the conditional
entropy
$H_{\SetS \condmid \SetS'}(\mu) \defeq H(\seqct_{\SetS} \condmid \seqct_{\SetS'}) = H_{\SetS \cup
  \SetS'}(\mu) - H_{\SetS'}(\mu)$.
By averaging the components of the entropy function with the same $\cardinal{\SetS}$, we obtain the
\emph{average entropy function} \cite{average-entropy-func}
\begin{equation}
  \label{eq:avg-entropy-func}
  h_k(\mu) = \binom{K}{k}^{-1} \sum_{\substack{\SetS \subseteq \{1,\dotsc,K\} \\ \cardinal{\SetS} = k}} H_{\SetS}(\mu).
\end{equation}
When $\mu$ is a random probability tuple, we can take the expectation and obtain the (average)
entropy function of its density.  It is obvious that $H_{\{1,\dotsc,K\}}(\mu) = h_K(\mu) = H(\mu)$
and $H_{\emptyset}(\mu) = h_0(\mu) = 0$.  Moreover, if $\mu$ has a symmetric density w.r.t.\ some
uniformly distributed random vector $\seqct^* \in \SetZ_2^K$, then $H_{\SetS}(\mu)$ is simply the conditional entropy
$H(\seqct^*_{\SetS} \condmid \mu)$, and the (average) entropy function then gives the amount of
correlation among the bits in $\seqct^*$ in the conditional distribution $p(\seqct^* \condmid \mu)$.

\subsection{Code Construction and the Quantization Algorithm}
Given a symmetric source coding
problem with $\cardinal{\SetG} = 2^K$, we thus construct the codebook
\begin{multlinefinal}
  \label{eq:SetU-Mary}
  \mathcal{U} = \mathcal{U}(\seqa) = \finalcr \{ \sequ = \sequ(\seqb,\seqa) \defeq
  \seqphi(\seqc) \defeq \seqphi(\seqb\mat{G} \oplus \seqa) \condmid \seqb \in \SetZ_2^{\nb} \},
\end{multlinefinal}
where $\mat{G} = (g_{ij})_{\nb\times \nc}$ is a \emph{binary} sparse generator matrix, now with
$\nc \defeq nK$ and $\nb \defeq nR$, and the scrambling sequence $\seqa \in \SetZ_2^{\nc}$.  For each
$\seqc = \seqb\mat{G} \oplus \seqa \in \SetZ_2^{\nc}$, a codeword
$\sequ = \seqphi(\seqc) \in \SetG^n$ is obtained by mapping every $K$ consecutive bits
$\seqct_j \defeq (c_{j_1}, \dotsc, c_{j_K})$, where $j_k \defeq K(j-1)+k$, into
$u_j \defeq \phi_j(\seqct_j)$ for $j=1,\dotsc,n$.  Each $\phi_j$ is an independently dithered
version of a fixed modulation mapping $\phi$, that is,
$\phi_j(\seqct_j) \defeq \phi(\seqct_j \oplus \seqeps_j) \oplus \delta_j$, with each $\seqeps_j$ and
$\delta_j$ chosen i.i.d.\ uniform from resp.\ $\SetZ_2^K$ and $\SetG$ and known to both the encoder
and the decoder, and the combined dithering sequences are denoted by
$\seqeps \defeq (\seqeps_j)_{j=1}^n \in \SetZ_2^{nK}$ and
$\seqdelta \defeq (\delta_j)_{j=1}^n \in \SetG^n$.\footnote{Although required for analysis, the
  $\seqeps_j$'s are in fact not necessary in the actual quantization algorithm, since in the
  \node{b}-steps actually performed, $\seqa$ can play the same role.}  In particular, when
$\SetG = \SetZ_M$ with $M=2^K$, $\phi(\cdot)$ can (but not forced to) be the Gray mapping, and the
resulting $\mathcal{U}$ can be periodically extended into $\Lambda = \mathcal{U} + M\SetZ^n$ for use
in $M$-ary MSE quantization.

\begin{figure}[!t]
  \centering
  \includegraphics{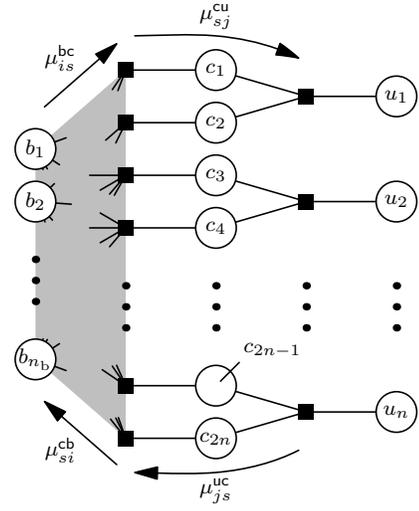}%
  \caption{The factor graph of the $2^K$-ary LDGM quantizer when $K=2$.
    The variable nodes $a_j$ are omitted here; they can also be shown explicitly during analysis of \node{a}-steps,
    similar to \prettyref{fig:binary-fg-a}.}%
  \label{fig:mary-fg}%
\end{figure}

Since every possible $\sequ \in \SetG^n$ occurs $2^{\nb}$ times over the $2^{\nc}$
$\mathcal{U}(\seqa)$'s (each for one $\seqa$), the discussion in \prettyref{sec:outline-quant-ana}
remains applicable.  Specifically, given $t>0$ and under a fixed $\mat{G}$, each $\seqy \in \SetY^n$
still gives a probability distribution
$q(\seqb, \seqa \condmid \seqy) = \e^{-nt d(\sequ(\seqb,\seqa), \seqy)}$ over all
$(\seqb, \seqa)$'s, and the quantization algorithm can still be regarded as an implementation of
BPPQ as defined in \prettyref{sec:outline-quant-ana}, which gives the same average distortion
$D_0(t)$ as TPQ when the synchronization conditions are satisfied.  Each $\trueextprob{b}{i}$ in
\eqref{eq:trueextb} can be expressed by the factor graph in \prettyref{fig:mary-fg}, where the variable nodes corresponding to $\seqa$,
$\seqeps$ and $\seqdelta$ have been omitted due to them being constant during the algorithm; the
factor nodes between variable node $b_{i'}$'s and $c_s$'s give the relationship
$\seqc = \seqb \mat{G} \oplus \seqa$, and they are thus called \emph{check nodes} like the binary
case, while each factor node between variable nodes $c_{j_1}, \dotsc, c_{j_K}$ and $u_j$ corresponds
to $u_j = \phi_j(\seqct_j)$.  The priors are also the same as the binary case: for any $i'$, $s$ and
$j$, $\priprob{b}{i'} = \constmsg{b_{i'}}$ if $b_{i'}$ has been determined (decimated) and
$\constmsg{*}$ otherwise, $\priprob{c}{s} = \constmsg{*}$, while $\priprob{u}{j}$, now a probability
tuple over $\SetG$, is still given by \eqref{eq:priprobs-u} according to $\seqy$ (possibly adjusted by the
recovery algorithm).  By following the BP rules on this factor graph, we thus yield the quantization algorithm
in \prettyref{alg:mary}, which is essentially the same as that used in \cite{flip-ml} except that
the computation of the $\msg{cu}{j_kj}$'s has been moved to the beginning of each iteration in order
to simplify the presentation of the analysis below.  Like the binary case, there remains the choice
between GD and PD in decimation as well as the decimation algorithm, which are dealt with in \cite{flip-ml} and will not be discussed in detail here.

\begin{figure}[!t]
  \centering\footnotesize
   \begin{algorithmic}
    \REQUIRE Quantizer parameters $d(\cdot,\cdot)$, $\mat{G}$, $\phi(\cdot)$, $\seqeps$, $\seqdelta$, $\seqa$, $t$, source sequence $\seqy$
    \ENSURE Quantized codeword $\sequ$ and the corresponding $\seqb$
    \STATE $\priprobx{u}{j}{u} \assign \e^{-t d(u, y_j)}$, $j=1,\dotsc,n$, $u \in \SetG$
    \vspace{0.5mm}%
    \STATE $\msg{bc}{is} \assign \constmsg{*}$, $i=1,\dotsc,\nb$, $s \in \neighbor{bc}{i\cdot}$
    \STATE $\priprob{b}{i} \assign \constmsg{*}$, $i=1,\dotsc,\nb$
    \STATE $\mathcal{E} \assign \{1,2,\dotsc,\nb\}$ \COMMENT{the bits in $\seqb$ not yet decimated}
    \REPEAT[belief propagation iteration]
      \FOR[BP computation of $\msg{cu}{sj}$]{$s=j_k=1$ to $\nc$}
        \STATE $\msg{cu}{sj} \assign \constmsg{a_s} \oplus \left( \oplusl_{i'\in\neighbor{bc}{\cdot s}} \msg{bc}{i's} \right) $
      \ENDFOR
      \STATE Adjust the $\priprob{u}{j}$'s with the recovery algorithm using $\msg{cu}{sj}$
      \FOR[BP computation of $\msg{uc}{jj_k}$]{$j=1$ to $n$}
        \vspace{0.5mm}
        \STATE $\msgx{uc}{jj_k}{c} \assign \sum_{\seqs{\ct}:\ct_k=c}
        \priprobx{u}{j}{\phi_j(\seq{\ct})} \prod_{k'\ne k} \msgx{cu}{j_{k'}j}{\ct_{k'}}$, \\
          $k=1,\dotsc,K$, $c=0,1$
      \ENDFOR
      \FOR[BP computation of $\msg{cb}{si}$]{$s=j_k=1$ to $\nc$}
        \STATE $\msg{cb}{si} \assign (\msg{uc}{js} \oplus \constmsg{a_s}) \oplus
          \left(\oplusl_{i'\in\neighbor{bc}{\cdot s} \excluding{i}} \msg{bc}{i's}\right)$, $i\in\neighbor{cb}{s\cdot}$
      \ENDFOR
      \FOR[BP computation at variable node $b_i$]{$i=1$ to $\nb$}
        \STATE $\msg{bc}{is} \assign \priprob{b}{i}
          \odot \left(\odotl_{s' \in\neighbor{cb}{\cdot i} \excluding{s}} \msg{cb}{s' i} \right)$,
          $s \in \neighbor{bc}{i\cdot}$
        \STATE $\extprob{b}{i} \assign \odotl_{s'\in\neighbor{cb}{\cdot i}} \msg{cb}{s'i}$
      \ENDFOR
      \WHILE{$\mathcal{E} \ne \emptyset$ and more decimation is necessary in this iteration}
        \STATE Choose the bit index $i^*$ to decimate and its value $b^*$
        \STATE $\priprob{b}{i^*} \assign \constmsg{b^*}$,
        $\msg{bc}{i^*s} \assign \constmsg{b^*}$, $s\in\neighbor{bc}{i^*\cdot}$ \COMMENT{decimate $b_i$ to $b^*$}
        \STATE $\mathcal{E} \assign \mathcal{E} \excluding{i^*}$
      \ENDWHILE
    \UNTIL{$\mathcal{E}=\emptyset$}
    \STATE $b_i \assign 0$ (resp.\ $1$) if $\priprob{b}{i}=\constmsg{0}$ (resp.\ $\constmsg{1}$), $i=1,\dotsc,\nb$
    \STATE $\sequ \assign \seqphi(\seqb \mat{G} \oplus \seqa)$
  \end{algorithmic}
  \caption{The quantization algorithm for a symmetric source coding problem over $\SetG$ with $\cardinal{\SetG} = 2^K$}
  \label{alg:mary}
\end{figure}

\subsection{The Asymptotic Synchronization Conditions}
\label{sec:sync-cond-ml}
The synchronization conditions for BPPQ to yield the same distortion performance of TPQ at
asymptotically large $n$ can now be analyzed in essentially the same way as the binary case.
$\mat{G}$ is still chosen to be the generator matrix of a variable-regular check-irregular LDGM
code, with all $\nb$ rows of $\mat{G}$ having $\db\ge 2$ 1's, i.e.\ every variable node $b_i$ in the
factor graph has the same degree $\db$.  To simplify analysis, for each $j = 1,\dotsc,n$, the columns
$j_1, \dotsc, j_K$ corresponding to the bits mapped to the same $u_j$ are made to possess the same number $d$ of
1's each, and we use $w_d$ to denote the fraction of columns with this $d$, and
$v_d \defeq K d w_d / (R \db)$ to denote the fraction of 1's in such columns, which satisfy the
constraints
\begin{equation}
  \label{eq:dd-constr-ml}
  \sum_d w_d = 1,\quad \sum_d v_d = 1,\quad w_d\ge 0, \quad d=1,2,\dotsc.
\end{equation}
This $d$ is henceforth called the \emph{check-degree} of variable node $u_j$.  At each $n$, the set
of $\mat{G}$'s with some given $R$, $\db$ and $\seqw \defeq (w_1,w_2,\dotsc)$ (rounded so that $nR$
and $n\seqw$ contain only integers) is the LDGM code ensemble with this degree distribution, and is
denoted $\GdbwK$, over which $\mat{G}$ is uniformly distributed.  TPQ (or BPPQ) instances having different values of $\mat{G}$,
$\seqeps$, $\seqdelta$, $\seqy$, as well as random sources $\omegaas$ and $\omegabs$ in
the decimation steps of TPQ and BPPQ, thus form an ensemble over which probabilities can be defined.
The analysis of the synchronization conditions is again performed over the TPQ ensemble, and the
reference codeword $(\seqbr, \seqar)$ or the corresponding $\seqcr$ or $\sequr$ remain defined as
the TPQ result.  The reference variables for the BP priors, messages and extrinsic information are
the same as the binary case in \prettyref{sec:binary-def-results}, with the addition of $c^*_s$ for $\msg{uc}{js}$ and
$\msg{cu}{sj}$.

It is easy to prove that the non-binary version of \prettyref{prop:ref-dist} still holds when
conditioned on any fixed $\mat{G}$, $\seqeps$ and $\seqdelta$; in particular, $p(\sequr \condmid \mat{G}, \seqeps, \seqdelta)$ is uniform
and $p(\seqy \condmid \sequr, \mat{G}, \seqeps, \seqdelta) = \prod_j p_{y\condmid u}(y_j \condmid u_j^*)$ is determined by the
test channel, so both $\seqy$ and $\sequr$ are independent from $\seqeps$ and $\seqdelta$.  The test
channel's symmetry (\prettyref{prop:test-channel-sym}) then ensures via
\prettyref{prop:sym-chan-sym-likelihood-M} that each $\priprob{u}{j}$ has a symmetric density over
$\SetG$ w.r.t.\ $u^*_j$, and by \prettyref{prop:sym-M-sym-2K}, after averaging over $\seqeps$ and
$\seqdelta$ (i.e.\ over all TPQ instances in the ensemble with the given $\mat{G}$),
$\priprob{u}{j} \compose \phi_j$ has a symmetric density over $\SetZ_2^K$ w.r.t.\
$\seqct^*_j \defeq (c^*_{j_1}, \dotsc, c^*_{j_K})$.  Similar to the $\Infopri{u}$ in the binary case, we now define
\begin{equation}
  \label{eq:Iu-ml}
  \begin{split}
    \Infopri{u} &\defeq K-\Expe{H(\priprob{u}{j})} = K-H(u^*_j \condmid \priprob{u}{j}) \\
    &= K-H(u^*_j \condmid y_j) = I(u^*_j ; y_j) = I(u;y),
  \end{split}
\end{equation}
where we have used the symmetry of $\priprob{u}{j}$, as well as the fact that $\priprob{u}{j}$ is a
function of $y_j$ and a sufficient statistic for $\sequr$ given $y_j$, and $I(u;y)$ is defined for
the test channel.  Since a modulation mapping only permutes the components of a probability tuple
without changing its entropy, we have
\begin{equation}
  \label{eq:Hupri-Iu}
  \Infopri{u} = K-\Expe{H(\priprob{u}{j} \compose \phi_j)}
\end{equation}
as well.

\begin{figure}[!t]
  \centering
  \subfigure[one repetition unit]{\label{fig:neigh-ml-layer}\includegraphics{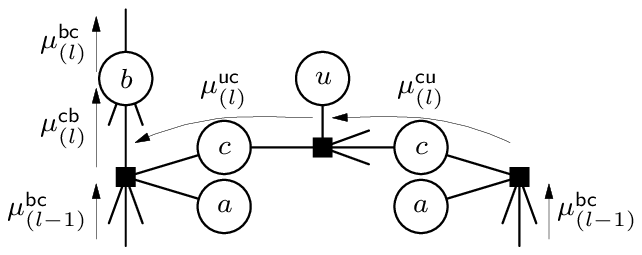}}\qquad
  \subfigure[with node subscripts]{\label{fig:neigh-ml-layer-idx}\includegraphics{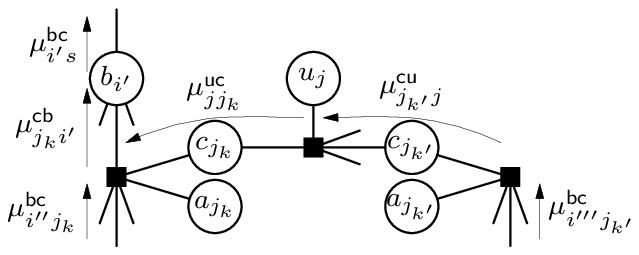}}\qquad
  \caption{Repetition units in each layer of the neighborhood $\neighiL$ in $2^K$-ary LDGM quantization.}%
\end{figure}

When analyzing the synchronization between TPQ and BPPQ in \node{b}-steps, similar to the binary case discussed in
\prettyref{sec:sync-b}, we can define $\extbil{L}$ and $\extbiu{L}$ as BP approximations to each
$\trueextprob{b}{i}$ using $L$ iterations, and thus affected by a depth-$L$ neighborhood $\neighiL$
of variable node $b_i$ in the factor graph.  Like the binary case depicted in
\prettyref{fig:neighi}, $\neighiL$ consists of repeated layers, but each layer is now as shown in
\prettyref{fig:neigh-ml-layer}; for clarity, the variable nodes $a_j$ previously omitted in
\prettyref{fig:mary-fg} are also included here.  If we consider a fixed $\mat{G}$ whose $\neighiL$
is loop-free, but still average the message densities over all $\seqeps$ and $\seqdelta$ (i.e.\
conditioned on $\mat{G}$ only), and define
$\SetCW \defeq \{ (\seqb, \seqa, \seqc) \condmid \seqc = \seqb\mat{G} \oplus \seqa \}$ like
\eqref{eq:SetCW}, then it is straightforward to show that \prettyref{prop:ext-nu-b} remains true
(with $\sequ$ replaced by $\seqc$ and each $\tpriprob{u}{j}$ in the theorem replaced by
$\tpriprob{u}{j} \compose \phi_j$ over $\SetZ_2^K$ and collectively denoted $\tprisphi{u}$),
i.e.\ $\trueextprob{b}{i}$, $\extbil{L}$ and $\extbiu{L}$ still possess the form
$\nu(\SetCW; \tprisx{b}{i}, \tpris{a}, \tprisphi{u})$; therefore, using the symmetry and the
physical degradation relationships among the priors $\tprisx{b}{i}$, $\tpris{a}$ and $\tprisphi{u}$,
as well as the fact that $\SetCW$ is a linear subspace (and thus a abelian subgroup) of
$\SetZ_2^{\nb+2\nc}$, we can apply \prettyref{prop:linear-sym-suff-stat-2K} and
\prettyref{prop:linear-pdeg-2K} to obtain the symmetry and physical degradation relationships among
$\trueextprob{b}{i}$, $\extbil{L}$ and $\extbiu{L}$ w.r.t.\ $b_i^*$, and these properties remain
true when averaged over all $\mat{G}$ with a loop-free $\neighiL$, which occur at high probability
as $n\to\infty$.  The synchronization error is thus still bounded by \eqref{eq:b-sync-bound}.  Similarly,
in \node{a}-steps the synchronization error can be bounded by \eqref{eq:a-sync-bound}.  Using
these results, it is straightforward to prove that conditions analogous to those in
\prettyref{prop:sync-lm} and \prettyref{thm:sync-ul} are still sufficient
for the synchronization conditions to be asymptotically satisfied, and the MI values used by these
conditions can be evaluated for a given degree distribution via density evolution, just like the
binary case.  In particular, if we adopt the notations in \prettyref{sec:sync-de}, e.g.\ $\dextblI{L}$,
to represent the densities of various binary message densities arising in DE, the DE rules at each
variable node $b_i$ remains the same, i.e.\ \eqref{eq:dextb-bc} and \eqref{eq:dextblI} in
\node{b}-steps and \eqref{eq:dexta-bc} and \eqref{eq:dextauI} in \node{a}-steps, while the
check-node rules \eqref{eq:dextb-cb} and \eqref{eq:dexta-cb} are now different.

For concreteness, we now take a look at the computation of $\dextblI{L}$; $\dextbuI{L}$ and
$\dextauI{L}$ can be obtained in an analogous manner.  Similar to \prettyref{prop:sync-lm} in the
binary case, we define a sequence $i = i(n) \in \{1,\dotsc,\nb\}$ that varies with $n$ with
$\lim_{n\to\infty} (i-1)/(\nb-1)$ equal to some $\Infopri{b}$.  Given $n$, $i=i(n)$ and a $\mat{G}$ whose
factor graph has a loop-free neighborhood $\neighi = \neighiL$ that can be further divided into $\neighib$ and $\neighio$, we initialize the BP messages $\msg{bc}{is}$
from $\neighib$ to $\neighio$ to be all-$\constmsg{*}$, and define the priors $\priprob{b}{i'}$ for
$i'\ne i$ to be $\constmsg{b^*_{i'}}$ when $i'<i$ and $\constmsg{*}$ otherwise, and $L$ BP
iterations then yield $\extprob{b}{i}$.  Now consider the density of this $\extprob{b}{i}$ w.r.t.\
$b^*_i$ over the entire TPQ ensemble with block length $n$; as $n\to\infty$, the difference
between $(i-1)/(\nb-1)$ and $\Infopri{b}$, the TPQ instances with loopy neighborhoods, and the
correlation among the nodes in the neighborhood in their degrees and $\priprob{b}{i'}$'s all have
vanishing influence, so the density will converge in distribution to $\dextblI{L}$.  This
$\dextblI{L}$ can now be obtained by performing DE iteratively corresponding to the BP computation
of $\extprob{b}{i}$, just like the binary case; in particular, the DE rules at variable nodes $b_{i'}$ remain
\eqref{eq:dextb-bc} and \eqref{eq:dextblI}, while the method to compute $\dmsgiter{cb}{l}$ from
$\dmsgiter{bc}{l-1}$ will now be shown.  For this purpose, we examine the
part of the factor graph around a variable node $u_j$ in the layer corresponding to iteration $l$ in
$\neighiL$, as shown in \prettyref{fig:neigh-ml-layer-idx}, where the subscripts of the nodes are
explicitly given for convenience of presentation.  Since $\neighiL$ is loop-free, the
$\msg{bc}{i''j_k}$'s (as well as the $\msg{bc}{i'''j_{k'}}$'s) from the leaves of \prettyref{fig:neigh-ml-layer-idx} can be regarded as
independent conditioned on $\seqbr$, with each
$\msg{bc}{i''j_k} \condmid b^*_{i''} \sim \dmsgiter{bc}{l-1}$.  Given the check-degree $d$ of
$u_j$, the conditional density of each $\msg{cu}{j_{k'}j}$, $\msg{uc}{jj_k}$ and $\msg{cb}{j_ki'}$
can be obtained, and averaging the density of $\msg{cb}{j_ki'}$ over $d$ then yields the desired
$\dmsgiter{cb}{l}$.

Like the binary case, DE can be performed numerically by discretizing the possible values of the
probability tuples.  As only binary probability tuples, whose possible values lie in a
one-dimensional space, is amenable to practical discretization, computing the density of
$\msg{uc}{jj_k}$ from that of the $\msg{cu}{j_{k'}j}$'s has to be done in a single step via table
lookup and is only practical when $K=2$.  For larger $K$, Monte-Carlo methods can be used for DE.

\subsection{The Case of Erasure-Like Problems}
\label{sec:mono-opt-ml-erasure}
In the binary case, BEQ is important due to its comparative simplicity of analysis, and the optimized
degree distributions of BEQ can serve as the starting point of degree distribution optimization in
more general problems.  For the $2^K$-ary LDGM code construction discussed here, this role is played
by quantization problems in the form of \prettyref{exm:erasure-quant-K}, henceforth called \emph{erasure-like} problems,
and again in the limit of $t\to\infty$ with the modulation mapping
$\phi: \SetZ_2^K \rightarrow \SetG = \SetZ_2^K$ chosen to be identity.  Like BEQ, when TPQ is run on
such a problem, we will prove that all probability tuples encountered in BP are erasure-like, making
analysis of the DE process substantially easier.

The analysis is similar to that in \prettyref{sec:beq-de}.  Recalling that the source alphabet
$\SetY$ is now the set of all affine subspaces of $\SetG = \SetZ_2^K$, the test channel
$p(u \condmid y)$ is, for any $y \in \SetY$, a uniform distribution over those $u \in y$.
Consequently, using the generalized version of \prettyref{prop:ref-dist} in \prettyref{sec:sync-cond-ml},
when conditioned on $\mat{G}$, $\seqeps$ and $\seqdelta$ (not explicitly shown as conditions),
$p(\seqy) = \prod_j p_y(y_j)$,
$p(\sequr \condmid \seqy) = \prod_j p_{u\condmid y}(u^*_j \condmid y_j)$ is a uniform distribution
over those $\sequr$ with $u^*_j \in y_j$ for all $j$ (the set of those $\sequr$, i.e. the Cartesian product
of all $y_j$'s, is henceforth denoted $\SetUy$), while $p(\seqbr, \seqar \condmid \sequr)$ is a uniform distribution over
those $(\seqbr, \seqar)$ with $\seqphi(\seqbr \mat{G} \oplus \seqar) = \sequr$.  In other words, given $\mat{G}$,
$\seqeps$, $\seqdelta$ and $\seqy$, TPQ yields any $(\seqbr, \seqar)$ in
\begin{equation}
  \SetCWy \defeq \{ (\seqb, \seqa) \condmid \seqphi(\seqb \mat{G} \oplus \seqa) \in \SetUy \},
\end{equation}
which is non-empty due to the freedom in choosing $\seqa$, with equal probability.  Since
$\phi(\cdot)$ is the identity map, we have
$\seqphi(\seqb \mat{G} \oplus \seqa) = \seqb \mat{G} \oplus \seqa \oplus \seqeps \oplus \seqdelta$,
where both $\seqeps$ and $\seqdelta$ are in $\SetZ_2^{nK}$ since $\SetG = \SetZ_2^K$.  Now as
$\SetUy$ is a Cartesian product of affine subspaces, it is itself an affine subspace of
$\SetZ_2^{nK}$, so $\SetCWy$ is an affine subspace of $\SetZ_2^{\nb+\nc}$ as well.

As a generalization to \prettyref{def:erasure-like-density}, we adopt the following definition for
erasure-like probability tuples over $\SetZ_2^K$:
\begin{definition}
  \label{def:erasure-like-density-K}
  A deterministic probability tuple $\mu$ over $\SetZ_2^K$ is said to be erasure-like w.r.t.\ some
  deterministic $\seqct \in \SetZ_2^K$, if there exists a (non-empty) affine subspace $\SetCWt$ of $\SetZ_2^K$
  such that $\seqct \in \SetCWt$ and
  $\mu(\seqct') = (1/\cardinal{\SetCWt}) \cdot \oneif{\seqct' \in \SetCWt}$.  A random probability tuple
  $\mu$ over $\SetZ_2^K$ is said to have an erasure-like density w.r.t.\ a random variable $\seqct$
  in $\SetZ_2^K$ if it is erasure-like w.r.t.\ $\seqct$ with probability 1.
\end{definition}

Each $\priprob{u}{j}$ from \eqref{eq:priprobs-u} is given by
$\priprobx{u}{j}{u} = (1/\cardinal{y_j}) \cdot \oneif{u\in y_j}$, where $\cardinal{y_j}$ is the
cardinality of affine space $y_j$.  Now that $\phi$ is identity and thus
$\phi_j(\seqct_j) = \seqct_j \oplus \seqeps_j \oplus \delta_j$ (the addition is over
$\SetG = \SetZ_2^K$), the probability tuple $\priprob{u}{j} \compose \phi_j$ over $\SetZ_2^K$ is
given by
$(\priprob{u}{j} \compose \phi_j)(\seqct) = (1/\cardinal{y_j}) \cdot \oneif{\seqct \in y_j \ominus
  (\seqeps_j \oplus \delta_j)}$.
Since $y_j \ominus (\seqeps_j \oplus \delta_j)$ is an affine subspace of $\SetZ_2^K$, and under TPQ
$\sequr \in \SetUy$ implies that $\seqct^*_j \defeq (c^*_{j_1}, \dotsc, c^*_{j_K})$ is its member,
we see that $\priprob{u}{j} \compose \phi_j$ is erasure-like w.r.t.\ $\seqct^*_j$.

The computation of $\msg{uc}{jj_k}$ in BP preserves erasure-likeness:
\begin{proposition}
  \label{prop:msg-uc-erasure}
  Let $(\seqbr, \seqar)$ and the corresponding $\seqcr = \seqbr \mat{G} \oplus \seqar$ and
  $\sequr = \seqphi(\seqcr)$ be random variables serving as the reference codeword.  For each $j$
  and $k$, if $\priprob{u}{j} \compose \phi_j$ is erasure-like w.r.t.\
  $\seqct^*_j \defeq (c^*_{j_1}, \dotsc, c^*_{j_K})$, and each $\msg{cu}{j_{k'}j}$ ($k' \ne k$) is
  erasure-like w.r.t.\ $\ct^*_{jk'} \defeq c^*_{j_{k'}}$, then $\msg{uc}{jj_k}$ given by
  \begin{equation}
    \msgx{uc}{jj_k}{c} = \sum_{\seqs{\ct}:\ct_k=c}
    \priprobx{u}{j}{\phi_j(\seq{\ct})} \prod_{k'\ne k} \msgx{cu}{j_{k'}j}{\ct_{k'}}, \quad c=0,1,
  \end{equation}
  is erasure-like as well w.r.t.\ $\ct^*_{jk} \defeq c^*_{j_k}$.
\end{proposition}
\begin{IEEEproof}
  Each possible erasure-like value of $\priprob{u}{j} \compose \phi_j$ has a corresponding affine subspace
  $\SetCWt_0$ of $\SetZ_2^K$ such that $\seqct^*_j \in \SetCWt_0$, and for any $\seqct \in \SetZ_2^K$,
  $(\priprob{u}{j} \compose \phi_j)(\seqct)$ is $1/\cardinal{\SetCWt_0}$ if $\seqct \in \SetCWt_0$ and 0 otherwise.
  Likewise, for every $k'\ne k$, each possible erasure-like value of $\msg{cu}{j_{k'}j}$ corresponds to an affine subspace $\SetCWt_{k'}$,
  given by $\{ \seqct \in \SetZ_2^K \condmid \ct_{k'} = c \}$ if $\msg{cu}{j_{k'}j} = \constmsg{c}$ ($c\in\SetZ_2$) and $\SetZ_2^K$ if $\msg{cu}{j_{k'}j} = \constmsg{*}$;
  obviously $\SetCWt_{k'}$ also contains $\seqct^*$ and, over $\seqct \in \SetZ_2^K$, $\msg{cu}{j_{k'}j}(\ct_{k'})$ is likewise equal
  to a positive constant normalization factor if $\seqct \in \SetCWt_{k'}$ and 0 otherwise.  Consequently, the intersection of $\SetCWt_0$ and all $\SetCWt_{k'}$ for $k'\ne k$,
  denoted by $\SetCWt$, is still an affine
  subspace containing $\seqct^*_j$, and $\priprobx{u}{j}{\phi_j(\seq{\ct})} \prod_{k'\ne k} \msgx{cu}{j_{k'}j}{\ct_{k'}}$ is a constant
  for $\seqct \in \SetCWt$ and zero otherwise.  If all $\seqct \in \SetCWt$ have the same $\ct_k$
  (which must be $\ct^*_{jk}$), then $\msg{uc}{jj_k} = \constmsg{\ct^*_{jk}}$; otherwise,
  $\SetCWt$ being an affine subspace implies that the number of $\seqct \in \SetCWt$ with
  $\ct_k = 0$ must be the same as those with $\ct_k = 1$, and $\msg{uc}{jj_k} = \constmsg{*}$.
  Therefore, $\msg{uc}{jj_k}$ is always erasure-like w.r.t.\ $\ct^*_{jk}$.
\end{IEEEproof}

We thus conclude that, under TPQ, all probability tuples involved in BP are indeed erasure-like, so
the densities of the binary ones can be characterized solely in terms of their MI.  Since
\eqref{eq:dextb-bc} and \eqref{eq:dextblI} are unchanged from the binary case, now the corresponding
MIs $\Infoiter{bc}{l}$ and $\Infoiter{cb}{l}$ still satisfy \eqref{eq:exit-binary-b} and
\eqref{eq:Ibext}, and only the relationship between $\Infoiter{bc}{l-1}$ and $\Infoiter{cb}{l}$
remains to be derived.

Following the discussion in \prettyref{sec:sync-cond-ml}, we consider the factor graph fragment in
\prettyref{fig:neigh-ml-layer-idx} with $u_j$ having check-degree $d$.  Given $\Info{bc} \defeq \Infoiter{bc}{l-1}$, we let
each incoming $\msg{bc}{i''j_k}$ and $\msg{bc}{i'''j_{k'}}$ from the bottom of \prettyref{fig:neigh-ml-layer-idx} be independently
$\constmsg{b^*_i}$ (with probability $\Info{bc}$) or $\constmsg{*}$, then each message
$\msg{cu}{j_{k'}j}$ in the figure, conditioned on the reference codeword, is also independent and erasure-like, being
$\constmsg{c^*_{j_{k'}}}$ with probability $\Infox{cu}{d} \defeq (\Info{bc})^d$ and $\constmsg{*}$
otherwise.

We now know from \prettyref{prop:msg-uc-erasure} that the $\msg{uc}{jj_k}$ obtained from the $\msg{cu}{j_{k'}j}$'s is erasure-like
as well.  The probability that $\msg{uc}{jj_k} = \constmsg{c^*_{j_k}}$ depends on $k$, but for the
purpose of computing $\Infoiter{cb}{l}$ only its average value over $k=1,\dotsc,K$ is needed, which
is denoted by $\Infox{uc}{d}$ and can be obtained from the entropy function of the density of
$\priprob{u}{j} \compose \phi_j$ w.r.t.\ $\seqct^*_j$.  Specifically, let
\begin{equation}
  \label{eq:S}
  \SetS \defeq \{ k' \in \{1,\dotsc,K\} \excluding{k} \mid \msg{cu}{j_{k'}j} = \constmsg{c^*_{j_{k'}}} \}
\end{equation}
be the set of ``known'' incoming messages, then each $\SetS$ with $\cardinal{\SetS} = l$ occurs with
probability $p_{d,l} \defeq \Info{bc}^{dl} \cdot (1-\Info{bc}^d)^{K-1-l}$, and given $\SetS$ and
$\priprob{u}{j} \compose \phi_j$, the probability that $\msg{uc}{jj_k} = \constmsg{c^*_{j_k}}$ is
simply $1-H_{\{k\} \condmid \SetS}(\priprob{u}{j} \compose \phi_j)$.  Taking expectations over
$\SetS$ and $\priprob{u}{j} \compose \phi_j$, and denoting the
$(\priprob{u}{j} \compose \phi_j)$-expectation of e.g.\
$H_{\{k\} \condmid \SetS}(\priprob{u}{j} \compose \phi_j)$ by just $H_{\{k\} \condmid \SetS}$, we
get
\begin{equation}
  \begin{split}
    \Infox{uc}{d} &= 1-\frac{1}{K} \sum_{k=1}^K \Expex{\SetS}{H_{\{k\} \condmid \SetS}} \\
    &= 1-\frac{1}{K} \sum_{k=1}^K
    \sum_{\SetS \subseteq \{1,\dotsc,K\} \excluding{k}} p_{d,\cardinal{\SetS}} \cdot H_{\{k\} \condmid \SetS} \\
    &= 1-\frac{1}{K} \sum_{l=0}^{K-1} p_{d,l}
    \sum_{k=1}^K \sum_{\substack{\SetS \subseteq \{1,\dotsc,K\} \excluding{k} \\ \cardinal{\SetS} = l}}
    H_{\{k\} \condmid \SetS} \\
    &= 1-\sum_{l=0}^{K-1} \binom{K-1}{l} \cdot p_{d,l} \cdot (h_{l+1} - h_l),
  \end{split}
\end{equation}
where $h_l$ is the average entropy function $h_l(\priprob{u}{j} \compose \phi_j)$ with expectation
taken over $\priprob{u}{j} \compose \phi_j$.

Finally, according to the BP rule computing $\msg{cb}{j_ki'}$ in \prettyref{fig:neigh-ml-layer-idx}, its MI averaged over $k$
and $d$ should be
\begin{equation}
  \label{eq:exit-c-ml}
  \Infoiter{cb}{l} = \sum_d v_d \Infox{uc}{d} \Info{bc}^{d-1}
  = \sum_{d,l} v_d \Infox{c}{l} \cdot \alpha_{d,l}(\Infoiter{bc}{l-1}),
\end{equation}
where we have defined for brevity
\begin{equation}
  \label{eq:alpha}
  \begin{split}
    \alpha_{d,l}(x) &\defeq \binom{K-1}{l} \cdot x^{d-1} p_{d,l} \\
    &= \binom{K-1}{l} \cdot x^{d(l+1)-1} (1-x^d)^{K-(l+1)}
  \end{split}
\end{equation}
and
\begin{equation}
  \label{eq:Ic-ml}
  \Infox{c}{l} \defeq 1-(h_{l+1} - h_l).
\end{equation}
In other words, when expressed in the form of \eqref{eq:exit-binary-c-de}, we now have
\begin{equation}
  \label{eq:exit-f-ea-ml}
  f(x) = \sum_{l=0}^{K-1} \frac{\Infox{c}{l}}{\Infopri{u}} \sum_d v_d \alpha_{d,l}(x),
\end{equation}
where the $\Infox{c}{l}$'s can be shown using \eqref{eq:Hupri-Iu} to satisfy
\begin{equation}
  \sum_{l=0}^{K-1} \Infox{c}{l} = K-h_K = K-\Expe{H(\priprob{u}{j} \compose \phi_j)} = \Infopri{u}.
\end{equation}


Having obtained the $f(x)$ of erasure-like problems, the optimization of degree distribution can
proceed using \prettyref{thm:sync-mono} just like the binary case.  For general symmetric source coding
problems, erasure approximation using the same entropy function (and thus the
same $\Infox{c}{l}$'s) and correction with DE results also allow degree distribution optimization to
proceed iteratively, which is essentially the optimization method in \cite{ldgm-vq-journal} and has
been shown to give good results in \cite{flip-ml}.  We have thus obtained a sound theoretical basis
for this optimization method.

\section{Conclusion and Future Work}
\label{sec:conclusion}
In this paper, considering the LDGM-based quantization codes for symmetric source coding problems
previously analyzed in \cite{ldgm-vq-journal} and \cite{flip-ml}, we have introduced the
synchronization conditions that allow the distortion performance of TPQ, namely $D_0(t)$, to be
achieved by the practically possible BPPQ, and then proved that degree distributions
satisfying certain criteria allow these synchronization conditions to be satisfied in an asymptotic
sense as block length $n$ and iteration count $L$ go to infinity.  By making use of the properties
of symmetric message densities, both binary ones and those over an abelian group, these results have
been obtained not only for binary code constructions but for $2^K$-ary BICM-like constructions as
well.  In this way, a firm theoretical basis for the optimization methods in
\cite{ldgm-vq-journal} has been established.

On the other hand, the asymptotic synchronization conditions are not able to analyze the impact of a
loss of synchronization between BPPQ and TPQ, sometimes called a \emph{decimation error}, which is
inevitable in practice due to finite $n$ and $L$.  Such decimation errors can be tackled in practice
with the recovery algorithm proposed in \cite{ldgm-vq-journal} and \cite{flip-ml}, and some ideas,
including the introduction of an idealized recovery algorithm in \cite{flip-ml}, have been proposed
to analyze the resulting performance.  However, except for the sometimes simpler BEQ case, the
analysis has yet to be made rigorous and should be improved accordingly.  Moreover, most analysis
work so far consider only the probabilistic decimator rather than the greedy decimator used in
practice, and all optimization methods are also based on them.  Some analysis of the characteristics
of GD, even empirical ones, would likely allow better optimization and a more thorough understanding
of the quantization process.

\appendices
\section{Proofs}

\subsection{Proof of \prettyref{prop:test-channel}}
\label{app:proof-test-channel}
Due to the symmetry of $p(y)$ and $d(u,y)$, the optimal test channel $p(u\condmid y)$ can be assumed
to give a uniform $p(u)$; otherwise the test channel
$p'(u\condmid y) \defeq (1/\cardinal{\SetG})\sum_{v\in\SetG} p_{u\condmid y}(u\ominus v \condmid \psi_v(y))$ would be
better as it gives the same $D$, the same or lower $R = I(u;y)$ (mutual information is convex
w.r.t.\ the channel transfer probabilities), and the corresponding
$p'(u)\defeq \sum_y p(y)p'(u\condmid y)$ is uniform.  Now $H(u)$ is a constant, so given $D$ the minimization of $I(u;y)$ is equivalent to the maximization of $H(u\condmid y)$, which is easily done with Lagrange multipliers and yields the results in \prettyref{prop:test-channel}.  It can be verified that the corresponding $p(u)$ is indeed uniform.
\qed

\subsection{Proof of \prettyref{prop:ref-dist}}
\label{app:proof-ref-dist}
Conditioned on a fixed $\mat{G}$, we have found in
\prettyref{sec:outline-quant-ana} that the TPQ yields any $(\seqbr, \seqar,
\sequr)$ satisfying $\seqbr\mat{G} \oplus \seqar = \sequr$ with probability proportional to $q(\seqbr, \seqar
\condmid \seqy) = \e^{-ntd(\sequr, \seqy)}$; in other words,
$p(\seqbr, \seqar, \sequr \condmid \seqy) = \e^{-ntd(\sequr, \seqy)} /
Q(\seqy)$, where the normalization factor
\begin{align}
  Q(\seqy) &\defeq \sum_{\seqbr, \seqar} \e^{-ntd(\sequr(\seqbr,\seqar), \seqy)} = 2^{\nb} \sum_{\sequr\in\SetZ_2^n} \e^{-ntd(\sequr, \seqy)} \\
  &= 2^{\nb} \prod_{j=1}^n \sum_{u\in\SetZ_2} \e^{-td(u,y_j)} = 2^{\nb} \prod_{j=1}^n Q(y_j)
\end{align}
due to each $\sequr \in \SetZ_2^n$ having $2^{\nb}$ combinations of
$(\seqbr, \seqar)$ with $\sequr = \seqbr\mat{G} \oplus
\seqar$.  We thus have
\begin{equation}
  p(\seqbr, \seqar, \sequr \condmid \seqy) = 2^{-\nb} \oneif{\seqbr\mat{G} \oplus \seqar = \sequr} \prod_{j=1}^n \frac{\e^{-td(u_j^*, y_j)}}{Q(y_j)},
\end{equation}
so the joint distribution of $(\seqbr, \seqar, \sequr, \seqy)$ given $\mat{G}$ is known, and the desired results immediately follow.
\qed

\subsection{Proof of \prettyref{prop:sym-chan-sym-likelihood}}
\label{app:proof-sym-chan-sym-likelihood}
Here we only consider the case where $y$ and thus
\begin{equation}
  \label{eq:z-def}
  z \defeq f(y) \defeq \mu(0) = \frac{\pyb(y\condmid 0)}{\pyb(y\condmid 0) + \pyb(y\condmid 1)}
\end{equation}
are continuous-valued.  Since $\psi_b(\cdot)$ is a group action of
$\SetZ_2$, $\psi_1$ must be a bijection with $\psi_1^{-1} = \psi_1$,
and using $\pyb(y\condmid 1) = \pyb(\psi_1(y)\condmid 0)$, we see that
$f(\psi_1(y)) = 1-f(y)$.

According to \prettyref{def:sym-density}, we need to prove that,
for any $z\in [0,1]$,
\begin{align}
  \label{eq:symz-1}
  \pzb(z\condmid 1) &= \pzb(1-z\condmid 0), \\
  \label{eq:symz-2}
  (1-z)\cdot \pzb(z\condmid 0) &= z\cdot \pzb(1-z\condmid 0).
\end{align}

Let $\dz$ be a small positive number.  Given an arbitrary $z_0\in [0,1-\dz]$, we can define
$\Zz \defeq [z_0,z_0+\dz]$, $\Zo \defeq 1-\Zz = [1-z_0-\dz,1-z_0]$, $\Yz \defeq f^{-1}(\Zz)$ and
$\Yo \defeq f^{-1}(\Zo)$.  Now for any $z=f(y)$, the events $z\in\Zz$, $1-z\in\Zo$, $y\in\Yz$ and
$\psi_1(y)\in\Yo$ are equivalent, and we can define $P_b$ as their probability conditioned on a
fixed $b=0,1$ and $P \defeq P_0+P_1$.  It can be observed that
\begin{equation}
  \begin{split}
    P_1 &= \Pr{z\in\Zz\condmid b=1} = \Pr{y\in\Yz\condmid b=1} \\
    &= \Pr{\psi_1(y)\in\Yz\condmid b=0} = \Pr{1-z\in\Zz\condmid b=0},
  \end{split}
\end{equation}
and by letting $\dz\to 0$, we get \eqref{eq:symz-1}.

To prove \eqref{eq:symz-2}, first note from \eqref{eq:z-def} that, for any $y\in\Yz$,
\begin{equation}
  z_0 P(y) \le f(y) P(y) = \pyb(y\condmid 0) \le (z_0+\dz) P(y),
\end{equation}
where $P(y) \defeq \pyb(y\condmid 0) + \pyb(y\condmid 1)$.  Integrating over $y\in\Yz$ we obtain
\begin{equation}
  z_0 P \le P_0 \le (z_0+\dz) P, \quad \text{i.e. } z_0 \le P_0/P \le z_0+\dz.
\end{equation}
As $\dz\to 0$, this becomes
\begin{equation}
\pzb(z_0\condmid 0) / (\pzb(z_0\condmid 0) + \pzb(z_0\condmid 1)) = z_0,
\end{equation}
which is equivalent to \eqref{eq:symz-2}. \qed

\subsection{Proof of \prettyref{prop:msg-compare-mi}}
\label{app:proof-msg-compare-mi}
We use $q_i \defeq \mu_i(0)$ ($i=1,2$) to uniquely represent each $\mu_i$.
$b\markov q_1\markov q_2$ thus forms a Markov chain.  As $b$ is
equiprobable and $\dmu_1$ and $\dmu_2$ are symmetric, we have
\begin{equation}
  \begin{split}
    q_2 &= p_{b\condmid q_2}(0\condmid q_2) = \int_0^1 p_{b\condmid q_1}(0\condmid q_1) p(q_1\condmid q_2) dq_1 \\
    &= \int_0^1 q_1 p(q_1\condmid q_2) dq_1 = \Expe{q_1\condmid q_2}.
  \end{split}
\end{equation}
Now let $f(q) = H_2(q)\cdot\ln2 + 2(q-q_2)^2$, which is
concave in the interval $[0,1]$ as $f''(q) = 4-(1/q + 1/(1-q)) \le 0$ for $0<q<1$, so by Jensen's
inequality $f(q_2) \ge \Expe{f(q_1)\condmid q_2}$, i.e.
\begin{equation}
  \label{eq:msg-compare-h2-q2}
  \Expe{H_2(q_1)\condmid q_2} + \frac{2}{\ln 2} \Expe{(q_1 - q_2)^2\condmid q_2} \le H_2(q_2).
\end{equation}
As $I(\dmu_i) = 1-\Expe{H(\mu_i)} = 1-\Expe{H_2(q_i)}$, $i=1,2$, taking
the expectation of \eqref{eq:msg-compare-h2-q2} over $q_2$ yields the
desired result. \qed

\subsection{Proof of \prettyref{prop:erasure-nu}}
\label{app:proof-erasure-nu}
By definition it suffices to prove that, given $\seqb \in \SetCW$, if each $\lambda_{i'}$ ($i'\ne i$) is either $\constmsg{b_{i'}}$ or $\constmsg{*}$, then
$\nu \defeq \nu(\SetCW; \lambda_{\except i})$ is either $\constmsg{b_i}$ or $\constmsg{*}$.
To prove this, we note that
\begin{equation}
  \label{eq:SetC1}
  \SetCW' \defeq \{ \seq{b'} \in \SetCW \condmid
  b'_{i'} = b_{i'} \text{ for all } i'\ne i \text{ with } \lambda_{i'} = \constmsg{b_{i'}} \}
\end{equation}
is a non-empty (since $\seq{b} \in \SetCW'$) affine subspace of $\SetCW$, so either all
vectors in $\mathcal{C'}$ have the same value at the $i$-th position (which is necessarily $b_i$),
or exactly half is 0 (or 1) at that position.  From the definition of $\nu$, it is
$\constmsg{b_i}$ in the former case and $\constmsg{*}$ in the latter. \qed

\subsection{Proof of \prettyref{prop:exit-prop}}
\label{app:proof-exit-prop}
\prettyref{enum:Ibext-order-L} and \prettyref{enum:Iaext-order-L} follow immediately from
respectively \eqref{eq:dextbi-degradation} and \eqref{eq:dextaj-degradation} using
\prettyref{prop:linear-ddeg} and the symmetry of the densities.  Alternatively,
properties of the MIs of DE results $\IbextlIL$ and $\IbextuIL$ can also be obtained by
noting that degradation relationships are preserved in every DE step.

In order to obtain properties \prettyref{enum:Ibext-order-Ib} and
\prettyref{enum:Iaext-order-Ia}, it is necessary to prove for a fixed $n$ and $l\le L$ that
$\dextbit$, $\dextbitL$, $\dextbil{l;L}$ and $\dextbiu{l;L}$ are respectively ordered by degradation in $i$,
while $\dextajt$, $\dextajtL$ and $\dextaju{l;L}$ are respectively ordered by degradation in $j$.  As the methods
are essentially the same, we only give the proof for $\dextbitL$.

Recall that for any $i$, $\trueextprob{b}{i} = \nu(\SetCW; \prisx{b}{i}, \pris{a}, \pris{u})$, with
$\SetCW$ defined in \eqref{eq:SetCW}, all $\priprob{a}{j} = \constmsg{a_j^*}$, and
$\priprob{b}{i''} = \constmsg{b_{i''}^*}$ if $i'' < i$ and $\constmsg{*}$ otherwise, while
$\dextbitL$ is the density of this $\trueextprob{b}{i}$ w.r.t.\ $b^*_i$ over $\mat{G}$ uniformly
distributed in $\GdbwbiL$.  In other words, $(b^*_i, \trueextprob{b}{i})$ can be viewed as random
variables defined on the probability space
\begin{multlinefinal}
  \label{eq:omega-perm}
  \Omega \defeq \{ (\mat{G}, \seqy, \omegaas, \omegabs) \condmid \mat{G} \in \GdbwbiL, \seqy \in \SetY^n, \finalcr
  \omegabs \in [0,1)^{\nb}, \omegaas \in [0,1)^{\nc} \}
\end{multlinefinal}
containing TPQ instances with $\mat{G}$ having loop-free neighborhoods, and $\dextbitL$ is their conditional
probability distribution.

Now for any $i'>i$, $\dextbiitL$ is the density of $\trueextprob{b}{i'}$ w.r.t.\
$b^*_{i'}$ over uniform $\mat{G} \in \GdbwbipL$, and the probability space $\Omega'$, over which the
random variables $b^*_{i'}$ and $\trueextprob{b}{i'}$ are defined, is given by \eqref{eq:omega-perm}
with $\GdbwbiL$ replaced by $\GdbwbipL$.  As $\dextbitL$ and $\dextbiitL$ are conditional
distributions of random variables defined on respectively $\Omega$ and $\Omega'$, for the purpose of
comparison we define a permutation $\pi$ of $\{1,\dotsc,\nb\}$ as
$\pi(i'') = (i''+(i'-i)) \bmod \nb$ (where the modulo operation is onto $\{1,\dotsc,\nb\}$), which
then gives a probability-preserving bijection from $\Omega$ to $\Omega'$ that renumbers every variable node
$b_{i''}$ in each TPQ instance in $\Omega$ into $b_{\pi(i'')}$; specifically, the TPQ instance
$(\mat{G}=(g_{i''j})_{\nb\times \nc}, \seqy, \omegabs, \omegaas) \in \Omega$ is mapped to
$(\mat{G}' = (g'_{i''j})_{\nb\times\nc}, \seqy, {\omegabs}', \omegaas) \in \Omega'$, where
$g'_{\pi(i''),j} \defeq g_{i''j}$ so that the factor graph remains unchanged apart from the
renumbering, and $\omegabs$ can be transformed into ${\omegabs}'$ in a probability-preserving manner such that the
each pre-transformation $b^*_{i''}$ is equal to the post-transformation
$b^*_{\pi(i'')}$.\footnote{Note that e.g.\ the pre-transformation $b^*_1$ is determined in the first
  \node{b}-step, while the corresponding post-transformation $b^*_{\pi(1)}$ is determined in the
  $\pi(1)$-th \node{b}-step, and the transformation from $\omegabs$ to ${\omegabs}'$ is meant to
  deal with this ordering difference.  By \prettyref{prop:ref-dist},
  $p(\seqbr, \seqar \condmid \mat{G}, \seqy)$ remains invariant when $\seqbr$ and $\mat{G}$ are
  simultaneously permuted with $\pi$; therefore, each possible pre-transformation $\seqbr$
  corresponds to a rectangular region of $\omegabs$ that yield it, while its transformed version
  corresponds to a rectangular region of ${\omegabs}'$, and both regions have the same volume equal
  to the probability, allowing a probability-preserving (i.e.\ measure-preserving) bijection to be defined between them.  Combining the
  bijections for each $\seqbr$ then yields the desired probability-preserving transformation from $\omegabs$ to
  ${\omegabs}'$.}  As $\mat{G} \in \GdbwbiL$ if and only if $\mat{G}' \in \GdbwbipL$, we have indeed
obtained an probability-preserving bijection from $\Omega$ to $\Omega'$.  With this bijection, the random
variable $b^*_{i'}$ on $\Omega'$ becomes $b^*_i$ on $\Omega$, and $\trueextprob{b}{i'}$ on $\Omega'$
becomes $\nu' \defeq \nu(\SetCW; \tprisx{b}{i}, \pris{a}, \pris{u})$ defined on $\Omega$, where
each $\tpriprob{b}{i''}$ is $\constmsg{b_{i''}^*}$ when $i'' < i$ or $i'' > \nb-(i'-i)$ and is
$\constmsg{*}$ otherwise, i.e.\ $\tprisx{b}{i}$ contains $(i'-i)$ extra ``known'' probability tuples
compared to $\prisx{b}{i}$.  Consequently, the density of $\nu'$ w.r.t.\ $b^*_i$ on $\Omega$ is the same as that of $\trueextprob{b}{i'}$ w.r.t.\ $b^*_{i'}$, i.e.\ $\dextbiitL$,
and by \prettyref{prop:linear-pdeg} we also have $\trueextprob{b}{i} \lepd \nu'$
w.r.t.\ $b^*_i$, hence $\dextbitL$ is a degraded version of $\dextbiitL$.  \qed

\subsection{Proof of \prettyref{prop:sync-lm}}
\label{app:proof-sync-lm}
\emph{Direct part}: To prove \eqref{eq:asympt-sync-b}, we start from \eqref{eq:b-sync-bound-t}.
For any $l\le L$ and sufficiently large $n$ (such that $\GdbwbiL$ is non-empty), \eqref{eq:b-sync-bound-t} can be reexpressed as
\begin{multlinefinal}
  \label{eq:b-sync-bound-t1}
  \Expe{(\extbil{l}(0) - \trueextprobx{b}{i}{0})^2 \Bigcondmid \mat{G} \in \GdbwbiL} \finalcr
  \le \frac{\ln 2}{2} \left( \IbextmInLx{\Ibprin} - \IbextlInlLx{\Ibprin} \right);
\end{multlinefinal}
As $\Pr{\mat{G} \notin \GdbwbiL} = \PloopbnL$, the unconditional expectation (over all $\mat{G}\in\Gdbwb$) can also be bounded as
\begin{multlinefinal}
  \label{eq:b-sync-bound-t2}
  \Expe{(\extbil{l}(0) - \trueextprobx{b}{i}{0})^2} \finalcr
  \le \frac{\ln 2}{2} \left( \IbextmInLx{\Ibprin} - \IbextlInlLx{\Ibprin} \right) + \PloopbnL.
\end{multlinefinal}
For any $\epsilon>0$, let $\Ibpril \defeq \max(0, \Ibpriz-\epsilon)$,
$\Ibpriu \defeq \min(1, \Ibpriz+\epsilon)$, then $\Ibpril \le \Ibprin \le \Ibpriu$ for all $n$
larger than some threshold $n_0(\epsilon)$, so we can use the monotonicity of $\IbextmInL$ and
$\IbextlInlL$ w.r.t.\ $\Infopri{b}$ to transform \eqref{eq:b-sync-bound-t2} into
\begin{multlinefinal}
  \Expe{(\extbil{l}(0) - \trueextprobx{b}{i}{0})^2} \finalcr
  \le \frac{\ln 2}{2} \left( \IbextmInLx{\Ibpriu} - \IbextlInlLx{\Ibpril} \right) + \PloopbnL,
\end{multlinefinal}
and taking the $n\to\infty$ limit then yields, for any $\epsilon>0$,
\begin{equation}
  \limsup_{n\to\infty} \Expe{(\extbil{l}(0) - \trueextprobx{b}{i}{0})^2}
  \le \frac{\ln 2}{2} \left( \IbextmIux{\Ibpriu} - \IbextlIlx{\Ibpril} \right).
\end{equation}
Now $\IbextmIux{\Ibpriu}
= \IbextlIx{\Ibpriu}$, so $\IbextmIux{\Ibpriu} -
\IbextlIlx{\Ibpril}$ is the sum of $\IbextlIx{\Ibpriu} - \IbextlIx{\Ibpril}$ and $\IbextlIx{\Ibpril}
- \IbextlIlx{\Ibpril}$.  Since we have assumed that $\IbextlI$ is continuous at
$\Ibpriz$,
the former can be made arbitrarily small by choosing a sufficiently small $\epsilon$,
and the latter then vanishes as well when $l\to\infty$.
We have thus proved \eqref{eq:asympt-sync-b} as desired, and \eqref{eq:asympt-sync-a} can be proved
similarly.

\emph{Converse part}: Assuming that \eqref{eq:b-sync-lm} is unsatisfied, then for a certain
$\Ibpril \in [0,1]$ we have $\IbextmIux{\Ibpril} - \IbextlIx{\Ibpril} = \delta > 0$.  By
\prettyref{prop:msg-compare-mi-converse}, there exists $\epsilon > 0$ such that for any $n$,
$l$, $L$ and $i$ satisfying $l\le L$, $\PloopbnL \le 1/2$ and $\IbextmInL - \IbextlInlL \ge \delta/4$ (where
$\Infopri{b} = (i-1)/(\nb-1)$), we have
\begin{equation}
  \label{eq:asympt-nsync-cond}
  \Expe{(\extbil{l}(0) - \trueextprobx{b}{i}{0})^2 \Bigcondmid \mat{G} \in \GdbwbiL} \ge 2\epsilon,
\end{equation}
and \eqref{eq:asympt-nsync-b} thus holds.  Now we just have to find, for any given $l$ and $n_0$, some $n\ge n_0$, $L\ge l$ and $i$ with $\PloopbnL \le 1/2$ and
$\IbextmInL - \IbextlInlL \ge \delta/4$.  Firstly, since $\IbextlIl \le \IbextlI$ for any $\Infopri{b}$ and in particular $\Ibpril$, we have
\begin{equation}
  \label{eq:syncc-L}
  \IbextmIux{\Ibpril} - \IbextlIlx{\Ibpril} \ge \delta.
\end{equation}
Using the continuity of $\IbextlIl$ w.r.t.\ $\Infopri{b}$, an $\Ibpriu \in (\Ibpril,1]$ (except that $\Ibpriu$ is allowed to be 1 when $\Ibpril = 1$) can be found
that makes
\begin{equation}
  \label{eq:syncc-lIL-cont}
  \IbextlIlx{\Ibpriu} \le \IbextlIlx{\Ibpril} + \delta/4.
\end{equation}
Now let ${\nb}_1 \defeq 1 + 1/(\Ibpriu - \Ibpril)$ (or 2 when $\Ibpril = 1$), and choose $n_1$ such
that any $n\ge n_1$ has the corresponding $\nb \ge {\nb}_1$,\footnote{Recall that we have defined $\nb=n\Rn$ for each $n$ with $\lim_{n\to\infty} \Rn=R$.} then there exists, for any $n\ge n_1$,
integer $i \in \{1,\dotsc,\nb\}$ such that $(i-1)/(\nb-1) \in [\Ibpril, \Ibpriu]$.  If we further
choose any $L\ge l$, then by $\IbextlIlx{\Ibpriu} = \lim_{n\to\infty} \IbextlInlLx{\Ibpriu}$ and
\eqref{eq:Ploopb-lim}, there also exists $n_2$ such that for any $n\ge n_2$,
\begin{equation}
  \label{eq:syncc-lIL-n}
  \IbextlInlLx{\Ibpriu} \le \IbextlIlx{\Ibpriu} + \delta/4, \quad \PloopbnL \le 1/2.
\end{equation}
On the other hand, since $\IbextmIux{\Ibpril} = \limsup_{n\to\infty} \IbextmInLx{\Ibpril}$, for the
given $n_0$ we can find $n \ge \max(n_0, n_1, n_2)$ such that
\begin{equation}
  \label{eq:syncc-mIu-n}
  \IbextmInLx{\Ibpril} \ge \IbextmIux{\Ibpril} - \delta/4.
\end{equation}
Combining \eqref{eq:syncc-L}, \eqref{eq:syncc-lIL-cont}, \eqref{eq:syncc-lIL-n} and
\eqref{eq:syncc-mIu-n} and using the monotonicity of $\IbextlInlL$ and $\IbextmInL$ w.r.t.\
$\Infopri{b}$, we conclude that, for any $\Infopri{b} \in [\Ibpril, \Ibpriu]$,
\begin{equation}
  \label{eq:syncc-boundn}
  \IbextmInL - \IbextlInlL \ge \IbextmInLx{\Ibpril} - \IbextlInlLx{\Ibpriu} \ge \delta/4.
\end{equation}
As the $n$ chosen above satisfies $n \ge n_1$, an $i \in \{1,\dotsc,\nb\}$ can be found such that
$(i-1)/(\nb-1) \in [\Ibpril, \Ibpriu]$, and \eqref{eq:syncc-boundn} is then satisfied at
$\Infopri{b} = (i-1)/(\nb-1)$, in which case $\IbextmInL - \IbextlInlL \ge \delta/4$, making
\eqref{eq:asympt-nsync-b} satisfied.

The part of the result when \eqref{eq:a-sync-lm} fails to hold can be proved similarly.
\qed

\subsection{Proof of \prettyref{prop:map-area}}
\label{app:proof-map-area}
Since the probabilities here are defined over the TPQ ensemble, by the arguments in
\prettyref{sec:outline-quant-ana}, given $\seqy$ and $\mat{G}$ each reference codeword
$(\seqbr, \seqar)$ occurs with probability
$p(\seqbr, \seqar \condmid \seqy, \mat{G}) = C \cdot q(\seqbr, \seqar \condmid \seqy)$ with $C$
being a normalization factor.  Substituting this into \eqref{eq:trueexta} and \eqref{eq:trueextb},
we see that
\begin{align}
  \trueextprobx{a}{j}{a} &= p(a_j^* = a \condmid a_1^*, \dotsc, a_{j-1}^*, \seqy, \mat{G}), \\
  \trueextprobx{b}{i}{b} &= p(b_i^* = b \condmid \seqar, b_1^*, \dotsc, b_{i-1}^*, \seqy, \mat{G}).
\end{align}
Therefore, in each TPQ instance, we have
\begin{align}
H(\trueextprob{a}{j}) &= H(a^*_j \condmid a^*_1, \dotsc, a^*_{j-1}, \seqy, \mat{G}), \\
H(\trueextprob{b}{i}) &= H(b^*_i \condmid \seqar, b_1^*, \dotsc, b_{i-1}^*, \seqy, \mat{G}),
\end{align}
where no expectation has been taken over the conditions in the entropy.  Now take the expectation
over all TPQ instances (i.e.\ over $\seqy$, $\mat{G}$, $\seqbr$ and $\seqar$) and sum over $i$ and
$j$, and we get
\begin{equation}
  \label{eq:map-area-sumH}
  \sum_{j=1}^{\nc} H(\dextajt) + \sum_{i=1}^{\nb} H(\dextbit) = H(\seqbr, \seqar \condmid \seqy, \mat{G}).
\end{equation}
On the other hand, from \prettyref{prop:ref-dist} we have
\begin{equation}
  \begin{split}
    H(\seqbr, \seqar \condmid \seqy, \mat{G}) &= H(\seqbr, \seqar \condmid \sequr, \mat{G}) + H(\sequr \condmid \seqy, \mat{G}) \\
    &= \nb + n H(u\condmid y),
  \end{split}
\end{equation}
where $p(u\condmid y)$ is the test channel.  Consequently, \eqref{eq:map-area-sumH} implies that
\begin{equation}
  \label{eq:map-area-sumI}
  \begin{split}
    \sum_{j=1}^{\nc} I(\dextajt) + \sum_{i=1}^{\nb} I(\dextbit)
    &= \sum_{j=1}^{\nc} \IaextmInx{\Infoprix{a}{j}} + \sum_{i=1}^{\nb} \IbextmInx{\Infoprix{b}{i}} \\
    &= n I(u; y) = n\Infopri{u},
  \end{split}
\end{equation}
which concludes the proof of \eqref{eq:map-area-n}.

In order to prove \eqref{eq:map-area}, we note that each summation in \eqref{eq:map-area-n} is
approximately proportional to the integral of $\IbextmIn$ or $\IaextmIn$ after linear interpolation; for example,
\begin{equation}
  \begin{split}
    &\quad \int_0^1 \IbextmIn \,d\Infopri{b} \\
    &= \sum_{i=1}^{\nb-1} \int_{\Infoprix{b}{i}}^{\Infoprix{b}{i+1}} \IbextmIn \,d\Infopri{b} \\
    &= \frac{1}{\nb-1} \sum_{i=1}^{\nb-1} \left( \frac{\IbextmInx{\Infoprix{b}{i}} + \IbextmInx{\Infoprix{b}{i+1}}}{2} \right) \\
    &= \frac{1}{\nb-1} \left( \sum_{i=1}^{\nb} \IbextmInx{\Infoprix{b}{i}} - \frac{\IbextmInx{0} + \IbextmInx{1}}{2} \right) \\
    &= \frac{1}{nR} \sum_{i=1}^{\nb} \IbextmInx{\Infoprix{b}{i}} + \asymptO\left( \frac{1}{n} \right).
  \end{split}
\end{equation}
Consequently,
\begin{equation}
  \begin{split}
    &\quad \int_0^1 \IaextmIn \,d\Infopri{a} + R\int_0^1 \IbextmIn \,d\Infopri{b} \\
    &= \frac{1}{n} \left( \sum_{j=1}^{\nc} \IaextmInx{\Infoprix{a}{j}} + \sum_{i=1}^{\nb} \IbextmInx{\Infoprix{b}{i}} \right)
    + \asymptO\left( \frac{1}{n} \right) \\
    &= \Infopri{u} + \asymptO(1/n).
  \end{split}
\end{equation}
Taking the $n\to\infty$ limit and applying Fatou's lemma yields \eqref{eq:map-area}. \qed

\subsection{Proof of \prettyref{prop:ebp-prop}}
\label{app:proof-ebp-prop}
Eqs.~\eqref{eq:Ib-range} and \eqref{eq:Ibext0} follow immediately from respectively \eqref{eq:Ib}
and \eqref{eq:Ibext-Ibc}.  For \eqref{eq:ebp-area}, first note from \eqref{eq:exit-f-beq}--\eqref{eq:exit-h-beq} that, under BEQ,
\begin{equation}
  \int_0^1 f(x)\,dx = \sum_d \frac{v_d}{d} x^d \big|_{x=0}^1 = \frac{1}{R\db},
\end{equation}
and $h'(y)/g(y) = \db$.  Therefore, letting $y = 1-\Infopri{u} f(x)$, we have
\begin{equation}
  \begin{split}
    &\quad \int_0^1 (1-\Infopri{b}) \frac{d\Ibext}{dx} \,dx \\
    &= \int_0^1 \frac{1-x}{g(y)} \cdot \Infopri{u} h'(y) f'(x) \,dx \\
    &= \db\Infopri{u} \int_0^1 (1-x) f'(x) \,dx \\
    &= \db\Infopri{u} \left( (1-x) f(x) |_{x=0}^1 + \int_0^1 f(x) \,dx \right) \\
    &= \db\Infopri{u} (-v_1 + 1/(R\db)) = \Infopri{u}/R - \db\Infopri{u}v_1. \ \qed
  \end{split}
\end{equation}

\subsection{Proof of \prettyref{prop:linear-sym-suff-stat-2K}}
\label{app:proof-linear-sym-suff-stat-2K}
In the proof we will use $\sequ'_{\exi}$ to denote
$(u'_1, \dotsc, u'_{i-1}, u'_{i+1}, \dotsc, u'_m)$, and $\SetZW_{\exi}$ to denote the direct product of $(\SetZW_j)_{j\ne i}$.

As $\SetCW$ is a coset, we have $\mathcal{C} = \mathcal{X} \oplus \sequ_0$ where
$\mathcal{X}$ is the corresponding subgroup of $\SetZW$ and $\sequ_0 \in \SetZW$.  For each
$d \in \SetZW_i$, we define
$\mathcal{X}_{d} \defeq \{ \seqd \in \mathcal{X} \condmid d_i = d \}$, then
$\mathcal{X}_0$ is in turn a subgroup of $\mathcal{X}$, and any other $\mathcal{X}_d$
is either empty or equal to $\mathcal{X}_0 \oplus \seqd$ where $\seqd$ is any element in
$\mathcal{X}_d$.  We can also define
$\SetD_i \defeq \{ d \in \SetZW_i \condmid \mathcal{X}_d \ne \emptyset \}$, which is
a subgroup of $\SetZW_i$, and it is easy to prove that
$\mathcal{X}_d \oplus \mathcal{X}_{d'} = \mathcal{X}_{d \oplus d'}$ for all
$d, d' \in \SetD_i$.  Note that since $\sequ$ is distributed over $\mathcal{C}$, for
$p(\nu \condmid \sequ)$ it is only necessary to consider $\sequ \in \SetCW$, and similarly for
$p(\nu \condmid u_i)$ only $u_i \in \SetCW_i \defeq \SetD_i \oplus u_{0i}$ is relevant,
as $u_i$ never takes other values.

By linearity, we may first assume that each $\lambda_j$ is discrete and has a conditional pmf
in the form of \eqref{eq:sym-basis-MK}, i.e.
\begin{equation}
  \label{eq:dsym-q1-2K}
  p(\lambda_j \condmid u_j) = \sum_{u'_j \in \SetZW_j} \lambdar_j(u_j \ominus u'_j) \cdot
  \oneif{\lambda_j = \lambdar_j \oplus \constmsg{u'_j}}
\end{equation}
for some deterministic probability tuple $\lambdar_j$ over $\SetZW_j$.
As a result, given $\sequ$, the probability that $\lambda_j = \lambdar_j \oplus \constmsg{u_j'}$
(here $\lambdar_j \oplus \constmsg{u_j'}$ for different values of $u_j'$ are safely viewed as distinct) for all
$j\ne i$ is $\lambdarexi(\sequ_{\exi} \ominus \sequ'_{\exi}) \defeq \prod_{j\ne i} \lambdar_j(u_j \ominus u'_j)$, and
the corresponding value of $\nu \defeq \nu(\SetCW; \lambda_{\except i})$ is denoted by $\nu_{\sequ'_{\exi}}$, which is given by (without normalization)
\begin{equation}
  \label{eq:linear-sym-2K-nu0}
  \begin{split}
  \nu_{\sequ'_{\exi}}(u) &= \sum_{\sequ''\in\SetCW: u''_i=u} \prod_{j\ne i} \lambda_j(u''_j) \\
  &= \sum_{\sequ''\in\mathcal{X}_d \oplus \sequ_0} \prod_{j\ne i} \lambdar_j(u''_j \ominus u'_j) \\
  &= \sum_{\sequ''\in\mathcal{X}_d \oplus \sequ_0} \lambdarexi(\sequ''_{\exi} \ominus \sequ'_{\exi}),
  \end{split}
\end{equation}
where we have let $u = d \oplus u_{0i}$ and $u_{0i} \in \SetZW_i$ is the $i$-th component of
$\sequ_0$ as usual.  Clearly, \eqref{eq:linear-sym-2K-nu0} is nonzero only for $d \in \SetD_i$ or
equivalently $u \in \SetCW_i$, and for any $\seqd \in \mathcal{X}_d$, it is easy to show that
\begin{equation}
  \label{eq:nu-cd-2K}
  \nu_{\sequ'_{\exi} \oplus \seqd_{\exi}} = \nu_{\sequ'_{\exi}} \oplus \constmsg{d};
\end{equation}
in other words,
\begin{equation}
  \label{eq:pnuc-2K}
  p(\nu \condmid \sequ) = \sum_{\sequ'_{\except i} \in \SetZW_{\exi}} \lambdarexi(\sequ_{\exi} \ominus \sequ'_{\exi}) \cdot
  \oneif{\nu = \nu_{\sequ'_{\exi}}}
\end{equation}
satisfies the invariant
\begin{equation}
  \label{eq:pnuc-2K-inv}
  p(\nu \condmid \sequ) = p_{\nu \condmid \sequ}(\nu \oplus \constmsg{d} \condmid \sequ \oplus \seqd), \quad \forall d\in \SetD_i,\ \seqd \in \mathcal{X}_{d}.
\end{equation}
From a fixed $\sequ \in \mathcal{C}$, as $\seqd$ ranges over
$\mathcal{X} = \cup_{d \in \SetD_i} \mathcal{X}_{d}$, $\sequ \oplus \seqd$ covers all
possible codewords in $\mathcal{C}$, thus \eqref{eq:pnuc-2K-inv} allows the entire
$p_{\nu \condmid \sequ}(\cdot \condmid \cdot)$ to be derived from its value for a single $\sequ$;
we can see that this $p(\nu \condmid \sequ)$ depends only on the $u_i$ component of $\sequ$, with
\begin{equation}
  \label{eq:pnuc-2K-markov}
  p(\nu \condmid \sequ) = p(\nu \condmid u_i) = p_{\nu \condmid u_i}(\nu \oplus \constmsg{d} \condmid u_i \oplus d),\quad
  \forall d\in\SetD_i.
\end{equation}
As $u_i$ can only take values in $\SetCW_i$, we may conclude from \eqref{eq:pnuc-2K-markov} that
both the Markov property for $\sequ \markov u_i \markov \nu$ and the symmetry condition
\eqref{eq:msg-sym-MK} are satisfied.  Moreover, note from \eqref{eq:nu-cd-2K} that $\nu_{\sequ'_{\exi}} = \nu_{\sequ'_{\exi} \oplus \seqd_{\exi}}$ for
any $\seqd \in \mathcal{X}_0$; without loss of generality, we may additionally assume that different $\nu_{\sequ'_{\exi}}$'s do not coincide when the difference in $\sequ'$
does not lie in $\mathcal{X}_0$ (otherwise only the normalization factor is affected), then from \eqref{eq:pnuc-2K} we can obtain the total conditional probability
of $\nu$ being a given $\nu_{\sequ'_{\exi}}$ as
\begin{equation}
  \begin{split}
    p(\nu = \nu_{\sequ'_{\exi}} \condmid u_i) &= p(\nu = \nu_{\sequ'_{\exi}} \condmid \sequ) \\
    &= \sum_{\seqd \in \mathcal{X}_0} \lambdarexi (\sequ_{\exi} \ominus \sequ'_{\exi} \oplus \seqd_{\exi}),
  \end{split}
\end{equation}
whose r.h.s.\ is simply (from \eqref{eq:linear-sym-2K-nu0}) $\nu_{\sequ'_{\exi}}(u)$ with
$u = d \oplus u_{0i} \in \SetCW_i$ if $\sequ \in \mathcal{X}_d \oplus \sequ_0$ or equivalently
$u_i = u$, hence the other symmetry condition \eqref{eq:msg-consistent-MK} is satisfied as well.
We have thus proved that $\nu$ has a symmetric density w.r.t.\ $u_i$ and $\sequ \markov u_i \markov \nu$ forms a Markov chain when the $\lambda_j$'s
have densities in the form of \eqref{eq:sym-basis-MK}.  As both properties are preserved in convex combinations, they remain true when the $\lambda_j$'s
have general symmetric densities.

Finally we prove that $\nu$ is a sufficient statistic for $u_i$, i.e.\
$p(u_i \condmid \nu) = p(u_i \condmid \lambda_{\exi})$ (note that the r.h.s.\ is equal to
$p(u_i \condmid \nu, \lambda_{\exi})$ because $\nu$ is a function of $\lambda_{\exi}$).  This
is where we need to use the uniformity of $p(\sequ)$ over $\SetCW$, which implies that
$p(u_i)$ is also uniform over $\SetCW_i$; under this condition, for any $u_i \in \SetCW_i$,
\begin{align}
  \label{eq:suff-stat-2K-0}
  p(u_i\condmid \nu) &\propto p(\nu\condmid u_i) \\
  \label{eq:suff-stat-2K-1}
  &\propto \nu(u_i) \\
  \label{eq:suff-stat-2K-2}
  &= \sum_{\sequ'\in\SetCW: u'_i=u_i} \prod_{j\ne i} \lambda_j(u'_j) \\
  \label{eq:suff-stat-2K-3}
  &\propto \sum_{\sequ'\in\SetCW: u'_i=u_i} \prod_{j\ne i} p_{\lambda_j\condmid u_j}(\lambda_j\condmid u'_j) \\
  \label{eq:suff-stat-2K-4}
  &\propto \sum_{\sequ'\in\SetCW: u'_i=u_i} p_{\sequ}(\sequ') \prod_{j\ne i} p_{\lambda_j\condmid u_j}(\lambda_j\condmid u'_j) \\
  \label{eq:suff-stat-2K-5}
  &= \sum_{\sequ'\in\SetCW: u'_i=u_i} p_{\sequ, \lambda_{\except i}}(\sequ', \lambda_{\except i}) \\
  \label{eq:suff-stat-2K-6}
  &= p(u_i, \lambda_{\except i}) \propto p(u_i\condmid \lambda_{\except i}),
\end{align}
where ``$\propto$'' means ``equal up to a factor that is the same for all
$u_i \in \SetCW_i$'', \eqref{eq:suff-stat-2K-1} and \eqref{eq:suff-stat-2K-3} use the
symmetry of resp.\ $\nu$ and $\lambda_{\exi}$'s density, while
\eqref{eq:suff-stat-2K-0} and \eqref{eq:suff-stat-2K-4} use the uniformity
of $u_i$ and $\sequ$ over respectively $\SetCW_i$ and $\SetCW$. \qed

\subsection{Proof of \prettyref{prop:linear-pdeg-2K}}
\label{app:proof-linear-pdeg-2K}
The known Markov-chain relationships among the random variables can be expressed as
\begin{equation}
  {\setlength{\arraycolsep}{0.222em}\begin{array}{ccccc}
    \sequ &\markov & \lambda_{\except i} &\markov & \lambda'_{\except i} \\
    | & & | & & | \\
    u_i & & \nu_i & & \nu'_i
  \end{array}},
\end{equation}
where every simple path in the graph forms a Markov chain.  Therefore, we can formally write (the
summations over $\lambda_{\except i}$ may represent integrals)
\begin{equation}
  \begin{split}
    p(\nu'_i \condmid \nu_i, u_i) &= \sum_{\lambda_{\except i}} p(\nu'_i, \lambda_{\except i} \condmid \nu_i, u_i) \\
    &= \sum_{\lambda_{\except i}} p(\nu'_i \condmid \lambda_{\except i}, \nu_i, u_i)
    p(\lambda_{\except i} \condmid \nu_i, u_i),
  \end{split}
\end{equation}
where $p(\nu'_i \condmid \lambda_{\except i}, \nu_i, u_i) = p(\nu'_i
\condmid \lambda_{\except i})$ is evident from the figure above, while
$p(\lambda_{\except i} \condmid \nu_i, u_i) = p(\lambda_{\except i}
\condmid \nu_i)$ comes from \prettyref{prop:linear-sym-suff-stat-2K}.  We have thus shown
that $p(\nu'_i \condmid \nu_i, u_i)$ does not depend on the value of
$u_i$, making $u_i \markov \nu_i \markov \nu'_i$ a Markov chain. \qed

\subsection{Proof of \prettyref{prop:sym-chan-sym-likelihood-M}}
\label{app:proof-sym-chan-sym-likelihood-M}
%
As $\psi_u(\cdot)$ is a group action, it is a bijection for any $u \in \SetG$ and partitions
$\SetY$ into orbits $\SetY=\cup_{\alpha}\SetYa$, where each orbit $\SetYa$ is a discrete set
$\{\psi_u^{-1}(y_{0\alpha}) \condmid u\in \SetG\}$ for some deterministic $y_{0\alpha} \in \SetY$.

We can first consider the case where $\SetY$ is a discrete set containing a single orbit
$\{ y_u \defeq \psi_u^{-1}(y_0) \condmid u\in\SetG \}$ for some $y_0$, such that
the conditional pmf has the form $\pyu(y\condmid 0) = \sum_{u'} p_{u'} \cdot \oneif{y = y_{u'}}$ (with $\sum_{u'} p_{u'} = 1$),
and by \eqref{eq:chan-sym-M}, $p(y\condmid u) = \sum_{u'} p_{u'\ominus u} \cdot \oneif{y = y_{u'}}$.
The use of $\oneif{\cdot}$ here allows for duplications among the $y_u$'s; such duplications can be characterized
by the stabilizer subgroup $\SetH$ of the group action, which is the same over the entire orbit since
$\SetG$ is abelian.  The normalized $\lambda$ corresponding to a given $y$ is then
$\lambda(u) = (1/\cardinal{\SetH}) \sum_{u'} p_{u'\ominus u} \cdot \oneif{y = y_{u'}}$, and for each $u''\in\SetG$, when
$y=y_{u''}$ this $\lambda$ is denoted $\lambda_{u''}$.  It is easy to find that
$\lambda_0(u) = (1/\cardinal{\SetH}) \sum_{u'\in \SetH} p_{u'\ominus u}$,
$\lambda_{u''} = \lambda_0 \oplus \constmsg{u''}$, and for any $u'\in\SetH$ we also have $y_{u''} = y_{u''\oplus u'}$ and thus
$\lambda_{u''} = \lambda_{u''\oplus u'}$.  Consequently,
\begin{equation}
  \label{eq:plu-chan0}
  \begin{split}
  p(\lambda \condmid u) &= \sum_{u'} p_{u'\ominus u} \cdot \oneif{\lambda = \lambda_{u'}} \\
    &= (1/\cardinal{\SetH}) \sum_{u''\in \SetH} \sum_{u'} p_{u'\ominus u \oplus u''} \cdot \oneif{\lambda = \lambda_{u'}} \\
    &= \sum_{u'} \lambda_0(u\ominus u') \cdot \oneif{\lambda = \lambda_{u'}},
  \end{split}
\end{equation}
which has the form of \eqref{eq:sym-basis-MK}, so $\lambda$ has a symmetric density w.r.t.\ $u$.

For more general $\SetY$ and channel $p(y\condmid u)$ satisfying \eqref{eq:chan-sym-M}, we can let
$\Eventa$ be the event that $y\in \SetYa$, and define
$p_{\alpha}(y\condmid u) \defeq p_{y\condmid u, \Eventa}(y\condmid u)$ as the pmf conditioned on
each $\Eventa$, so that $p(y\condmid u)$ can be viewed as a convex combination (or time-sharing) of
channels $p_{\alpha}(y\condmid u)$, each with a discrete output alphabet $\SetYa$; here summation of \eqref{eq:chan-sym-M} over $y\in\SetYa$
gives $p(\Eventa\condmid u) = p(\Eventa\condmid 0)$ (both viewed as pdfs), so the required independence between $\Eventa$ and $u$ is satisfied.
For any $y\in\SetYa$, $p_{\alpha}(y \condmid u) = p(y\condmid u) / p(\Eventa\condmid u)$ with $p(\Eventa\condmid u)$ not varying with $u$, so the $\lambda$ computed from
$p(y\condmid u)$ and from $p_{\alpha}(y\condmid u)$ are identical.  By the above argument, each
$p_{\alpha}(\cdot \condmid \cdot)$ yields a symmetric density for $\lambda$, while the overall
density of $\lambda$ is a convex combination of these densities and thus symmetric as well. \qed

\subsection{Proof of \prettyref{prop:sym-M-sym-2K}}
\label{app:proof-sym-M-sym-2K}
We only need to consider the case that $p(\lambda \condmid u)$ has the form of \eqref{eq:sym-basis-MK}, i.e.
\begin{equation}
  p(\lambda \condmid u) = \sum_{u'} \lambdar(u \ominus u') \cdot \oneif{\lambda = \lambdar \oplus \constmsg{u'}}.
\end{equation}
Transforming $\lambda$ and $u$ into $\mu \defeq \lambdaphi$ and $\seqct \defeq \invphi(u)$, then they are still independent from $\seqeps$, and
\begin{equation}
  \label{eq:M-2K-pmuc}
  \begin{split}
    p(\mu \condmid \seqct) &= \sum_{u'} \lambdar(\phi(\seqct) \ominus u') \cdot
    \oneif{\mu = (\lambdar \oplus \constmsg{u'}) \compose \phi} \\
    &= \sum_{u'} \mur_{u'}(\seqct) \cdot \oneif{\mu = \mur_{u'}},
  \end{split}
\end{equation}
where we have defined $\mur_{u'} \defeq (\lambdar \oplus \constmsg{u'}) \compose \phi$.
Eq.~\eqref{eq:M-2K-pmuc} shows that $\mu$ does not necessarily have a symmetric density w.r.t.\
$\seqct$, thus the necessity of $\seqeps$.  On the other hand, now
$\mu_1 \defeq \lambdaphi_1 = \mu \ominus \constmsg{\seqeps}$ (here $\mu_1$, $\mu$ and $\constmsg{\seqeps}$ are probability tuples over $\SetZ_2^K$) and
$\seqct_1 \defeq \invphi_1(u) = \seqct \ominus \seqeps$, and since $u$ is uniformly distributed
over $\SetG$, we also have $p(\seqct) = p(\seqct_1) = 1/\cardinal{\SetG}$, so
\begin{equation}
  \begin{split}
  p(\mu_1, \seqct_1 \condmid \seqeps)
  &= p_{\mu, \seqct}(\mu_1 \oplus \constmsg{\seqeps}, \seqct_1 \oplus \seqeps) \\
  &= \frac{1}{\cardinal{\SetG}} \sum_{u'} \mur_{u'}(\seqct_1 \oplus \seqeps) \cdot \oneif{\mu_1 = \mur_{u'} \ominus \constmsg{\seqeps}},
  \end{split}
\end{equation}
and marginalizing over $\seqeps$ yields
\begin{equation}
  p(\mu_1 \condmid \seqct_1)
  = \frac{1}{\cardinal{\SetG}} \sum_{u'} \sum_{\seqeps} \mur_{u'}(\seqct_1 \oplus \seqeps) \cdot \oneif{\mu_1 = \mur_{u'} \ominus \constmsg{\seqeps}}.
\end{equation}
The symmetry of $\mu_1$ w.r.t.\ $\seqct_1$ is now obvious, as each term in the summation over $u'$
corresponds to a symmetric density in the form of \eqref{eq:sym-basis-MK}, and the summation creates a convex
combination of these densities. \qed

\subsection{Proof of \prettyref{prop:sym-2K-sym-M}}
\label{app:proof-sym-2K-sym-M}
It is only necessary to consider the case that
\begin{equation}
  p(\mu \condmid \seqct) = \sum_{\seqct'} \mur(\seqct \ominus \seqct') \cdot \oneif{\mu = \mur \oplus \constmsg{\seqct'}}.
\end{equation}
Now transform $\mu$ and $\seqct$ into respectively $\lambda \defeq \mu \compose \invphi$ and $u \defeq \phi(\seqct)$ such that
$\lambda_1 \defeq \mu \compose \invphi_1 = \lambda \oplus \constmsg{\delta}$ and
$u_1 \defeq \phi_1(\seqct) = u \oplus \delta$.  $\lambda$ and $u$ thus remain independent from
$\delta$, with
\begin{equation}
  \begin{split}
    p(\lambda \condmid u) &= \sum_{\seqct'} \mur(\invphi(u) \ominus \seqct')
    \cdot \oneif{\lambda = (\mur \oplus \constmsg{\seqct'}) \compose \invphi} \\
    &= \sum_{\seqct'} \lambdar_{\seqct'}(u) \cdot \oneif{\lambda = \lambdar_{\seqct'}},
  \end{split}
\end{equation}
where $\lambdar_{\seqct'} \defeq (\mur \oplus \constmsg{\seqct'}) \compose \invphi$.  Since
$\seqct$ is uniformly distributed over $\SetZ_2^K$, we have $p(u) = p(u_1) = 1/\cardinal{\SetG}$, so
\begin{equation}
  \begin{split}
    p(\lambda_1, u_1 \condmid \delta) &= p_{\lambda,u}(\lambda_1 \ominus \constmsg{\delta}, u_1 \ominus \delta) \\
    &= \frac{1}{\cardinal{\SetG}} \sum_{\seqct'} \lambdar_{\seqct'}(u_1\ominus \delta) \cdot \oneif{\lambda_1 = \lambdar_{\seqct'} \oplus \constmsg{\delta}},
  \end{split}
\end{equation}
and marginalizing over $\delta$ yields
\begin{equation}
  p(\lambda_1 \condmid u_1) = \frac{1}{\cardinal{\SetG}} \sum_{\seqct'} \sum_{\delta}
  \lambdar_{\seqct'}(u_1\ominus \delta) \cdot \oneif{\lambda_1 = \lambdar_{\seqct'} \oplus \constmsg{\delta}},
\end{equation}
which is a convex combination of symmetric densities and thus symmetric. \qed

\bibliographystyle{IEEEtran}
\bibliography{IEEEabrv,ldgm-vq}



\end{document}